%% file: RationalKernels_v3.tex
\documentclass[12pt]{article}
\pdfoutput=1
\usepackage{jheppub}

\usepackage{fontawesome5}
\usepackage{comment}
\usepackage[normalem]{ulem}
\usepackage{hyperref,amsfonts}
\usepackage{amsmath}
\hypersetup{
 colorlinks=true,
 urlcolor=blue,
 anchorcolor=blue,
 citecolor=blue,
 filecolor=blue,
 linkcolor=blue,
 menucolor=blue,
 pagecolor=blue,
 linktocpage=true,}
\usepackage{graphicx,color,subfig}
\usepackage{esint}
\usepackage{empheq}
\usepackage{bbold}
\usepackage[usenames,dvipsnames,table]{xcolor}
\graphicspath{{./figures/}}
\usepackage[dvipsnames]{xcolor}
\usepackage{tikz}
\usepackage{braket}
\usepackage{cancel}
\usepackage{mathtools}

\usepackage{pifont}

\usepackage{float}

\renewcommand{\arraystretch}{1.3} 
\newcommand{\be}{\begin{equation}}
\newcommand{\ee}{\end{equation}}
\newcommand{\bea}{\begin{aligned}}
\newcommand{\eea}{\end{aligned}}
\newcommand{\sixjnorm}[6]{\begin{vmatrix}
#1 & #2 & #3 \\
#4 & #5 & #6
\end{vmatrix}
}
\definecolor{light-gray}{gray}{0.75}
\definecolor{mygreen}{RGB}{220,230,220}

\usepackage{wasysym}

\newcommand{\D}{\mathcal{D}}

\definecolor{myforestgreen}{RGB}{24, 150, 144}

\begin{document}

\def\b{\mathsf{b}}
\def\s{\mathsf{s}}

\def\Re{\mathrm{Re}}
\def\Im{\mathrm{Im}} 
 \def\M{{\mathcal M}}
\def\hat{\widehat}
 \def\tilde{\widetilde}
\def\V{{\mathcal V}}
\def\C{{\mathbb C}}
\def\CC{{\mathcal C}}
\def\F{{\mathcal F}}
\def\O{{\mathcal O}} 
\def\Z{{\mathbb Z}}
\def\Tr{{\mathrm{Tr}}}
\def\R{{\mathbb R}}
\def\RR{{\mathcal R}}
\def\S{{\mathcal S}}
\def\I{{\mathcal I}}
\def\D{{\mathcal D}}
\def\A{{\mathcal A}}
\def\bar{\overline}
\def\T{{\mathcal T}}
\def\J{{\mathcal J}}
\def\tilde{\widetilde}
\newcommand{\modS}{\mathbb{S}}	 
\newcommand{\fusion}{\mathbb{F}}	 
\newcommand{\dd}{\text{d}}

\newcommand{\JR}[1]{{\color{red}{#1}}}

\title{On the Virasoro Crossing Kernels \\ at Rational Central Charge}
\author[\clubsuit]{Julien Roussillon}
\author[\spadesuit,\eighthnote]{Ioannis Tsiares}
\affiliation[\clubsuit]{Department of Mathematics and Systems Analysis,
P.O. Box 11100, FI-00076, Aalto University, Finland}
\affiliation[\spadesuit]{Universit\'e Paris-Saclay, CNRS, CEA, Institut de Physique Th\'eorique, \\91191, Gif-sur-Yvette, France}
\affiliation[\eighthnote]{Laboratoire de Physique, \'Ecole Normale Sup\'erieure, \\
   Universit{\'e} PSL, CNRS, Sorbonne Universit{\'e}, Universit{\'e} Paris Cit{\'e}, \\
   24 rue Lhomond, F-75005 Paris, France}
\emailAdd{jujuroussillon@gmail.com, ioannis.tsiares@phys.ens.fr}
\abstract{We report novel analytic results for the Virasoro modular and fusion kernels relevant to 2d conformal field theories (CFTs), 3d topological field theories (TQFTs), and the representation theory of certain quantum groups. For all rational values of the parameter $b^2\in\mathbb{Q}^{\times}$ -- corresponding in 2d CFT to all rational central charge values in the domain $(-\infty,1]\cup[25,\infty)$ -- we establish two main results. First, in the domain $c\in\mathbb{Q}_{[25,\infty)}$ we show that the modular and fusion kernels derived by Teschner and Teschner-Vartanov respectively can be expressed as a linear combination of two functions, which (i) are themselves admissible crossing kernels, (ii) have square-root branch point singularities in the Liouville momenta, (iii) are not reflection-symmetric in the Liouville momenta. These features illustrate that the space of solutions to the basic shift relations determining these kernels is broader than previously assumed. Second, in the domain $c\in\mathbb{Q}_{(-\infty,1]}$ we derive for the first time the physical modular and fusion kernels for generic values of the Liouville momenta. These can again be written as a linear combination of two other admissible kernels but overall, and unlike the Teschner and Teschner-Vartanov solutions for $c\geq 25$, they possess square-root branch point singularities. As a corollary, we demonstrate that timelike Liouville theory at $c\in\mathbb{Q}_{(-\infty,1]}$ is crossing symmetric and modular covariant. Surprisingly, the crossing kernels at any $b^2\in\mathbb{Q}^{\times}$ behave as if they were semiclassical and one-loop exact, and we discuss the interpretation of this fact in the context of the 2d conformal bootstrap and the 3d TQFT that captures pure 3d gravity with negative cosmological constant.}
\maketitle
\section{Introduction and Summary} 

Two-dimensional Conformal Field Theories (CFTs)  lie at a rare confluence of physics and mathematics. On the physics side, they capture a wide range of phenomena: from the universality of statistical systems at criticality,  to the dynamics of strings in the worldsheet formulation of string theory. In the AdS/CFT correspondence \cite{Maldacena:1997re}, 2d CFTs encode in a highly non-trivial fashion the quantum gravitational degrees of freedom in a three-dimensional spacetime with negative cosmological constant. On the mathematics side, the presence of (the infinite dimensional) Virasoro symmetry has revealed many interesting connections with the representation theory of certain quantum groups \cite{Faddeev:1999fe}, the algebraic geometry of Riemann surfaces, surprising relations with $\mathcal{N}=2$ gauge theories in four dimensions \cite{Alday:2009aq} and topological field theories in three dimensions \cite{Witten:1988hf,Collier:2023fwi,Hartman:2025cyj}, as well as more recently with probabilistic constructions for Virasoro conformal blocks and the actual path integral for certain 2d CFTs \cite{Guillarmou:2020wbo,Ghosal:2020sis,Guillarmou:2024lqk}. 
\par The purpose of the present work is to study a particular quantity that plays a pivotal role primarily in the Virasoro conformal bootstrap -- namely, the crossing kernel of Virasoro conformal blocks. We will be interested in the crossing kernels on two particular Riemann surfaces: the torus with one marked point (modular kernel) and the sphere with four marked points (fusion kernel). The precise definitions of these kernels are given in sections \ref{sec:defkern1} and \ref{sec:defkern2}. 
\par The Virasoro modular and fusion kernels are mostly known through integral representations, due to the seminal works of Teschner \cite{Teschner:2003at}, Ponsot-Teschner \cite{Ponsot:2000mt}, and Teschner-Vartanov \cite{Teschner:2012em}. As we will review later (also in Appendix \ref{appendixA}), those formulas are build out of special meromorphic functions such as the Barnes double gamma function $\Gamma_b$ and the double sine function $S_b$ -- the latter being closely related to Faddeev's quantum dilogarithm \cite{Faddeev:1999fe} -- and are realized as solutions to particular shift relations (or, difference equations). They also depend in a specific way on the central charge via $c=1+6(b+b^{-1})^2$. On the other hand, non-integral representations for the Virasoro kernels are known only for the cases $c=1$ and $c=25$. At $c=1$, there is an interesting relation to the Painlev\'e VI connection constant \cite{Iorgov:2013uoa,DelMonte:2025vqv}. More recently, Ribault and one of the present authors showed that the fusion kernel at $c=25$ is directly related to the one at $c=1$ via a particular symmetry of the shift relations called \emph{Virasoro-Wick Rotation} (see section \ref{sec:VWR})\cite{Ribault:2023vqs}. Roughly speaking, this symmetry maps a solution of the corresponding shift relations valid at a given $c\in[25,\infty)$ to another solution valid at $26-c \in (-\infty,1]$. 
\par The aim of this paper is to generalize the $c=1$ and $c=25$ results and provide non-integral representations for both the modular and fusion kernels at \emph{any rational} central charge in the domain $c \in (-\infty,1] \cup [25,+\infty)$. As we will explain in detail, at rational central charge the Barnes double gamma function and the double sine function reduce to (ratios and products of) the Barnes' $G$ function, which captures the analytic continuation of the superfactorial. Schematically, 
\be
\bea
\Gamma_b(z)&\longrightarrow \prod_{r=0}^{m-1}\prod_{s=0}^{n-1}G\left(\frac{z}{\sqrt{mn}}+\frac{r}{m}+\frac{s}{n}\right)^{-1},\\
S_b(z)&\longrightarrow \prod _{k=0}^{m-1} \prod _{l=0}^{n-1} \frac{G\left(\frac{k+1}{m}+\frac{l+1}{n}-\frac{z}{\sqrt{m n}}\right)}{G\left(\frac{k}{m}+\frac{l}{n}+\frac{z}{\sqrt{m n}}\right)}.
\eea
\ee
Here $m,n$ is a pair of coprime positive integers that determine the (rational) central charge as $c=13+6\left(\frac{m}{n}+\frac{n}{m}\right)$ for $c\geq25$, or as $c=13-6\left(\frac{m}{n}+\frac{n}{m}\right)$ for $c\leq1$. See Appendix \ref{appendixA} for more precise definitions and explanations.
\par What is intricate about the appearance of products of Barnes' $G$ functions in the expressions for the kernels is that they obey a certain (quasi-)periodicity property in their argument. This allows us to start from the original integral representation of the modular and fusion kernels and compute these integrals \textit{in closed form}. Indeed, our computations commence with a critical, though very simple, Lemma first presented by Garoufalidis and Kashaev in \cite{Garoufalidis:2014ifa} in relation to the so-called state integrals that appear in abundance in quantum topology. We state it here for completeness\footnote{Onwards, we will be referring to it as ``GK lemma''. }.
\\
\cite[Lemma 2.1]{Garoufalidis:2014ifa}: Let $a \in \mathbb{C} \backslash \{0\}$ and $\mathcal U$ be a translationally invariant open set, that is $\mathcal U = a + \mathcal U$. Moreover, let $f: \mathcal U \to \mathbb C$ be an analytic function which satisfies the following ``quasi-periodicity'' relation
\begin{equation}
    f(z-a) f(z+a) = f(z)^2.
\end{equation}
Then, for an oriented path $\mathcal C \subset \mathcal U$ such that $f(z)(f(z) - f(z+a)) \neq 0$ for all $z \in \mathcal{C}$, we have the identity
\begin{equation}
    \int_\mathcal{C} f(z)dz = \left(\int_{\mathcal C} - \int_{a + \mathcal C} \right) \frac{f(z)}{1 - \frac{f(a+z)}{f(z)}} dz. \qquad \square
\end{equation}
\par The proof is straightforward and is described in \cite{Garoufalidis:2014ifa}. We can illustrate the utility of this lemma with a familiar example\footnote{IT thanks Davide Saccardo for discussions on this.}. Consider the usual Gaussian integral
\be
\int_{i\mathbb{R}} \frac{\dd x}{i} \ e^{x^2}=\sqrt{\pi}.
\ee
Another way to calculate it is to realize that the integrand is a quasi-periodic function satisfying
\be
f(x+a)f(x-a)=f(x)^2\, \quad \quad \text{with} \ a\equiv \sqrt{\frac{\pi}{2}}\left(1+i\right).
\ee
Applying the GK lemma we get
\be
\bea
\int_{i\mathbb{R}} \frac{\dd x}{i} \ e^{x^2}=\left(\int_{i\mathbb{R}} -\int_{i\mathbb{R}+\sqrt{\frac{\pi}{2}}}\right)\frac{\dd x}{i} \ \frac{e^{x^2}}{1-e^{a^2+2xa}}.
\eea
\ee
We can now evaluate the integral by closing the contour at infinity and picking up the relevant poles which in general are located at $x=\frac{a}{2}\times (2m+1), \ m\in\mathbb{Z}$.
It is easy to see that there is actually only a \textit{single} pole located inside the strip of interest, namely for $m=0$. Therefore we get
\be
\bea
\ointclockwise_{\mathcal{C}} \frac{\dd x}{i} \ \frac{e^{x^2}}{1-e^{a^2+2xa}}=-2\pi  \text{Res}_{x=\frac{a}{2}}\frac{e^{x^2}}{1-e^{a^2+2xa}}=\sqrt{\pi}.
\eea
\ee
\par Our calculations with the modular and fusion kernels are essentially more sophisticated instances of this simple example. However, the physical interpretations of the results have far-reaching implications, and we dedicate a great portion of the paper exploring these consequences in detail. 
\par We continue the rest of this section by introducing the notations and definitions of the quantities that we will use throughout the paper. The summary of results and the organization of the paper are presented in section \ref{sec:summaryandorg}.

\subsection{Notations}

We will use the following notations throughout the paper. \\
\\
\textit{Central charge}:\\
As usual, for $Q\equiv b+b^{-1}$ we parametrize
\begin{equation}
\begin{aligned}
c=1+6Q^2=\begin{cases}
13+6\left(b^2+b^{-2}\right)\ , \ \ \ \ \ \ c\geq25\\
13-6\left(\hat{b}^{2}+\hat{b}^{-2}\right)\ , \ \ \ \ \ c\leq1
\end{cases}.
\end{aligned}
\end{equation}
\begin{itemize}
\item For $c\in\mathbb{C}\backslash(-\infty,1]$ the parameter $b$ takes values in $\mathbb{C}\backslash i\mathbb{R}$. We will mostly focus on the range $c\in[25,\infty)$, and hence we choose $b\in\mathbb{R}_{(0,1]}$. \item For $c\in(-\infty,1]$ the parameter $b$ is purely imaginary, and hence we write $b=i\hat{b}$, with $\hat{b}\in\mathbb{R}_{(0,1]}$.
\end{itemize}
\
\\
\textit{Chiral conformal dimensions}:\\
\be
\bea
h=\frac{Q^2}{4}-P^2=\begin{cases} \frac{(b+b^{-1})^2}{4}-P^2 \ , \ \ \ \ \ \ c\geq25\\
\\
-\frac{(\hat{b}^{-1}-\hat{b})^2}{4}-P^2\ , \ \ \ \ c\leq1
\end{cases} .
\eea
\ee
\\
\textit{Tetrahedron notation for the fusion kernel}:
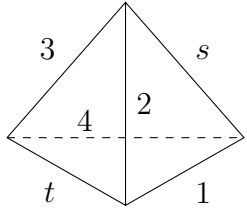
\begin{align}
 \begin{tikzpicture}[baseline = (current bounding box.center), scale = .45]
  \draw (0, 0) -- node[below right]{$1$} (3.5, 2)  -- node[above right] {$s$}(0, 6) -- node[right]{$2$} (0, 0) -- node[below left] {$t$} (-3.5, 2) -- node[above left] {$3$} (0, 6);
  \draw[dashed] (-3.5, 2) -- (3.5, 2);
  \node at (-1.2, 2.5) {$4$};
 \end{tikzpicture}
\quad \quad \quad \     
\renewcommand{\arraystretch}{1.3}
 \begin{array}{|l|c|l|}
  \hline 
  \text{Name} & \text{Notation} & \text{Value}
  \\
  \hline\hline 
  \text{Edges} & E & \{1,2,3,4,s,t\} 
  \\
  \text{Pairs of opposite edges} & P & \{13,24,st\}
  \\
  \text{Faces} & F & \{12s,34s,23t,14t\}
  \\
  \text{Vertices} & V &  \{14s,12t,34t,23s\}
  \\
  \hline 
 \end{array}
 \label{tetra}
\end{align}
Formulas will involve assigning signs to edges. We use the notations:
\begin{itemize}
 \item $\sigma\in\mathbb{Z}_2^E$ is an assignment of a sign $\sigma_i\in\{+,-\}$ for any $i\in E$, and $\sigma\in \mathbb{Z}_2^f$ for a triple of signs on a face $f\in F$.
 \item $\sigma_E,\sigma_v,\sigma_f,\sigma_p$ for products of $6,3,3$ or $2$ signs on all edges, a vertex, a face, or two opposite edges.
 \item $\sigma\in\mathbb{Z}_2^E|\sigma_V=1$ for sign assignments whose products are $1$ at each vertex. There are $8$ such assignments, and they can be split in two halves according to $\sigma_E=\pm 1$.
 \item The indicator function $\eta_i\in \mathbb{Z}_2^F$ is $\eta_i(f)=1$ if the edge $i$ belongs to the face $f$, and $\eta_i(f)=-1$ otherwise. 
\end{itemize}
To be more explicit, below is the set $\sigma\in\mathbb{Z}_2^E|\sigma_V=1$:
\begin{align}\label{tbl:2kalos}
\renewcommand{\arraystretch}{0.8}
 \begin{array}{|cccccc|c|}
 \hline 
  s & t & 1 & 2 & 3 & 4 & \sigma_E 
  \\
  \hline 
   - & + & - & - & + & + & - 
  \\
 - & + & + & + & - & - & - 
\\
 + & - & - & + & + & - & -  
\\
 + & - & + & - & - & + & -  
\\
\hline 
 + & + & - & - & - & - & +  
\\
 + & + & + & + & + & + & + 
\\
 - & - & - & + & - & + & + 
\\
 - & - & + & - & + & - & + 
\\
\hline 
 \end{array}
\end{align}
Example: When we write (c.f. (\ref{eq:prefintegdF}))
\be
\prod_{\substack{\sigma\in \mathbb{Z}_2^E|\\ \sigma_V=1}} 
 S_b\left(u+\tfrac{Q}{2} +\tfrac{Q}{4}\sigma_E+\tfrac12\textstyle{\sum}_{i\in E}\sigma_iP_i\right)^{-\sigma_E},
\ee
this is equal to
\be
\bea
\frac{S_b\left(u+\frac{Q}{4}+\frac{1}{2}P_{34t|12s}\right)S_b\left(u+\frac{Q}{4}+\frac{1}{2}P_{12t|34s}\right)S_b\left(u+\frac{Q}{4}+\frac{1}{2}P_{23s|14t}\right)S_b\left(u+\frac{Q}{4}+\frac{1}{2}P_{14s|23t}\right)}{S_b\left(u+\frac{3Q}{4}+\frac{1}{2}P_{st|1234}\right)S_b\left(u+\frac{3Q}{4}+\frac{1}{2}P_{st1234}\right)S_b\left(u+\frac{3Q}{4}+\frac{1}{2}P_{24|st13}\right)S_b\left(u+\frac{3Q}{4}+\frac{1}{2}P_{13|st24}\right)},
\eea
\ee
where we adopted the convenient notation 
$$P_{I|J}\equiv \sum_{i\in I}P_i-\sum_{j\in J}P_j.
$$
\subsection{Definition of Crossing Kernels for $c\in\mathbb{C}\backslash(-\infty,1]$}\label{sec:defkern1}

\paragraph{Modular Kernel.} The modular kernel implements the S-transform of torus one-point Virasoro conformal blocks according to the following defining relation:
\be\label{eq:defcrossingM}
\bea
 (-i\tau)^{h_0}\mathcal{F}^{(b), \tau}_{P_s} &= \int_{i\mathbb{R}}\frac{dP_t}{i}\ \mathbf{M}^{(b)}_{P_s,P_t}[P_0] \ \mathcal{F}^{(b), -1/\tau}_{P_t}, 
\eea
\ee
where $\tau$ is the modular parameter on the torus, and the blocks are normalized as $\mathcal{F}^{(b), \tau}_{P_s}=q^{h_{s}-\frac{c}{24}}\left(1+o(q)\right)$ with $q=e^{2\pi i \tau}$. 
\par Teschner \cite{Teschner:2003at} gave a very explicit formula for this kernel, valid at central charge $c\in\mathbb{C}\backslash(-\infty,1]$, as a meromorphic and even (i.e. reflection-symmetric) function of the various momenta
\be
\bea \label{eq:defmodularkernel}
 & \mathbf{M}^{(b)}_{P_s,P_t}[P_0] :=  M_b(P_0|P_s,P_t)
 \int_{i\mathbb{R}} \frac{du}{i} \ m_b(u)\ .
\eea
\ee
The prefactor and integrand\footnote{The integrand is also a function of $P_0,P_s,P_t$ but we suppress this dependence for brevity.} are expressed in terms of the Barnes double gamma function $\Gamma_b(x)$ and the double sine function $S_b(x)$ (c.f. Appendix \ref{appendixA}) as follows
\be\label{eq:prefintegdM}
\bea
M_b(P_0|P_s,P_t)&=\frac{\rho^{(b)}_0(P_t)}{2S_b(\frac{Q}{2}+P_0)}
\prod_\pm \frac{\Gamma_b(Q\pm 2P_s)}{\Gamma_b(Q\pm 2P_t)} \frac{\Gamma_b(\frac{Q}{2}-P_0\pm 2P_t)}{\Gamma_b(\frac{Q}{2}-P_0\pm 2P_s)} , \\
m_b(u)&=e^{4\pi iP_su}\prod_{\pm}\frac{S_b\left(u+\frac{Q}{4}+\frac{1}{2}\left(\pm 2P_t+P_0\right)\right)}{S_b\left(u+\frac{3Q}{4}+\frac{1}{2}\left(\pm 2P_t-P_0\right)\right)} \ .
\eea
\ee
Here $\rho^{(b)}_0(P)\equiv -4\sqrt{2}\sin{(2\pi b P)\sin{(2\pi b^{-1}P)}}$ is the Plancherel measure of the modular double of the quantum group $U_q(\mathfrak{sl}_2)$\cite{Ponsot:2000mt}. The integral (\ref{eq:defmodularkernel}) converges so long as\footnote{Outside this range, we define the kernel via its shift relations.}
\be\label{eq:condmodkernelintegral}
\bea
|\text{Re}P_s|<\frac{\text{Re}a_0}{2}, \quad a_0\equiv \frac{Q}{2}-P_0.
\eea
\ee
For $P_0,P_s,P_t\in i\mathbb{R}$ the kernel (\ref{eq:defmodularkernel}) is real. It can be shown that the expression (\ref{eq:defmodularkernel}) is the \textit{unique meromorphic} solution to the modular kernel shift relations for $b\in\mathbb{R}_{(0,1]}$ \cite{Eberhardt:2023mrq} which arise a consequence of the Moore-Seiberg consistency conditions \cite{Moore:1988uz}.
\paragraph{Fusion Kernel.}The fusion kernel implements crossing transformations of sphere four-point Virasoro conformal blocks according to the following defining relation:
\be\label{eq:defcrossingF}
\bea
 \mathcal{F}^{(b), s-\text{channel}}_{P_s} = \int_{i\mathbb{R}} \frac{dP_t}{i}\ \mathbf{F}^{(b)}_{P_s,P_t}\left[\begin{smallmatrix} P_2 & P_3 \\ P_1 & P_4 \end{smallmatrix}\right]  \mathcal{F}^{(b), t-\text{channel}}_{P_t} \ ,
\eea
\ee
where the blocks are normalized as $\mathcal{F}^{(b), s-\text{channel}}_{P_s}(z) = z^{h_s-h_1-h_2}\left(1+ O(z)\right)$.
\par Teschner and Vartanov \cite{Teschner:2012em} gave a very explicit formula for this kernel, valid at central charge $c\in\mathbb{C}\backslash(-\infty,1]$, as a meromorphic and even function of the various momenta\footnote{Note that compared to the standard definition (see e.g. \cite[(3.50)]{Eberhardt:2023mrq}) we have shifted the integration variable in the integral as $p_{\text{\cite{Eberhardt:2023mrq}}}\equiv u+\frac{Q}{4}+\frac{1}{2}P_{1234st}$ (using the convenient notation described above). }
\be
\bea \label{eq:deffusionkernel}
 & \mathbf{F}^{(b)}_{P_s,P_t}\left[\begin{smallmatrix} P_2 & P_3 \\ P_1 & P_4 \end{smallmatrix}\right] :=  F_b(P_i|P_s,P_t)
 \int_{i\mathbb{R}} \frac{du}{i} \ f_b(u)\ .
\eea
\ee
The prefactor and integrand read
\be\label{eq:prefintegdF}
\bea
F_b(P_i|P_s,P_t)&=\frac{\Gamma_b(Q\pm 2P_s)}{2\Gamma_b(\pm 2P_t)}\prod_{f\in F} \prod_{\substack{\sigma\in \mathbb{Z}_2^f| \\ \sigma_f = \eta_t(f)}} \Gamma_b\left(\tfrac{Q}{2}+\textstyle{\sum}_{i\in f}\sigma_iP_i\right)^{\sigma_f} ,\\
f_b(u)&=\prod_{\substack{\sigma\in \mathbb{Z}_2^E|\\ \sigma_V=1}} 
 S_b\left(u+\tfrac{Q}{2} +\tfrac{Q}{4}\sigma_E+\tfrac12\textstyle{\sum}_{i\in E}\sigma_iP_i\right)^{-\sigma_E}.
\eea
\ee
This normalization of the fusion kernel makes manifest its (almost) tetrahedral symmetry but obscures the reflection symmetries in the momenta. We also have the obvious symmetries
\be\label{eq:symmF}
\bea
\mathbf{F}^{(b)}_{P_s,P_t}\left[\begin{smallmatrix} P_2 & P_3 \\ P_1 & P_4 \end{smallmatrix}\right] =\mathbf{F}^{(b)}_{P_s,P_t}\left[\begin{smallmatrix} P_1 & P_4 \\ P_2 & P_3 \end{smallmatrix}\right] =\mathbf{F}^{(b)}_{P_s,P_t}\left[\begin{smallmatrix} P_3 & P_2 \\ P_4 & P_1 \end{smallmatrix}\right].
\eea
\ee 
\par The integral (\ref{eq:deffusionkernel}) is well-defined and convergent with an exponential suppression. Indeed, as e.g. described in \cite{Eberhardt:2023mrq},  for $u=ix$ the integrand behaves as
\be\label{eq:condfuskernelintegral}
\bea
\left(e^{2\pi i \left(P_1P_4+P_2P_s+P_3P_t\right)-\pi i Q\sum_{j=\{1,\cdots,4,s,t\}}P_j}\right)\times e^{-2\pi Qx}, \quad \quad \Re(x)\rightarrow+\infty
\eea
\ee
and analogously for $\Re(x)\rightarrow-\infty$. It can be shown that the expression (\ref{eq:deffusionkernel}) is the \textit{unique meromorphic} solution to the fusion kernel shift relations for $b\in\mathbb{R}_{(0,1]}$ \cite{Eberhardt:2023mrq} as a consequence of the Moore-Seiberg consistency conditions.
\subsection{Definition of Crossing Kernels for $c\in(-\infty,1]$}\label{sec:defkern2}

For $c\leq 1$ there are three important remarks that differentiate the crossing kernels compared to the ones in the complement regime. 
\begin{itemize}
\item The $c\leq1$ crossing kernels are \textit{not} the analytic continuations of the modular and fusion kernels from the range $c\in\mathbb{C}\backslash(-\infty,1]$. This is easily seen from the fact that all the special functions entering in (\ref{eq:defmodularkernel}), (\ref{eq:deffusionkernel}) have a natural boundary of analyticity exactly when $b\in i  \mathbb{R}$. We will denote the kernels valid at $c\leq1$ as $\hat{\mathbf{M}}^{(\hat{b})},\hat{\mathbf{F}}^{(\hat{b})}$. 
\item It was shown explicitly in \cite{Ribault:2023vqs} that, contrary to the kernels valid in $c\in\mathbb{C}\backslash(-\infty,1]$, the kernels $\hat{\mathbf{M}}^{(\hat{b})},\hat{\mathbf{F}}^{(\hat{b})}$ \textit{cannot} be meromorphic functions of the various momenta $P_i$. The reason is the following: for $c\leq 1$ there exist unique meromorphic solutions $\mathfrak{R}\mathbf{M},\mathfrak{R}\mathbf{F}$ to the shift relations for the modular and fusion kernels respectively, that take the following form
\be\label{eq:meromcleq1}
\bea
\mathfrak{R}\mathbf{M}_{P_s,P_t}[P_0] &:=\frac{P_t}{P_s}\mathbf{M}^{(\hat{b})}_{iP_t,iP_s}[iP_0],\\
\mathfrak{R}\mathbf{F}_{P_s,P_t}\left[\begin{smallmatrix} P_2 & P_3 \\ P_1 & P_4 \end{smallmatrix}\right] &:=\frac{P_t}{P_s}\mathbf{F}^{(\hat{b})}_{iP_t,iP_s}\left[\begin{smallmatrix} iP_2 & iP_1 \\ iP_3 & iP_4 \end{smallmatrix}\right] , \ \ \ \ \ \ \ \  \hat{b}\in\mathbb{R}_{(0,1]},
\eea
\ee
where on the RHS the kernels are given by (\ref{eq:defmodularkernel}), (\ref{eq:deffusionkernel}), except readily evaluated at rotated values of the momenta and appropriate permutations as indicated. The important point is that, even though $\mathfrak{R}\mathbf{M},\mathfrak{R}\mathbf{F}$ are the \textit{unique meromorphic} solutions to the shift relations for $c\leq1$, they do \textit{not} satisfy the crossing relations (\ref{eq:defcrossingMcleq1}), (\ref{eq:defcrossingFcleq1}) with the corresponding $c\leq1$ blocks\cite{Ribault:2023vqs}\footnote{In particular, it can be argued that the integral of these kernels against a conformal block vanishes identically, simply from the parity property in the corresponding integration variable $P$ .}. One is then led to interpret $\mathfrak{R}\mathbf{M}$ and $\mathfrak{R}\mathbf{F}$ merely as ``unphysical'' solutions of the shift relations, and should instead search for different kinds of solutions—most likely, after dropping the meromorphicity assumption— which are ``physical'', i.e. they also satisfy the crossing transformations of the corresponding blocks. The latter will be denoted $\hat{\mathbf{M}}^{(\hat{b})},\hat{\mathbf{F}}^{(\hat{b})}$, as mentioned above.
\item Finally, a small technical comment is that for $c\leq1$ the contour of integration over which Virasoro blocks on a different channel branch into a block at the original channel should be shifted by an amount $\Lambda\neq0$ compared to the contour in the case $c\in\mathbb{C}\backslash(-\infty,1]$. This is due to the presence of poles coming from the blocks which lie exactly on the imaginary axis (in our notation).
\end{itemize}
The defining relations of the physical crossing kernels $\hat{\mathbf{M}}^{(\hat{b})},\hat{\mathbf{F}}^{(\hat{b})}$ are the following:
\paragraph{Modular kernel.}
\be\label{eq:defcrossingMcleq1}
\bea (-i\tau)^{h_0}\mathcal{F}^{(i\hat{b}), \tau}_{P_s} &= \int_{i\mathbb{R}+\Lambda}\frac{dP_t}{i}\ \hat{\mathbf{M}}^{(\hat{b})}_{P_s,P_t}[P_0] \ \mathcal{F}^{(i\hat{b}), -1/\tau}_{P_t}. 
\eea
\ee
\paragraph{Fusion kernel.}
\be\label{eq:defcrossingFcleq1}
\bea
\mathcal{F}^{(i\hat{b}), s-\text{channel}}_{P_s} = \int_{i\mathbb{R}+\Lambda} \frac{dP_t}{i}\ \hat{\mathbf{F}}^{(\hat{b})}_{P_s,P_t}\left[\begin{smallmatrix} P_2 & P_3 \\ P_1 & P_4 \end{smallmatrix}\right]  \mathcal{F}^{(i\hat{b}), t-\text{channel}}_{P_t}.
\eea
\ee
Unlike the kernels in the complement central charge regime, no explicit general expressions are known to date for $\hat{\mathbf{M}}^{(\hat{b})},\hat{\mathbf{F}}^{(\hat{b})}$ for generic values of $\hat{b}\in\mathbb{R}_{(0,1]}$ and momenta $P_i$. In the present work, we will provide explicit expressions for these kernels for the case $\hat{b}=\sqrt{m/n}$ with $(m,n)$ any pair of coprime positive integers. Our expressions are valid for general values of the various momenta $P_i$ and display a particular kind of non-meromorphicity (square root branch points) as we will discuss extensively.

\subsection{Virasoro-Wick Rotation}\label{sec:VWR}
The Virasoro-Wick Rotation (VWR) is defined as the following pair of maps $\mathfrak{R}_m,\mathfrak{R}_f$ respectively on the modular and fusion kernels
\be\label{eq:VWRdefn}
\bea
\mathfrak{R}_m:& \ \mathbf{M}^{(b)}_{P_s,P_t}[P_0] \longrightarrow \frac{P_t}{P_s}\mathbf{M}^{(ib)}_{iP_t,iP_s}[iP_0]=:\mathfrak{R}\mathbf{M}^{(b)}_{P_s,P_t}[P_0],\\
\mathfrak{R}_f:& \ \mathbf{F}^{(b)}_{P_s,P_t}\left[\begin{smallmatrix} P_2 & P_3 \\ P_1 & P_4 \end{smallmatrix}\right] \longrightarrow \frac{P_t}{P_s}\mathbf{F}^{(ib)}_{iP_t,iP_s}\left[\begin{smallmatrix} iP_2 & iP_1 \\ iP_3 & iP_4 \end{smallmatrix}\right]=:\mathfrak{R}\mathbf{F}^{(b)}_{P_s,P_t}\left[\begin{smallmatrix} P_2 & P_3 \\ P_1 & P_4 \end{smallmatrix}\right].
\eea
\ee
Note that 
\be
\bea
\mathfrak{R}_m^2\left(\mathbf{M}^{(b)}_{P_s,P_t}[P_0]\right)&= \mathbf{M}^{(-b)}_{-P_s,-P_t}[-P_0]  \ \ \text{and} \ \
\mathfrak{R}_f^2\left(\mathbf{F}^{(b)}_{P_s,P_t}\left[\begin{smallmatrix} P_2 & P_3 \\ P_1 & P_4 \end{smallmatrix}\right]\right)&=\mathbf{F}^{(-b)}_{-P_s,-P_t}\left[\begin{smallmatrix} -P_2 & -P_3 \\ -P_1 & -P_4 \end{smallmatrix}\right]
\eea
\ee
which implies that $\mathfrak{R}_m,\mathfrak{R}_f$ are involutions when acting on even functions of the momenta $P_i$ and the parameter $b$. This is indeed the case for the kernels given by (\ref{eq:defmodularkernel}), (\ref{eq:deffusionkernel}).
\par In \cite{Ribault:2023vqs} it was shown that the basic shift relations that determine the modular and fusion kernels are invariant respectively under the maps $\mathfrak{R}_m,\mathfrak{R}_f$. In other words, if $\mathbf{M}^{(b)}_{P_s,P_t}[P_0]$ and $\mathbf{F}^{(b)}_{P_s,P_t}\left[\begin{smallmatrix} P_2 & P_3 \\ P_1 & P_4 \end{smallmatrix}\right]$ are some given solutions to the corresponding shift relations, then so are $c_m \mathfrak{R}\mathbf{M}^{(b)}_{P_s,P_t}[P_0]$ and $c_f \mathfrak{R}\mathbf{F}^{(b)}_{P_s,P_t}\left[\begin{smallmatrix} P_2 & P_3 \\ P_1 & P_4 \end{smallmatrix}\right]$ for some momentum-independent constants $c_m,c_f$. Due to the transformation $b\rightarrow ib$, the VWR symmetry has the feature of mapping solutions valid at central charge $c$ to solutions valid at central charge $26-c$.
\subsection{Main results at rational central charge}\label{sec:summaryandorg}
We will now summarize the main novel contributions of the present work. Let us define
\be
\mathsf{b}=\sqrt{m/n}, \ \s=\sqrt{mn} , \ \mathsf{Q}=\b+\b^{-1} \ , \ \ \ \text{for }  (m,n) \ \text{pair of coprime positive integers}.
\ee
When $b=\mathsf{b}$, these correspond to the rational central charge values
\be
c=13+6\left(\frac{m}{n}+\frac{n}{m}\right)\geq25.
\ee
Our first main result is that the modular and fusion kernels  $\mathbf{M},\mathbf{F}$ given by (\ref{eq:defmodularkernel}), (\ref{eq:deffusionkernel}) split into two natural functions
\be\label{eq:linphysicalc25}
\bea
\mathbf{M}^{(\mathsf{b})}_{P_s,P_t}[P_0]&=\frac{1}{2}\left(\mathbf{M}^{(+)}_{P_s,P_t}[P_0]+\mathbf{M}^{(-)}_{P_s,P_t}[P_0]\right),\\
\mathbf{F}^{(\mathsf{b})}_{P_s,P_t}\left[\begin{smallmatrix} P_2 & P_3 \\ P_1 & P_4 \end{smallmatrix}\right]&=\frac{1}{2}\left(\mathbf{F}^{(+)}_{P_s,P_t}\left[\begin{smallmatrix} P_2 & P_3 \\ P_1 & P_4 \end{smallmatrix}\right]+\mathbf{F}^{(-)}_{P_s,P_t}\left[\begin{smallmatrix} P_2 & P_3 \\ P_1 & P_4 \end{smallmatrix}\right]\right),
\eea
\ee
when evaluated at $b=\mathsf{b}$. In particular, the functions $\mathbf{M}^{(\pm)},\mathbf{F}^{(\pm)}$ are themselves admissible crossing kernels, i.e. they satisfy (\ref{eq:defcrossingM}) and (\ref{eq:defcrossingF}) respectively
\begin{align}
(-i\tau)^{h_0}\mathcal{F}^{(\mathsf{b}), \tau}_{P_s} &= \int_{i\mathbb{R}}\frac{dP_t}{i}\ \mathbf{M}^{(\epsilon)}_{P_s,P_t}[P_0] \ \mathcal{F}^{(\mathsf{b}), -1/\tau}_{P_t},\label{eq:mpmmod} \ \ \ \ \ \ \ \ \ \ \ \ \ \epsilon=\pm \\
 \mathcal{F}^{(\mathsf{b}), s-\text{channel}}_{P_s} &= \int_{i\mathbb{R}} \frac{dP_t}{i}\ \mathbf{F}^{(\eta)}_{P_s,P_t}\left[\begin{smallmatrix} P_2 & P_3 \\ P_1 & P_4 \end{smallmatrix}\right]  \mathcal{F}^{(\mathsf{b}), t-\text{channel}}_{P_t}\label{eq:fpmfus} \ , \ \ \ \ \eta=\pm.
\end{align}
Their explicit expressions are given in (\ref{eq:splitmodmain2}), (\ref{eq:splitfusmain2}). As we explain, these new solutions have novel analytic properties (square root branch points) compared to the original kernels $\mathbf{M},\mathbf{F}$, which highlights the fact that the space of solutions to the basic shift relations (arising from the Moore Seiberg consistency conditions) is richer when one drops the assumption of meromorphicity in the momenta, at least for the special cases when $b=\mathsf{b}$. In addition they are \textit{not} invariant under reflections in the internal momenta\footnote{Due to these non-trivial properties of $\mathbf{M}^{(\pm)},\mathbf{F}^{(\pm)}$, we highlight that it is essential to take the range of integration to be the full imaginary line for (\ref{eq:mpmmod}), (\ref{eq:fpmfus}) to hold. This is contrary to the integration contour for the full kernels $\mathbf{M}^{(\mathsf{b})}, \mathbf{F}^{(\mathsf{b})}$ in (\ref{eq:defcrossingM}), (\ref{eq:defcrossingF}) which can also be taken as $i\mathbb{R}_+$ due to the reflection symmetry in the internal momenta. Analogous statements hold for $\hat{\mathbf{M}}^{(\pm)},\hat{\mathbf{F}}^{(\pm)}$.}. 
\par The second main result of our work is to provide explicit expressions for the so-far unknown kernels $\hat{\mathbf{M}}^{(\hat{b})},\hat{\mathbf{F}}^{(\hat{b})}$ when $\hat{b}=\mathsf{b}$, and for generic values of the momenta $P_i$. These correspond to the rational central charge values
\be
c=13-6\left(\frac{m}{n}+\frac{n}{m}\right)\leq1.
\ee
Using the symmetry of the shift relations under VWR, it is straightforward to show that
\be\label{eq:linphysical}
\bea
\hat{\mathbf{M}}^{(\mathsf{b})}_{P_s,P_t}[P_0]&=\frac{1}{2}\left(\hat{\mathbf{M}}^{(+)}_{P_s,P_t}[P_0]+\hat{\mathbf{M}}^{(-)}_{P_s,P_t}[P_0]\right),\\
\hat{\mathbf{F}}^{(\mathsf{b})}_{P_s,P_t}\left[\begin{smallmatrix} P_2 & P_3 \\ P_1 & P_4 \end{smallmatrix}\right]&=\frac{1}{2}\left(\hat{\mathbf{F}}^{(+)}_{P_s,P_t}\left[\begin{smallmatrix} P_2 & P_3 \\ P_1 & P_4 \end{smallmatrix}\right]+\hat{\mathbf{F}}^{(-)}_{P_s,P_t}\left[\begin{smallmatrix} P_2 & P_3 \\ P_1 & P_4 \end{smallmatrix}\right]\right)
\eea
\ee
where each of $\hat{\mathbf{M}}^{(\pm)},\hat{\mathbf{F}}^{(\pm)}$ are admissible crossing kernels satisfying (\ref{eq:defcrossingMcleq1}) and (\ref{eq:defcrossingFcleq1}) respectively
\be
\bea
(-i\tau)^{h_0}\mathcal{F}^{(i\mathsf{b}), \tau}_{P_s} &= \int_{i\mathbb{R}+\Lambda}\frac{dP_t}{i}\ \hat{\mathbf{M}}^{(\epsilon)}_{P_s,P_t}[P_0] \ \mathcal{F}^{(i\mathsf{b}), -1/\tau}_{P_t}, \ \ \ \ \ \ \ \ \ \ \ \ \ \ \ \ \ \ \ \ \ \epsilon=\pm,\\
\mathcal{F}^{(i\mathsf{b}), s-\text{channel}}_{P_s} &= \int_{i\mathbb{R}+\Lambda} \frac{dP_t}{i}\ \hat{\mathbf{F}}^{(\eta)}_{P_s,P_t}\left[\begin{smallmatrix} P_2 & P_3 \\ P_1 & P_4 \end{smallmatrix}\right]  \mathcal{F}^{(i\mathsf{b}), t-\text{channel}}_{P_t}, \ \ \ \ \ \ \ \ \ \ \ \ \ \eta=\pm
.
\eea
\ee
Their explicit expressions are given by the Virasoro-Wick Rotation of the corresponding kernels at $c\in[25,\infty)$, namely\footnote{As we make clear in section \ref{sec:cleq1kernels}, by $\mathfrak{R}$ on the RHS of (\ref{eq:VWRMPMFPM}) we mean that we perform the VWR maps (\ref{eq:VWRdefn}) on $\mathbf{M}^{(\pm)},\mathbf{F}^{(\pm)}$ \textit{without} the $b\rightarrow ib$ transformation. In other words, we permute $s\leftrightarrow t$ and $1\leftrightarrow 3$ (in the fusion kernel case), rotate all the momenta by a factor of $i$, multiply by an overall $P_t/P_s$, but we do \textit{not} alter the dependence of $\mathbf{M}^{(\pm)},\mathbf{F}^{(\pm)}$ on the $m,n$ co-prime integers. These just go along for the ride, defining $\hat{b}=\sqrt{m/n}$.}
\be\label{eq:VWRMPMFPM}
\bea
\hat{\mathbf{M}}^{(\pm)}_{P_s,P_t}[P_0]&=\mp i \mathfrak{R}\mathbf{M}^{(\pm)}_{P_s,P_t}[P_0],\\
\hat{\mathbf{F}}^{(\pm)}_{P_s,P_t}\left[\begin{smallmatrix} P_2 & P_3 \\ P_1 & P_4 \end{smallmatrix}\right]&=\mp i \mathfrak{R}\mathbf{F}^{(\pm)}_{P_s,P_t}\left[\begin{smallmatrix} P_2 & P_3 \\ P_1 & P_4 \end{smallmatrix}\right].
\eea
\ee
 Their explicit expressions are given in (\ref{eq:splitmodleq1main2}), (\ref{eq:splitfusleq1main2}). The novelty of the physical kernels (\ref{eq:linphysical}) is that they define overall a \textit{non-meromorphic} function in the Liouville momenta, which is in contrast with the physical kernels (\ref{eq:linphysicalc25}) in the regime $c\in[25,\infty)$. Despite that fact, we will show that the full linear combinations (\ref{eq:linphysical}) define an even (i.e. reflection-symmetric) function in all the involved momenta, just like the Teschner and Teschner-Vartanov solutions. More explicitly, we will see that the kernels (\ref{eq:linphysical}) can be written as
 \be
 \bea
 \hat{\mathbf{M}}^{(\mathsf{b})}_{P_s,P_t}[P_0]&=\frac{1}{2}\text{disc}_{P_t}\left[\hat{\mathbf{M}}^{(+)}_{P_s,P_t}[P_0]\right],\\
\hat{\mathbf{F}}^{(\mathsf{b})}_{P_s,P_t}\left[\begin{smallmatrix} P_2 & P_3 \\ P_1 & P_4 \end{smallmatrix}\right]&=\frac{1}{2}\text{disc}_{P_t}\left[\hat{\mathbf{F}}^{(+)}_{P_s,P_t}\left[\begin{smallmatrix} P_2 & P_3 \\ P_1 & P_4 \end{smallmatrix}\right]\right]
 \eea
 \ee
 where the discontinuity is of square-root type in the integration variable $P_t$.
 \par We note finally that our formulas for rational $c\leq 1$ show that the previously known meromorphic kernels (\ref{eq:meromcleq1}) are realized as the \textit{opposite} linear combinations of $\hat{\mathbf{M}}^{(\pm)},\hat{\mathbf{F}}^{(\pm)}$, namely 
 \be
 \bea
 \frac{1}{2}\left(\hat{\mathbf{M}}^{(+)}_{P_s,P_t}[P_0]-\hat{\mathbf{M}}^{(-)}_{P_s,P_t}[P_0]\right)&=\frac{P_t}{P_s}\mathbf{M}^{(\b)}_{iP_t,iP_s}[iP_0],\\
 \frac{1}{2}\left(\hat{\mathbf{F}}^{(+)}_{P_s,P_t}\left[\begin{smallmatrix} P_2 & P_3 \\ P_1 & P_4 \end{smallmatrix}\right]-\hat{\mathbf{F}}^{(-)}_{P_s,P_t}\left[\begin{smallmatrix} P_2 & P_3 \\ P_1 & P_4 \end{smallmatrix}\right]\right)&=\frac{P_t}{P_s}\mathbf{F}^{(\b)}_{iP_t,iP_s}\left[\begin{smallmatrix} iP_2 & iP_1 \\ iP_3 & iP_4 \end{smallmatrix}\right].
 \eea
 \ee
These particular linear combinations define a non trivial, but overall \textit{meromorphic} function of the momenta, as it is clear from the RHS. However they are overall \textit{odd} functions in the internal momenta, which demonstrates that when integrated against the $c\leq1$ blocks they obviously yield zero.
\\
\paragraph{Organization of the paper.} The present work is structured as follows. In section \ref{sec:cgeq25rationalkernels} we discuss in detail the kernels $\mathbf{M}^{(\pm)}$ (section \ref{sec:modkernelcgeq25}) and $\mathbf{F}^{(\pm)}$ (section \ref{sec:fusionkernelcgeq25}) for $c\in\mathbb{Q}_{[25,\infty)}$. Their analytic derivation starting from the Teschner modular kernel and the Teschner-Vartanov fusion kernel at $b=\b$ and using the GK lemma is presented in Appendix \ref{AppendixB}. In section \ref{sec:properties} we discuss their analytic and reflection properties, and in section \ref{sec:shiftrel} (and Appendix \ref{app:shiftrelFpm}) we prove that they satisfy the corresponding shift relations. We also show in section \ref{sec:liouvillec25} that the new kernels implement crossing symmetry and modular covariance in Liouville theory. In section \ref{sec:cleq1kernels} we turn to the case of $c \in \mathbb{Q}_{(-\infty,1]}$ where, using the Virasoro-Wick Rotation symmetry of the shift relations, we derive for the first time the physical modular (section \ref{sec:modkerncleq1}) and fusion (section \ref{sec:fuskerncleq1}) kernels for any value of the central charge in this range. We use this knowledge later in section \ref{sec:liouvillecless1} to prove the long anticipated crossing symmetry and modular covariance of timelike Liouville theory in the said central charge range. We conclude in section \ref{sec:discussion} with a discussion of an intriguing feature of our results: the modular and fusion kernels at $b^2\in\mathbb{Q}^{\times}$ exhibit a semiclassical and one-loop exact form, which naturally leads us to propose a conjecture about their general transformation properties with respect to the parameter $b^2$. Appendix \ref{appendixA} contains a compendium of the special functions (and their properties) used throughout the paper.

\section{Virasoro Kernels at rational $c\in[25,\infty)$}\label{sec:cgeq25rationalkernels}
In the present section we show analytically that the original expressions for the modular and fusion kernels (\ref{eq:defmodularkernel}), (\ref{eq:deffusionkernel}) at $b^2\in\mathbb{Q}_{>0}$ can be expressed as linear combinations of two qualitatively new solutions to the shift relations, each of which thereby defines a modular and fusion kernel in its own right. The main novelty of these kernels is that they possess square-root branch point singularities and are exchanged under reflections in the internal Liouville momenta. After discussing their properties, we also show that the new kernels implement crossing symmetry and modular covariance in Liouville theory at rational $c\geq 25$.

\subsection{Two novel solutions for the Modular Kernel}\label{sec:modkernelcgeq25}
The modular kernel (\ref{eq:defmodularkernel}), (\ref{eq:prefintegdM}), when evaluated at $b=\b$ becomes
\be
\bea \label{eq:modkenrbrat}
 & \mathbf{M}^{(b=\b)}_{P_s,P_t}[P_0] =  M_\b(P_0|P_s,P_t)
 \int_{i\mathbb{R}} \frac{du}{i} \ m_\b(u)
\eea
\ee
where
\be\label{eq:prefintegdMrational}
\bea
M_\b(P_0|P_s,P_t)&=\frac{\s \left(2\pi\right)^{\s P_0-m-n}\mathcal{N}_{m,n}(P_s)\rho^{(\b)}_0(P_t)}{2\tilde{G}_{m,n}\left(-\frac{P_0}{\s}\right)}\\
&\quad  \times 
\frac{\prod_\pm G_{m,n}\left(m^{-1}+n^{-1}\pm\frac{2P_t}{\s}\right)G_{m,n}\left(\frac{m^{-1}+n^{-1}}{2}-\frac{P_0\pm2P_s}{\s}\right)}{\prod_\pm G_{m,n}\left(m^{-1}+n^{-1}\pm\frac{2P_s}{\s}\right)G_{m,n}\left(\frac{m^{-1}+n^{-1}}{2}-\frac{P_0\pm2P_t}{\s}\right)},\\
m_\b(u)&=\frac{e^{-4\pi i\s P_s u}}{\mathcal{N}^{(\text{m})}_{m,n}(P_s)} \prod_{\pm}\frac{\tilde{G}_{m,n}\left(u+\frac{m^{-1}+n^{-1}}{4}-\frac{1}{2}\left(\pm\frac{2P_t}{\s}+\frac{P_0}{\s}\right)\right)}{\tilde{G}_{m,n}\left(u-\frac{m^{-1}+n^{-1}}{4}-\frac{1}{2}\left(\pm\frac{2P_t}{\s}-\frac{P_0}{\s}\right)\right)}.
\eea
\ee
The definitions of $G_{m,n}(x)$ and $\tilde{G}_{m,n}(x)$ in terms of the Barnes' $G$ function as well as their relations with $\Gamma_b$ and $S_b$ when $b=\b$ are given in detail in Appendix \ref{appendixA}. In going from (\ref{eq:defmodularkernel}), (\ref{eq:prefintegdM}) to (\ref{eq:modkenrbrat}), (\ref{eq:prefintegdMrational}) we have additionally made two modifications\footnote{In other words, and for the sake of clarity, we emphasize that $M_{\b},m_{\b}$ here are \textit{not} exactly equal to $M_{b},m_b$ defined in (\ref{eq:prefintegdM}) when evaluated at $b=\b$.}: first we multiplied and divided by a convenient $P_s-$dependent constant $$
\mathcal{N}^{(\text{m})}_{m,n}(P_s)\equiv (-1)^{mn}(2\pi)^{2mn}\s^{-2}e^{2\pi i \s P_s}.
$$
Second, for the integral we rescaled the original integration variable by $$u\rightarrow -\s u.
$$
This is natural since $S_\b(x)\propto \tilde{G}_{m,n}(-x/\s+Q/(2\s))$ (c.f. (\ref{eq:sbmn})). It is also obvious that this rescaling has no effect on the integration contour. 
\par We can now state the first main result of this section. The modular kernel (\ref{eq:modkenrbrat}) can be written as
\be\label{eq:splitModmain}
\bea
\mathbf{M}^{(b=\b)}_{P_s,P_t}[P_0] =\frac{1}{2}\left(\mathbf{M}^{(+)}_{P_s,P_t}[P_0]+\mathbf{M}^{(-)}_{P_s,P_t}[P_0]\right)
\eea
\ee
with
\be\label{eq:splitmodmain2}
\boxed{
\bea
\mathbf{M}^{(+)}_{P_s,P_t}[P_0]&=\frac{M_\b(P_0|P_s,P_t)}{2i\sin{\left(2\pi \s P_t\right)}\sin{\left(2\pi \s P_s\right)}}\times \frac{1}{\sqrt{\mathfrak{D}^{(m)}}}\sum_{k=0}^{\s^2-1}\mathcal{M}_{\b}\left(\frac{\log{z^{(\text{m})}_1}}{2\pi i \s^2}+\frac{k}{\s^2}\right),\\
\mathbf{M}^{(-)}_{P_s,P_t}[P_0]&=-\frac{M_\b(P_0|P_s,P_t)}{2i\sin{\left(2\pi \s P_t\right)}\sin{\left(2\pi \s P_s\right)}}\times \frac{1}{\sqrt{\mathfrak{D}^{(m)}}}\sum_{k=0}^{\s^2-1}\mathcal{M}_{\b}\left(\frac{\log{z^{(\text{m})}_2}}{2\pi i \s^2}+\frac{k}{\s^2}\right)
\eea
}.
\ee
\par The various terms are defined as follows. The summand function $\mathcal{M}_\b$ is almost equal to the original integrand $m_\b$, namely
\be\label{eq:modularaction}
\bea
\mathcal{M}_{\b}(u)&:=m_{\b}(u)\times \frac{e^{\pi i \s \left(P_0+2P_s-\frac{m+n}{2\s}\right)}}{\s^2}\prod_{\pm}\left(e^{-\pi i \s^2 u}+e^{\pi i \s^2\left[u+1+ 2\left(\frac{m^{-1}+n^{-1}}{4}-\frac{1}{2}\left(\pm\frac{2P_t}{\s}+\frac{P_0}{\s}\right)\right)\right]}\right)\\
&=\frac{e^{-4\pi i \s P_s u}\prod_{\pm}\tilde{G}_{m,n}\left(u-1+\frac{m^{-1}+n^{-1}}{4}-\frac{1}{2}\left(\pm\frac{2P_t}{\s}+\frac{P_0}{\s}\right)\right)}{\prod_{\pm}\tilde{G}_{m,n}\left(u-\frac{m^{-1}+n^{-1}}{4}-\frac{1}{2}\left(\pm\frac{2P_t}{\s}-\frac{P_0}{\s}\right)\right)}.
\eea
\ee
The data $\{z_{1}^{(\text{m})},z_{2}^{(\text{m})},\mathfrak{D}^{(\text{m})}\}$ originate from a specific degree-two polynomial --which from now on we dub \textit{quantum modular polynomial}-- defined as
\be\label{eq:modpoly}
\boxed{
\bea
P^{(\text{m})}_{mn}(z;\vec{P})&:=\alpha^{(\text{m})}_{mn}z^2+\beta^{(\text{m})}_{mn}z+\gamma^{(\text{m})}_{mn}
\eea
}
\ee
where, after denoting $c_{mn}(x)\equiv \cos{(2\pi \s x)}, s_{mn}(x)\equiv \sin{(2\pi \s x)}$, the coefficients are
\be\label{eq:modcoeff}
\bea
\alpha^{(\text{m})}_{mn}&=s_{mn}{\left(P_s-\frac{P_0}{2}+\frac{m+n}{4\s}\right)},\\
\beta^{(\text{m})}_{mn}&=2(-1)^{mn}c_{mn}{\left(P_t\right)}s_{mn}{\left(P_s\right)},\\
\gamma^{(\text{m})}_{mn}&=s_{mn}{\left(P_s+\frac{P_0}{2}-\frac{m+n}{4\s}\right)}.
\eea
\ee
The discriminant reads
\be
\bea
\Delta^{(m)}&=\left(\beta^{(\text{m})}_{mn}\right)^2-4\alpha^{(\text{m})}_{mn}\gamma^{(\text{m})}_{mn}=4\left[\sin{\left(\pi\s \left(P_0-\frac{m+n}{2\s}\right)\right)}^2-\sin{\left(2\pi \s P_s\right)}^2\sin{\left(2\pi \s P_t\right)}^2\right]\\
&\equiv \left(2i\times s_{mn}{\left(P_t\right)}s_{mn}{\left(P_s\right)}\right)^2\mathfrak{D}^{(\text{m})}.
\eea
\ee
We will refer to the quantity
\be
\bea
\mathfrak{D}^{(\text{m})}:=1-\left(\frac{s_{mn}{\left(\frac{P_0}{2}-\frac{m+n}{4\s}\right)}}{s_{mn}{\left(P_t\right)}s_{mn}{\left(P_s\right)}}\right)^2
\eea
\ee
as the \textit{quantum modular determinant}, and we will express various quantities in terms of it. There are of course other equivalent formulas for the discriminant, and we will next present two of them, each with an interesting \textit{geometric} interpretation that we will discuss shortly. 
\par Consider the following symmetric $4\times 4$ matrices:
\be\label{eq:grammodular1}
\bea
\mathcal{O}^{\text{(m)}}_{m,n}:=\begin{pmatrix}
1 &  -c_{mn}\left(P_s\right) & 0 &  0 \\
 -c_{mn}\left(P_s\right) & 1 & s_{mn}\left(\frac{P_0}{2}-\frac{m+n}{4\s}\right) & 0 \\
0 & s_{mn}\left(\frac{P_0}{2}-\frac{m+n}{4\s}\right) & 1 & -c_{mn}\left(P_t\right) \\
0  & 0 & -c_{mn}\left(P_t\right) & 1
\end{pmatrix},
\eea
\ee
and
\be\label{eq:grammodular2}
\bea
\mathcal{G}^{\text{(m)}}_{m,n}:=\begin{pmatrix}
1 &  s_{mn}\left(\frac{P_0}{2}-\frac{m+n}{4\s}\right) & s_{mn}\left(\frac{P_0}{2}-\frac{m+n}{4\s}\right) &  -c_{mn}\left(2P_s\right) \\
 s_{mn}\left(\frac{P_0}{2}-\frac{m+n}{4\s}\right) & 1 & -c_{mn}\left(2P_t\right) & s_{mn}\left(\frac{P_0}{2}-\frac{m+n}{4\s}\right) \\
s_{mn}\left(\frac{P_0}{2}-\frac{m+n}{4\s}\right) & -c_{mn}\left(2P_t\right) & 1 & s_{mn}\left(\frac{P_0}{2}-\frac{m+n}{4\s}\right)\\
-c_{mn}\left(2P_s\right)  & s_{mn}\left(\frac{P_0}{2}-\frac{m+n}{4\s}\right) & s_{mn}\left(\frac{P_0}{2}-\frac{m+n}{4\s}\right) & 1
\end{pmatrix}.
\eea
\ee
It is straightforward to check that
\be
\bea
\text{det}\left[\mathcal{O}^{\text{(m)}}_{m,n}\right]=-\frac{\Delta^{(m)}}{4}, \quad \quad  \text{det}\left[\mathcal{G}^{\text{(m)}}_{m,n}\right]=-\left(1+c_{mn}\left(2P_s\right)\right)\left(1+c_{mn}\left(2P_t\right)\right)\Delta^{(m)}.
\eea
\ee
\par We recognize the matrix $\mathcal{O}^{\text{(m)}}_{m,n}$ as the standard \textit{Gram matrix} that encodes the six (interior) dihedral angles $\psi_i$ of a three-dimensional \textit{orthoscheme}\footnote{An orthoscheme is a simplex that basically generalizes the right triangle of two dimensions (in either euclidean, hyperbolic, or spherical geometries). In three dimensions it is a bounded 3-simplex where two of its edges are orthogonal (under a given geometry) to two respective planes (see \cite{chokim}). Its importance comes from the fact that any $n-$polyhedron can be represented as an algebraic sum of $n-$orthoschemes.} -- denoted $OT(\psi_1,\psi_2,\psi_3)$ -- from the following general form for the Gram matrix of dihedral angles for a three-dimensional tetrahedron  \cite{chokim,murakamiyano}:
\begin{align}\label{eq:genangleGram}
\mathcal{G}_{\text{angles}}&= \begin{pmatrix}
1 & -\cos \psi_1 & -\cos \psi_4 & -\cos \psi_6 \\
-\cos \psi_1 & 1 & -\cos \psi_2 & -\cos \psi_5 \\
-\cos \psi_4& -\cos \psi_2 & 1 & -\cos \psi_3 \\
-\cos \psi_6 & -\cos \psi_5 & -\cos \psi_3 & 1
\end{pmatrix} \ . 
\end{align}
By definition, a 3-orthoscheme has three right dihedral angles and hence in (\ref{eq:genangleGram}) we get $\cos \psi_4=\cos \psi_5=\cos \psi_6=0$. The remaining three in our case are identified as follows
\be\label{eq:anglesOT}
\bea
\psi_1\equiv2\pi \s P_s , \qquad \psi_2\equiv\pi \s P_0+\frac{\pi}{2}(1-m-n) , \qquad \psi_3\equiv2\pi \s P_t.
\eea
\ee 
On the other hand, the matrix $\mathcal{G}^{\text{(m)}}_{m,n}$ resembles the standard \textit{vertex Gram matrix} that encodes the inner products of the vertices of a three-dimensional tetrahedron, and therefore also its corresponding \textit{edge lengths} $\ell^{(\text{m})}_i$, for $i=1,\cdots,6$, via \cite{chokim,murakamiyano,Ushijima}
\begin{align}\label{eq:genlengthGram}
\mathcal{G}_{\text{lengths}}&= \begin{pmatrix}
1 & -\cosh \ell_4 & -\cosh \ell_5 & -\cosh \ell_3 \\
-\cosh \ell_4 & 1 & -\cosh \ell_6 & -\cosh \ell_2 \\
-\cosh \ell_5 & -\cosh \ell_6 & 1 & -\cosh \ell_1 \\
-\cosh \ell_3 & -\cosh \ell_2 & -\cosh \ell_1 & 1
\end{pmatrix} \ . 
\end{align}
Denoting the relevant tetrahedron as $T(\ell^{(\text{m})}_i)$, we read the following edge lengths from (\ref{eq:grammodular2}) and (\ref{eq:genlengthGram}): 
\be\label{eq:lengthsmod}
\bea
\ell^{(\text{m})}_1&=\ell^{(\text{m})}_2=\ell^{(\text{m})}_4=\ell^{(\text{m})}_5\equiv \ell^{(\text{m})}_0\equiv  \pi i \s P_0+\frac{i\pi}{2}(1-m-n),  \\
\ell^{(\text{m})}_3&\equiv\ell^{(\text{m})}_s\equiv 4\pi i \s P_s, \qquad \ell^{(\text{m})}_6\equiv\ell^{(\text{m})}_t\equiv 4\pi i \s P_t.
\eea
\ee
The interpretation in terms of angles or lengths here requires some restriction on the values of the Liouville momenta. Indeed, we need $P_s,P_t,P_0\in\mathbb{R}$ such that $\psi_i\in(0,\pi)$ for the orthoscheme case, and $P_s,P_t,P_0\in \mathbb{C}$ with $\Re{P_s},\Re{P_t}= \frac{\mathbb{Z}}{2\s}$, and $\Re{P_0}=\frac{m+n-1}{2\s}+\frac{2\mathbb{Z}}{\s}$ for the tetrahedron case\footnote{General necessary and sufficient conditions for \textit{hyperbolic} orthoschemes and tetrahedra to exist are given in \cite{chokim,murakamiyano}. Among those, a distinguished one is the requirement that the determinants (\ref{eq:genangleGram}) (for the hyperbolic orthoscheme), (\ref{eq:genlengthGram}) (for the hyperbolic tetrahedron) should be strictly negative.}. So far our expressions for the modular kernel have been mostly meromorphic (away from square-root branch cuts and poles) and hence there is no a priori indication that we should restrict to those values of the momenta in our analysis. Nevertheless, it is a striking and rather intriguing feature that three-polyhedra emerge in the formulas for the modular kernel at $b=\b$. We will return to discuss this point in Section \ref{sec:discussion}.
\par Finally, the roots of the quantum modular polynomial read 
\be\label{eq:modularroots}
\bea
z^{(\text{m})}_{1,2}&=\frac{-\beta^{(\text{m})}_{mn}\pm\sqrt{\Delta^{(m)}}}{2\alpha^{(\text{m})}_{mn}}\\
&=\frac{s_{mn}{\left(P_s\right)}}{s_{mn}{\left(P_0/2-P_s-\frac{m+n}{4\s}\right)}}\left[(-1)^{mn}c_{mn}{\left(P_t\right)}\mp i s_{mn}{\left(P_t\right)}\left(\mathfrak{D}^{(\text{m})}\right)^{1/2}\right].
\eea
\ee
It is therefore manifest that each $\mathbf{M}^{(\pm)}$ is also invariant under the exchange of the two co-prime integers $m\leftrightarrow n$ (or $\b\rightarrow \b^{-1}$).
\par The analytic derivation of the result (\ref{eq:splitModmain}), (\ref{eq:splitModmain}) uses basic complex analysis and is explained in detail in Appendix \ref{appb1}. The main idea is that the integrand $m_\b(u)$ in (\ref{eq:modkenrbrat}) defines a quasi-periodic function with quasi-period $1$. Hence, we can apply the GK lemma and write the integral as a difference of two translated vertical integrals, which we can then close and pick up the residues of the relevant poles. As it turns out, the only relevant poles in this domain come from the roots of the quantum modular polynomial (\ref{eq:modpoly}). This explains the factors of $\frac{\pm1}{\sqrt{\Delta^{(\text{m})}}}$ in the definitions of $\mathbf{M}^{(\pm)}$. The rest is simply evaluation of the residues.
\par It is remarkable that we can describe the modular kernel at all rational values of central charge $c\in[25,\infty)$ in the universal form (\ref{eq:splitModmain}), (\ref{eq:splitmodmain2}), without actually having to perform an integral. This is especially useful for numerical implementations of the kernel. What is even more striking, as we explain in detail later, is that the kernels $\mathbf{M}^{(+)}, \mathbf{M}^{(-)}$ individually obey the modular transformation of the toric conformal blocks (\ref{eq:mpmmod}), despite possessing non-trivial analytic structure as a function of the integration momentum $P_t$ as well as non-trivial properties under reflections of the internal momenta.
\par A similar story holds for the fusion kernel which we describe next.
\subsection{Two novel solutions for the Fusion Kernel}\label{sec:fusionkernelcgeq25}
The fusion kernel (\ref{eq:deffusionkernel}), (\ref{eq:prefintegdF}), when evaluated at $b=\b$ becomes
\be
\bea \label{eq:fuskenrbrat}
 & \mathbf{F}^{(b=\b)}_{P_s,P_t}\left[\begin{smallmatrix} P_2 & P_3 \\ P_1 & P_4 \end{smallmatrix}\right] =  F_\b(P_1,P_2,P_3,P_4|P_s,P_t)
 \int_{i\mathbb{R}} \frac{du}{i} \ f_\b(u)
\eea
\ee
where
\be\label{eq:prefintegdFrational}
\bea
&F_\b(P_1,P_2,P_3,P_4|P_s,P_t)=\frac{i (2\pi)^{4\s^2-(m+n)} \ G_{m,n}\left(\pm\frac{2P_t}{\s}\right)}{\s \ G_{m,n}\left(m^{-1}+n^{-1}\pm\frac{2P_s}{\s}\right)}\\
&\qquad \quad \qquad \qquad \times \prod_{f\in F} \prod_{\substack{\sigma\in \mathbb{Z}_2^f| \\ \sigma_f = \eta_t(f)}}G_{m,n}\left(\frac{m^{-1}+n^{-1}}{2}+\textstyle{\sum}_{i\in f}\sigma_i\frac{P_i}{\s}\right)^{-\sigma_f},\\
&\qquad f_\b(u)=\frac{1}{\mathcal{N}^{(\text{f})}_{m,n}}\prod_{\substack{\sigma\in \mathbb{Z}_2^E|\\ \sigma_V=1}}\tilde{G}_{m,n}\left(u-\sigma_{E}\frac{m^{-1}+n^{-1}}{4}-\frac{1}{2}\sum_{i\in E} \sigma_i \frac{P_i}{\s}\right)^{-\sigma_E}.
\eea
\ee
Similar to the case of the modular kernel, in going from (\ref{eq:deffusionkernel}), (\ref{eq:prefintegdF}) to (\ref{eq:fuskenrbrat}), (\ref{eq:prefintegdFrational}) we have made two modifications: first, we multiplied and divided by a convenient constant
\be
\mathcal{N}^{(\text{f})}_{m,n}\equiv 2i \s^{-2}(2\pi)^{4\s^2},
\ee
and second, we again rescaled the integration variable by $u\rightarrow -\s u$.
\par We can now state the second main result of this section. The fusion kernel (\ref{eq:fuskenrbrat}) can be written as
\be\label{eq:splitfusmain}
\bea
 \mathbf{F}^{(b=\b)}_{P_s,P_t}\left[\begin{smallmatrix} P_2 & P_3 \\ P_1 & P_4 \end{smallmatrix}\right] = \frac{1}{2}\left(\mathbf{F}^{(+)}_{P_s,P_t}\left[\begin{smallmatrix} P_2 & P_3 \\ P_1 & P_4 \end{smallmatrix}\right]+\mathbf{F}^{(-)}_{P_s,P_t}\left[\begin{smallmatrix} P_2 & P_3 \\ P_1 & P_4 \end{smallmatrix}\right]\right)
\eea
\ee
with
\be\label{eq:splitfusmain2}
\boxed{
\bea
\mathbf{F}^{(+)}_{P_s,P_t}\left[\begin{smallmatrix} P_2 & P_3 \\ P_1 & P_4 \end{smallmatrix}\right]&=\frac{F_\b(P_1,P_2,P_3,P_4|P_s,P_t)}{2i\sin{\left(2\pi \s P_t\right)}\sin{\left(2\pi \s P_s\right)}}\times \frac{1}{\sqrt{\mathfrak{D}^{(f)}}}\sum_{k=0}^{\s^2-1}\mathcal{F}_{\b}\left(\frac{\log{z^{(\text{f})}_1}}{2\pi i \s^2}+\frac{k}{\s^2}\right),\\
\mathbf{F}^{(-)}_{P_s,P_t}\left[\begin{smallmatrix} P_2 & P_3 \\ P_1 & P_4 \end{smallmatrix}\right]&=-\frac{F_\b(P_1,P_2,P_3,P_4|P_s,P_t)}{2i\sin{\left(2\pi \s P_t\right)}\sin{\left(2\pi \s P_s\right)}}\times \frac{1}{\sqrt{\mathfrak{D}^{(f)}}}\sum_{k=0}^{\s^2-1}\mathcal{F}_{\b}\left(\frac{\log{z^{(\text{f})}_2}}{2\pi i \s^2}+\frac{k}{\s^2}\right) 
\eea
}.
\ee
\par The various terms are defined as follows. The summand function $\mathcal{F}_\b$ reads
\be\label{eq:fusionaction}
\bea
\mathcal{F}_\b(u)&:=f_\b(u)\times\frac{2i(-1)^{mn+1}}{\s^2}\prod_{J\in\{\sigma_E=-1\}}\left(e^{-\pi i \s^2 u}+e^{\pi i \s^2\left[ u+1+2\left(\frac{m^{-1}+n^{-1}}{4}-\frac{1}{2}\sum_{i\in E} \sigma^{(J)}_i \frac{P_i}{\s}\right)\right]}\right) \\
&=\frac{\prod_{J\in\{\sigma_E=-1\}}\tilde{G}_{m,n}\left(u-1+\frac{m^{-1}+n^{-1}}{4}-\frac{1}{2}\sum_{i\in E} \sigma^{(J)}_i \frac{P_i}{\s}\right)}{\prod_{I\in\{\sigma_E=+1\}}\tilde{G}_{m,n}\left(u-\frac{m^{-1}+n^{-1}}{4}-\frac{1}{2}\sum_{i\in E} \sigma^{(I)}_i \frac{P_i}{\s}\right)}.
\eea
\ee
The data $\{z_{1}^{(\text{f})},z_{2}^{(\text{f})},\mathfrak{D}^{(\text{f})}\}$ originate from a specific degree-two polynomial --which from now on we dub \textit{quantum fusion polynomial}-- defined as
\be\label{eq:fuspoly}
\boxed{
\bea
P^{(\text{f})}_{mn}(z;\vec{P})&:= \alpha^{(\text{f})}_{mn}z^2+\beta^{(\text{f})}_{mn}z+\gamma^{(\text{f})}_{mn}
\eea
}
\ee
where the coefficients are
\be\label{eq:fuscoeff}
\bea
\alpha^{(\text{f})}_{mn}&=\sum_{\substack{\sigma\in \mathbb{Z}_2^E|\\ \sigma_V=1}}\sigma_E \ e^{2\pi i \s^2\left(\sigma_E\frac{m^{-1}+n^{-1}}{4}+\frac{1}{2}\sum_{i\in E}\sigma_i\frac{P_i}{\s}\right)},\\
\beta^{(\text{f})}_{mn}&=4\sum_{ij\in P}s_{mn}{(P_i)}s_{mn}{(P_j)},\\
\gamma^{(\text{f})}_{mn}&=\sum_{\substack{\sigma\in \mathbb{Z}_2^E|\\ \sigma_V=1}}\sigma_E \ e^{-2\pi i \s^2\left(\sigma_E\frac{m^{-1}+n^{-1}}{4}+\frac{1}{2}\sum_{i\in E}\sigma_i\frac{P_i}{\s}\right)}.
\eea
\ee
It is straightforward to check that the coefficients obey the following symmetries:
\be\label{eq:symmetriescoeff}
\bea
\left.\alpha^{(\text{f})}_{mn}\right|_{P_s\leftrightarrow P_t, P_1\leftrightarrow P_3}&=\alpha^{(\text{f})}_{mn}, \quad \left.\beta^{(\text{f})}_{mn}\right|_{P_s\leftrightarrow P_t, P_1\leftrightarrow P_3}=\beta^{(\text{f})}_{mn}, \quad \left.\gamma^{(\text{f})}_{mn}\right|_{P_s\leftrightarrow P_t, P_1\leftrightarrow P_3}=\gamma^{(\text{f})}_{mn},\\
\left.\alpha^{(\text{f})}_{mn}\right|_{P_s\leftrightarrow P_t, P_2\leftrightarrow P_4}&=\alpha^{(\text{f})}_{mn}, \quad \left.\beta^{(\text{f})}_{mn}\right|_{P_s\leftrightarrow P_t, P_2\leftrightarrow P_4}=\beta^{(\text{f})}_{mn}, \quad \left.\gamma^{(\text{f})}_{mn}\right|_{P_s\leftrightarrow P_t, P_2\leftrightarrow P_4}=\gamma^{(\text{f})}_{mn}.
\eea
\ee
Consequently, we also expect that the roots of the polynomial will enjoy the same symmetries as we discuss shortly.
The discriminant of the quantum fusion polynomial can be written conveniently in terms of the determinant of a particular $4\times 4$ matrix. Setting
\be
a_i\equiv P_i-\frac{m+n}{2\s},
\ee
consider the symmetric matrix:
\be\label{eq:gramf}
\bea
\mathcal{G}^{\text{(f)}}_{m,n}:=\begin{pmatrix}
1 & -c_{mn}\left(a_2\right) & -c_{mn}(a_3) & c_{mn}(a_s) \\
 -c_{mn}(a_2) & 1 & c_{mn}(a_t) & -c_{mn}(a_1) \\
-c_{mn}(a_3) & c_{mn}(a_t) & 1 & -c_{mn}(a_4) \\
 c_{mn}(a_s) & -c_{mn}(a_1) & -c_{mn}(a_4) & 1
\end{pmatrix}.
\eea
\ee
Then,
\be
\bea
\Delta^{(f)}&=\left(\beta^{(\text{f})}_{mn}\right)^2-4\alpha^{(\text{f})}_{mn}\gamma^{(\text{f})}_{mn}=16 \ \text{det}\left[\mathcal{G}^{\text{(f)}}_{m,n}\right]\\
&\equiv \left(2i\times s_{mn}{\left(P_t\right)}s_{mn}{\left( P_s\right)}\right)^2\mathfrak{D}^{(\text{f})}.
\eea
\ee
We will refer to the quantity
\be\label{eq:fusdeterminant}
\bea
\mathfrak{D}^{(\text{f})}:=\frac{-4 \ \text{det} \left[\mathcal{G}^{\text{(f)}}_{m,n}\right]}{s_{mn}{\left( P_s\right)}^2s_{mn}{\left( P_t\right)}^2}
\eea
\ee
as the \textit{quantum fusion determinant}, and we will express various quantities in terms of it. It is a non-trivial observation is that this quantity is \textit{invariant} under simultaneous exchanges of $P_s,P_t$ and $P_1,P_3$, or simultaneous exchanges of $P_s,P_t$ and $P_2,P_4$, as expected from the symmetries of the coefficients that we discussed above. Indeed,
\be\label{eq:symmetriesD}
\bea
\left.\mathfrak{D}^{(\text{f})}\right|_{P_s\leftrightarrow P_t, P_1\leftrightarrow P_3}=\mathfrak{D}^{(\text{f})}, \quad \quad \left.\mathfrak{D}^{(\text{f})}\right|_{P_s\leftrightarrow P_t, P_2\leftrightarrow P_4}=\mathfrak{D}^{(\text{f})}.
\eea
\ee  
\par Similar to the case of the modular kernel, we recognize the matrix $\mathcal{G}^{\text{(f)}}_{m,n}$ as the standard \textit{vertex Gram matrix} that encodes the six edge lengths $\ell^{(\text{f})}_i$ of a \textit{3-tetrahedron}, denoted as $T(\ell^{(\text{f})}_i)$, $i=1,\cdots,6$. According to (\ref{eq:genlengthGram}), we read from (\ref{eq:gramf}) the following lengths:
\be\label{eq:lengthsfusion}
\bea
\ell^{(\text{f})}_1&\equiv 2\pi i \s P_4-i\pi(m+n), \ \ \ell^{(\text{f})}_2\equiv2\pi i \s P_1-i\pi(m+n), \ \ \ell^{(\text{f})}_3\equiv2\pi i \s P_s-i\pi(1+m+n), \\
\ell^{(\text{f})}_4&\equiv2\pi i \s P_2-i\pi(m+n) , \ \ \ell^{(\text{f})}_5\equiv 2\pi i \s P_3-i\pi(m+n), \ \ \ell^{(\text{f})}_6\equiv 2\pi i \s P_t-i\pi(1+m+n).
\eea
\ee
The interpretation in terms of lengths here requires $P_s, P_t, P_1,P_2,P_3,P_4\in\mathbb{C}$ with $\Re{P_i}=\frac{m+n}{2\s}+\frac{\mathbb{Z}}{\s}$ for $i=1,\cdots,4$, and $\Re{P_s},\Re{P_t}=\frac{1+m+n}{2\s}+\frac{\mathbb{Z}}{\s}$, though so far there is no indication from our formulas that we should restrict in this particular locus of momenta.
We will return to comment on this rather remarkable fact that there is an emergent (quantum) tetrahedron for the fusion kernel evaluated at $b=\b$ in Section \ref{sec:discussion}.
\par Finally, the roots of the quantum fusion polynomial read 
\be\label{eq:fusionroots}
\bea
z^{(\text{f})}_{1,2}&=\frac{-\beta^{(\text{f})}_{mn}\pm2is_{mn}{\left(P_t\right)}s_{mn}{\left( P_s\right)}\sqrt{\mathfrak{D}^{(f)}}}{2\alpha^{(\text{f})}_{mn}},
\eea
\ee
and hence it is by now evident that each $\mathbf{F}^{(\pm)}$ is invariant under $m\leftrightarrow n$ (or $\b\rightarrow \b^{-1}$).
Due to (\ref{eq:symmetriescoeff}), (\ref{eq:symmetriesD}), we also have the symmetry
\be\label{eq:symmetriesroots}
\bea
\left.z^{(\text{f})}_{1,2}\right|_{P_s\leftrightarrow P_t, P_1\leftrightarrow P_3}=z^{(\text{f})}_{1,2},\quad \quad \left.z^{(\text{f})}_{1,2}\right|_{P_s\leftrightarrow P_t, P_2\leftrightarrow P_4}=z^{(\text{f})}_{1,2},
\eea
\ee
as expected.
\par The analytic derivation of the result (\ref{eq:splitfusmain}), (\ref{eq:splitfusmain2}) is described in Appendix \ref{appb2} and follows the same logic as in the case of the modular kernel. The only differences are the involved functions. As we will explain in section \ref{sec:shiftrel} (and partly in Appendix \ref{app:shiftrelFpm}) it is an intricate fact that each of $\mathbf{F}^{(\pm)}$ obeys the crossing transformation of the sphere four-point conformal blocks (\ref{eq:fpmfus}), despite possessing non-trivial analytic structure as a function of the integration momentum $P_t$ as well as non trivial reflection properties in the internal momenta.
\par Let us remark also that in the special case $m=n=1$ our results reproduce exactly the ones discussed in \cite{Ribault:2023vqs} for the fusion kernel at $c=25$, and therefore prove a conjectural identity involving an integral of Barnes' $G$ functions in the last section of that paper. Here we have gone one step further and realize concretely that the philosophy of \cite{Ribault:2023vqs} applies to \textit{any} central charge of the form $c=13+6(\frac{m}{n}+\frac{n}{m})$ where $(m,n)$ is any co-prime pair of positive integers.
\subsection{Properties of the non-meromorphic kernels}\label{sec:properties}
Having defined the functions $\mathbf{M}^{(\pm)}, \mathbf{F}^{(\pm)}$ that build up the Teschner modular kernel and the Teschner–Vartanov fusion kernel respectively at $b=\b$, we next proceed to discuss their non-trivial properties.
\
\paragraph{Analytic structure.} The Teschner modular kernel $\mathbf{M}^{(b=\b)}$ and the Teschner-Vartanov fusion kernel $\mathbf{F}^{(b=\b)}$ are meromorphic functions in all the Liouville momenta. Their decompositions into $\mathbf{M}^{(\pm)}$ and $\mathbf{F}^{(\pm)}$ looks at first puzzling since the latter kernels appear to have both square-root and logarithmic branch point singularities. However we will show in this section that only half of this statement is true, i.e. there are \textit{no} logarithmic branch points and the two kernels $\mathbf{M}^{(\pm)},\mathbf{F}^{(\pm)}$ simply encode the two branches of the square-root, thereby resulting in a meromoprhic function for $\mathbf{M}^{(b=\b)}$, $\mathbf{F}^{(b=\b)}$ as expected.
The way this is achieved is non trivial, and relies on specific periodicity properties of the summand functions $\mathcal{M}_\b(u), \mathcal{F}_{\b}(u)$ when evaluated at the roots of the quantum modular and fusion polynomials.
\par To see that there are no logarithmic singularities we need to show that 
\be\label{eq:periodsumM}
\bea
\sum_{r=0}^{\s^2-1}\mathcal{M}_{\b}\left(\frac{\log{z^{(\text{m})}_{1,2}}}{2\pi i \s^2}+\frac{r+l}{\s^2}\right)&=\sum_{r=0}^{\s^2-1}\mathcal{M}_{\b}\left(\frac{\log{z^{(\text{m})}_{1,2}}}{2\pi i \s^2}+\frac{r}{\s^2}\right), \\
\quad \sum_{r=0}^{\s^2-1}\mathcal{F}_{\b}\left(\frac{\log{z^{(\text{f})}_{1,2}}}{2\pi i \s^2}+\frac{r+l}{\s^2}\right)&=\sum_{r=0}^{\s^2-1}\mathcal{F}_{\b}\left(\frac{\log{z^{(\text{f})}_{1,2}}}{2\pi i \s^2}+\frac{r}{\s^2}\right), \quad \quad \forall l\in\mathbb{Z}.
\eea
\ee
In fact we can prove an even stronger result which is going to be useful for us later when we prove the shift relations. We will show that
\begin{align}\label{eq:periodsumMStrong}
\sum_{r=0}^{\s^2-1}\mathcal{M}_{\b}\left(\frac{\log{z^{(\text{m})}_{1,2}}}{2\pi i \s^2}+\frac{r+l}{\s^2}\right)f\left(x+\frac{\mathfrak{c}\left(r+l\right)}{\s^2}\right)&=\sum_{r=0}^{\s^2-1}\mathcal{M}_{\b}\left(\frac{\log{z^{(\text{m})}_{1,2}}}{2\pi i \s^2}+\frac{r}{\s^2}\right)f\left(x+\frac{\mathfrak{c}r}{\s^2}\right),\\
\sum_{r=0}^{\s^2-1}\mathcal{F}_{\b}\left(\frac{\log{z^{(\text{f})}_{1,2}}}{2\pi i \s^2}+\frac{r+l}{\s^2}\right)f\left(x+\frac{\mathfrak{c}\left(r+l\right)}{\s^2}\right)&=\sum_{r=0}^{\s^2-1}\mathcal{F}_{\b}\left(\frac{\log{z^{(\text{f})}_{1,2}}}{2\pi i \s^2}+\frac{r}{\s^2}\right)f\left(x+\frac{\mathfrak{c}r}{\s^2}\right)\label{eq:periodsumFStrong},
\end{align}
$\forall l\in\mathbb{Z}$, and $x$ in the domain of a given periodic function $f$ satisfying
\be\label{eq:periodarbfn}
f\left(x+\mathfrak{c}\right)=f\left(x\right), \quad \quad \mathfrak{c}\in\mathbb{C}.
\ee
This is a general result, but as we will see later, only the case of $\mathbb{Z}\pi-$periodic functions $f$ (such as the usual trigonometric functions) will be relevant to our analysis.
\par The identities (\ref{eq:periodsumMStrong}), (\ref{eq:periodsumFStrong}) are true due to some specific periodicity properties of $\mathcal{M}_{\b},\mathcal{F}_{\b}$ which we now describe for the case of the modular kernel\footnote{In the following we will discuss in detail the case of the modular kernel; the case of the fusion kernel works in exactly the same way for reasons that will become clear.}. Using the identity (\ref{eq:relmM2}) it is obvious to derive the following relation for unit shifts
\be
\bea
\frac{\mathcal{M}_{\b}(u+l\pm1)}{\mathcal{M}_{\b}(u+l)}=\frac{\mathcal{M}_{\b}(u\pm1)}{\mathcal{M}_{\b}(u)}=\left[\frac{m_1(z(u))}{m_2(z(u))}\right]^{\pm1}, \quad \quad \forall l\in\mathbb{Z},
\eea
\ee
where $z(u)\equiv e^{2\pi i \s^2 u}$ and $m_1,m_2$ are given in (\ref{eq:m1m2}). This implies that for general shifts we get
\be
\bea
\frac{\mathcal{M}_{\b}(u+l)}{\mathcal{M}_{\b}(u)}=\left[\frac{m_1(z(u))}{m_2(z(u))}\right]^l, \quad \quad \forall l\in\mathbb{Z}.
\eea
\ee
The non-trivial observation now is that the difference of the functions $m_1,m_2$ is equal to the quantum modular polynomial (c.f. discussion around (\ref{eq:diffm1m2})) and hence when the last equation is evaluated at $z(u)=z_{1,2}^{\text{(m)}}$ we get
\be\label{eq:periodicityMb}
\bea
\mathcal{M}_{\b}\left(\frac{\log{z_{1,2}^{\text{(m)}}}}{2\pi i \s^2}+\frac{q}{\s^2}+l\right)=\mathcal{M}_{\b}\left(\frac{\log{z_{1,2}^{\text{(m)}}}}{2\pi i \s^2}+\frac{q}{\s^2}\right), \quad \quad \forall l\in\mathbb{Z}
\eea
\ee
where we chose an arbitrary monodromy for the logarithm, labelled by $q\in\mathbb{Z}$. For the same reasons (c.f. discussion around (\ref{eq:fuspolyapp})), for the fusion kernel we get a similar relation
\be\label{eq:periodicityFb}
\bea
\mathcal{F}_{\b}\left(\frac{\log{z_{1,2}^{\text{(f)}}}}{2\pi i \s^2}+\frac{q'}{\s^2}+l\right)=\mathcal{F}_{\b}\left(\frac{\log{z_{1,2}^{\text{(f)}}}}{2\pi i \s^2}+\frac{q'}{\s^2}\right), \quad \quad \forall l\in\mathbb{Z}.
\eea
\ee
It is now easy to see that (\ref{eq:periodsumMStrong}) follows. For $l>0$ (similar manipulations hold for $l<0$) we have
\begin{align*}
&\sum_{r=0}^{\s^2-1}\mathcal{M}_{\b}\left(\frac{\log{z^{(\text{m})}_{1,2}}}{2\pi i \s^2}+\frac{r+l}{\s^2}\right)f\left(x+\frac{\mathfrak{c}\left(r+l\right)}{\s^2}\right)=\sum_{q=l}^{l-1+\s^2}\mathcal{M}_{\b}\left(\frac{\log{z^{(\text{m})}_{1,2}}}{2\pi i \s^2}+\frac{q}{\s^2}\right)f\left(x+\frac{\mathfrak{c}q}{\s^2}\right)\\
&=\sum_{q=l}^{\s^2-1}\mathcal{M}_{\b}\left(\frac{\log{z^{(\text{m})}_{1,2}}}{2\pi i \s^2}+\frac{q}{\s^2}\right)f\left(x+\frac{\mathfrak{c}q}{\s^2}\right)+\sum_{q=\s^2}^{l-1+\s^2}\mathcal{M}_{\b}\left(\frac{\log{z^{(\text{m})}_{1,2}}}{2\pi i \s^2}+\frac{q}{\s^2}\right)f\left(x+\frac{\mathfrak{c}q}{\s^2}\right)
\end{align*}
\be
\bea
&=\sum_{q=0}^{\s^2-1}\mathcal{M}_{\b}\left(\frac{\log{z^{(\text{m})}_{1,2}}}{2\pi i \s^2}+\frac{q}{\s^2}\right)f\left(x+\frac{\mathfrak{c}q}{\s^2}\right)\\
&\ \ +\left\{\sum_{q=\s^2}^{l-1+\s^2}\mathcal{M}_{\b}\left(\frac{\log{z^{(\text{m})}_{1,2}}}{2\pi i \s^2}+\frac{q}{\s^2}\right)f\left(x+\frac{\mathfrak{c}q}{\s^2}\right)-\sum_{q=0}^{l-1}\mathcal{M}_{\b}\left(\frac{\log{z^{(\text{m})}_{1,2}}}{2\pi i \s^2}+\frac{q}{\s^2}\right)f\left(x+\frac{\mathfrak{c}q}{\s^2}\right)\right\}\\
&=\sum_{q=0}^{\s^2-1}\mathcal{M}_{\b}\left(\frac{\log{z^{(\text{m})}_{1,2}}}{2\pi i \s^2}+\frac{q}{\s^2}\right)f\left(x+\frac{\mathfrak{c}q}{\s^2}\right)\\
& \ +\sum_{q=0}^{l-1}\left\{\mathcal{M}_{\b}\left(\frac{\log{z^{(\text{m})}_{1,2}}}{2\pi i \s^2}+\frac{q}{\s^2}+1\right)f\left(x+\frac{\mathfrak{c}q}{\s^2}+\mathfrak{c}\right)-\mathcal{M}_{\b}\left(\frac{\log{z^{(\text{m})}_{1,2}}}{2\pi i \s^2}+\frac{q}{\s^2}\right)f\left(x+\frac{\mathfrak{c}q}{\s^2}\right)\right\}.
\eea
\ee
The last term in the bracket is \textit{zero} due to (\ref{eq:periodarbfn}), (\ref{eq:periodicityMb}), and hence this concludes the proof of (\ref{eq:periodsumMStrong}). An exactly identical result holds if we replace $\mathcal{M}_\b$ with $\mathcal{F}_\b$, and $z_{1,2}^{(\text{m})}$ with $z_{1,2}^{(\text{f})}$, which is the case of the fusion kernel (\ref{eq:periodsumFStrong}). The reason is simply because both $\mathcal{M}_\b$ and $\mathcal{F}_\b$ are built out of the same function $\tilde{G}_{m,n}$.
\par In conclusion, the new kernels are characterized overall by an obvious meromoprhic piece and, in addition, \textit{square-root branch point singularities} coming from the square root of the modular/fusion determinants $\mathfrak{D}^{\text{(m)}},\mathfrak{D}^{\text{(f)}}$. Understanding in detail these branch points is an extremely interesting problem. We make some very preliminary comments here, and leave more detailed study to the future. In the simpler case of the modular kernel the branch points are characterized by the solutions to the equation
\be\label{eq:branchptsmodular}
\bea
\mathfrak{D}^{(\text{m})}=0 \quad \Longrightarrow \quad \sin{(2\pi \s P_t^*)}=\pm \frac{\sin{(\pi \s a_0)}}{\sin{(2\pi \s P_s)}}.
\eea
\ee
We remind the reader that in our notation $P_t$ is the integration variable in the crossing transformation of the blocks, and hence $P_s,P_0$ should be thought of as given in (\ref{eq:branchptsmodular}). This immediately implies that for given $m,n$ there are infinite number of branch points labelled by the integers
\be
\bea
\pm P_t^*+\frac{\mathbb{Z}}{\s}.
\eea
\ee
Understanding better the location of these branch points is a subtle and elegant problem to analyze because it can lead to new integration contours for the crossing transformation of the blocks! Similar logic applies in the case of the fusion kernel where now the solutions are the more complicated zeroes of the determinant of the $4\times 4$ matrix $\mathcal{G}^{(\text{f})}$ (c.f. (\ref{eq:fusdeterminant})). We defer a detailed investigation of these questions to future work.
\paragraph{Reflections.} Let us next examine the kernels (\ref{eq:splitmodmain2}), (\ref{eq:splitfusmain2}) under reflections of the two internal momenta $P_s,P_t$. It is known that the Teschner modular kernel and Teschner-Vartanov fusion kernel are invariant under such reflections. However we will show here that the non-meromorphic kernels $\mathbf{M}^{(\pm)}$, $\mathbf{F}^{(\pm)}$ \textit{get exchanged} under such reflections.
\par We will describe in detail the case of the modular kernel. For reflections $P_t\rightarrow -P_t$ it is clear from the definitions that $M_{\b}$, $\mathcal{M}_{\b}$, and $\mathfrak{D}^{(\text{m})}$ are invariant. We also notice from (\ref{eq:modularroots}) that the roots of the polynomial get exchanged under such reflection, and hence overall
\be
\bea
\mathbf{M}^{(+)}_{P_s,-P_t}[P_0]= \mathbf{M}^{(-)}_{P_s,P_t}[P_0], \quad \text{and} \quad \mathbf{M}^{(-)}_{P_s,-P_t}[P_0]= \mathbf{M}^{(+)}_{P_s,P_t}[P_0], 
\eea
\ee
which is the argued behaviour.
\par For reflections $P_s\rightarrow -P_s$ things are slightly more involved. First, it is easy to see that $\mathfrak{D}^{(\text{m})}$ is again invariant. The functions $M_\b,\mathcal{M}_{\b}$ behave as
\begin{align}
M_\b(P_0|-P_s,P_t)&=e^{-4\pi i \s P_s}M_\b(P_0|P_s,P_t),\label{eq:Mbind}\\
\left.\mathcal{M}_\b(u)\right|_{P_s\rightarrow-P_s}&=e^{8\pi i \s P_su}\mathcal{M}_\b(u)=e^{4\pi i sP_s}\mathcal{M}_\b\left(-u+1\right).\label{MMbid}
\end{align}
In (\ref{MMbid}) the second equality follows from the definition (\ref{eq:modularaction}) and the property $\tilde{G}^{-1}_{m,n}(x)=\tilde{G}_{m,n}(-x)$. Finally, for the roots of the polynomial we find
\be
\bea
\left.z^{(\text{m})}_{1}\right|_{P_s\rightarrow-P_s}=\left(\frac{\alpha^{(\text{m})}_{mn}}{\gamma^{(\text{m})}_{mn}}\right) z^{(\text{m})}_{1}=\left(z^{(\text{m})}_{2}\right)^{-1},\\
\left.z^{(\text{m})}_{2}\right|_{P_s\rightarrow-P_s}=\left(\frac{\alpha^{(\text{m})}_{mn}}{\gamma^{(\text{m})}_{mn}}\right) z^{(\text{m})}_{2}=\left(z^{(\text{m})}_{1}\right)^{-1}.
\eea
\ee
Therefore, e.g. for $\mathbf{M}^{(+)}$ (analogously for $\mathbf{M}^{(-)}$) we get
\be
\bea
\mathbf{M}^{(+)}_{-P_s,P_t}[P_0]=&-\frac{M_\b(P_0|P_s,P_t)}{2i\sin{\left(2\pi \s P_t\right)}\sin{\left(2\pi \s P_s\right)}} \times\frac{1}{\sqrt{\mathfrak{D}^{(m)}}}\sum_{k=0}^{\s^2-1}\mathcal{M}_{\b}\left(\frac{\log{z^{(\text{m})}_2}}{2\pi i \s^2}+1-\frac{k}{\s^2}\right)\\
&=-\frac{M_\b(P_0|P_s,P_t)}{2i\sin{\left(2\pi \s P_t\right)}\sin{\left(2\pi \s P_s\right)}} \times\frac{1}{\sqrt{\mathfrak{D}^{(m)}}}\sum_{k=1}^{\s^2}\mathcal{M}_{\b}\left(\frac{\log{z^{(\text{m})}_2}}{2\pi i \s^2}+\frac{k}{\s^2}\right)\\
&=-\frac{M_\b(P_0|P_s,P_t)}{2i\sin{\left(2\pi \s P_t\right)}\sin{\left(2\pi \s P_s\right)}} \times\frac{1}{\sqrt{\mathfrak{D}^{(m)}}}\sum_{k=0}^{\s^2-1}\mathcal{M}_{\b}\left(\frac{\log{z^{(\text{m})}_2}}{2\pi i \s^2}+\frac{k}{\s^2}\right)\\
&=\mathbf{M}^{(-)}_{P_s,P_t}[P_0]
\eea
\ee
 where in the second line we reorganized the sum, and in the third line we used the periodicity property (\ref{eq:periodicityMb}). This concludes the derivation.
\par We will explain in plain terms how things work for the fusion kernel. On top of the periodicity tools that we have described so far, in order to prove the reflection properties of the fusion kernels $\mathbf{F}^{(\pm)}$ one has to invoke a very special (and a priori surprising) symmetry of the sums in (\ref{eq:splitfusmain2}) which is given by the Weyl group $W(D_6)\cong S_6\ltimes \mathbb{Z}_2^6$. This is exactly the same symmetry as the one enjoyed by the integral in the original Teschner-Vartanov expression for the fusion kernel (see \cite{Eberhardt:2023mrq} for a nice explanation and relevant proofs), except in our case one can see that the symmetry also exchanges the two roots of the quantum fusion polynomial $z^{(\text{f})}_{1}\leftrightarrow z^{(\text{f})}_{2}$. The remaining factors are then easily analyzed under $P_s(P_t)\rightarrow-P_s(-P_t)$, namely the fusion determinant $\mathfrak{D}^{\text{(f)}}$ is manifestly invariant, and the prefactor $F_\b(P_i|P_s,P_t)$ gets exactly the \textit{inverse} of the covariant factor\footnote{Explicitly, these factors are \be
\bea
F_{\b}(P_i|P_s,-P_t)&=F_{\b}(P_i|P_s,P_t)\times \prod_{f= \{23t\},\{14t\}} \prod_{\substack{\sigma\in \mathbb{Z}_2^f| \\ \sigma_f = 1}} \tilde{G}_{m,n}\left(\textstyle{\sum}_{i\in f}\sigma_i\frac{P_i}{\s}\right),\\
F_{\b}(P_i|-P_s,P_t)&=F_{\b}(P_i|P_s,P_t)\times \prod_{f= \{12s\},\{34s\}} \prod_{\substack{\sigma\in \mathbb{Z}_2^f| \\ \sigma_f = -1}} \tilde{G}_{m,n}\left(\textstyle{\sum}_{i\in f}\sigma_i\frac{P_i}{\s}\right)^{-1}.
\eea
\ee} that is acquired from the sums due to the aforementioned symmetry, thereby leading to
\be
\bea
\mathbf{F}^{(\pm)}_{P_s,-P_t}\left[\begin{smallmatrix} P_2 & P_3 \\ P_1 & P_4 \end{smallmatrix}\right]=\mathbf{F}^{(\mp)}_{P_s,P_t}\left[\begin{smallmatrix} P_2 & P_3 \\ P_1 & P_4 \end{smallmatrix}\right] \quad \text{and} \quad \mathbf{F}^{(\pm)}_{-P_s,P_t}\left[\begin{smallmatrix} P_2 & P_3 \\ P_1 & P_4 \end{smallmatrix}\right]=\mathbf{F}^{(\mp)}_{P_s,P_t}\left[\begin{smallmatrix} P_2 & P_3 \\ P_1 & P_4 \end{smallmatrix}\right].
\eea
\ee
\par The aforementioned behaviour of $\mathbf{M}^{(\pm)},\mathbf{F}^{(\pm)}$ under reflections in the internal momenta is expected to lead to the following \textit{idempotency relations}:
\begin{align} 
\int_{i\mathbb{R}}\text{d}P_t \ \mathbf{M}^{(\pm)}_{P_s,P_t}[P_0]\mathbf{M}^{(\pm)}_{P_t,P_s'}[P_0]&\overset{!}{=}\delta(P_s+P_s')  , \label{eq:modinv1} \\\int_{i\mathbb{R}}\text{d}P_t \ \mathbf{M}^{(\pm)}_{P_s,P_t}[P_0]\mathbf{M}^{(\mp)}_{P_t,P_s'}[P_0]&\overset{!}{=}\delta(P_s-P_s'),
\end{align}
and
\begin{align} 
\int_{i\mathbb{R}}\text{d}P_t \ \mathbf{F}^{(\pm)}_{P_s,P_t}\left[\begin{smallmatrix} P_2 & P_3 \\ P_1 & P_4 \end{smallmatrix}\right]\mathbf{F}^{(\pm)}_{P_t,P_s'}\left[\begin{smallmatrix} P_2 & P_1 \\ P_3 & P_4 \end{smallmatrix}\right]&\overset{!}{=}\delta(P_s+P_s')  , \\ \ \int_{i\mathbb{R}}\text{d}P_t \ \mathbf{F}^{(\pm)}_{P_s,P_t}\left[\begin{smallmatrix} P_2 & P_3 \\ P_1 & P_4 \end{smallmatrix}\right]\mathbf{F}^{(\mp)}_{P_t,P_s'}\left[\begin{smallmatrix} P_2 & P_1 \\ P_3 & P_4 \end{smallmatrix}\right]&\overset{!}{=}\delta(P_s-P_s').\label{eq:fusinv2}
\end{align}
It is an important task to prove concretely that the kernels $\mathbf{M}^{(\pm)}, \mathbf{F}^{(\pm)}$ satisfy conditions such as (\ref{eq:modinv1})-(\ref{eq:fusinv2}) or, more broadly, the general consistency conditions of the Virasoro crossing kernels as described by Moore and Seiberg \cite{Moore:1988uz} (see also \cite{Eberhardt:2023mrq}) . In the following we will provide strong evidence towards the validity of these conditions by proving concretely that our new kernels satisfy the basic shift relations for the modular and fusion kernels (which essentially follow from the Moore-Seiberg conditions).
\subsection{Shift relations}\label{sec:shiftrel}
We now turn to the shift relations and show that the non-meromorphic kernels $\mathbf{M}^{(\pm)}_{P_s,P_t}[P_0]$ and $\mathbf{F}^{(\pm)}_{P_s,P_t}\left[\begin{smallmatrix} P_2 & P_3 \\ P_1 & P_4 \end{smallmatrix}\right]$ \textit{individually} satisfy the basic shift relations for the modular and fusion kernel respectively. 
\paragraph{Modular kernel.} For the case of the modular kernel, and for general $b^2$ (not necessarily rational), the shift relations read (see e.g. \cite{Nemkov:2015zha, Ribault:2023vqs})
\begin{subequations}
\begin{align}
 \sum_\pm A_b(\pm P_s,P_0) e^{\pm \frac{b}{2}\partial_{P_s}} \mathbf{M}^{(b)}_{P_s,P_t}[P_0] &= 2\cos (2\pi bP_t) \mathbf{M}^{(b)}_{P_s,P_t}[P_0]\ ,
 \label{amm}
 \\
 \sum_\pm e^{\pm \frac{b}{2}\partial_{P_t}} A_b(\mp P_t,P_0) \mathbf{M}^{(b)}_{P_s,P_t}[P_0] &= 2\cos(2\pi bP_s) \mathbf{M}^{(b)}_{P_s,P_t}[P_0]\ ,
 \label{ammp}
 \\
 \sum_\pm e^{-b\partial_{P_0}} E_b(\pm P_s,P_t,P_0) e^{\pm \frac{b}{2} \partial_{P_s}} \mathbf{M}^{(b)}_{P_s,P_t}[P_0] &=  \mathbf{M}^{(b)}_{P_s,P_t}[P_0]\ .
 \label{ebpz}
\end{align}
\end{subequations}
The coefficients are the following combinations of Gamma functions
\begin{align}
 A_b(P_s,P_0) &= \frac{\Gamma(2bP_s)\Gamma(1+b^2+ 2bP_s)}{\prod_{\pm} \Gamma(\frac{1+b^2}{2} \pm bP_0 + 2bP_s)} \ , 
 \label{ab}
 \\
 E_b(P_s,P_t,P_0)  &= \frac{1}{2\pi} \Gamma(2bP_s)\Gamma(1+b^2+2bP_s) \frac{\prod_\pm \Gamma(\frac12 -\frac{b^2}{2} -bP_0 \pm 2bP_t)}{\prod_\pm \Gamma(\frac12\pm \frac{b^2}{2} -bP_0 +2bP_s)}\ .
 \label{eb}
\end{align}
Similar shift relations hold if we replace $b\rightarrow b^{-1}$. Since we already know that the Teschner kernel $\mathbf{M}^{(b=\b)}_{P_s,P_t}[P_0]$ satisfies the shift relations, we only need to show that one of the two non-meromoprhic kernels, say $\mathbf{M}^{(+)}$, satisfies it too. Then the result for $\mathbf{M}^{(-)}$ follows from linearity of the equations. Here we will show explicitly that $\mathbf{M}^{(+)}_{P_s,P_t}[P_0]$ satisfies the shift relation (\ref{amm}). The remaining two equations can be proved similarly with the tools that we describe below.
\
\\
\textit{Proof}: Without loss of generality in the following we consider one of the co-prime integers to be odd, i.e. $m\in \mathbb{Z}_{\text{odd}}\geq1$\footnote{Since $(m,n)$ are co-primes we necessarily have at least one of the two to be an odd integer. There is no loss of generality in picking $m$ to be so, since we have a symmetry $m\leftrightarrow n$ (i.e. $b\leftrightarrow b^{-1}$.)}. Let us then split $\mathbf{M}^{(+)}$ into two factors as
\be
\bea
\mathbf{M}^{(+)}_{P_s,P_t}[P_0]&=\mathcal{P}^{(\text{m})}(P_0,P_s,P_t)\times \mathcal{S}_1^{(\text{m})}(P_0,P_s,P_t)
\eea
\ee
with
\be
\bea
\mathcal{P}^{(\text{m})}(P_0,P_s,P_t)&\equiv \frac{M_\b(P_0|P_s,P_t)}{2i\sin{\left(2\pi \s P_t\right)}\sin{\left(2\pi \s P_s\right)}\sqrt{\mathfrak{D}^{(m)}}}, \\
\mathcal{S}_1^{(\text{m})}(P_0,P_s,P_t)&\equiv\sum_{k=0}^{\s^2-1}\mathcal{M}_{\b}\left(\frac{\log{z^{(\text{m})}_1}}{2\pi i \s^2}+\frac{k}{\s^2}\right).
\eea
\ee 
It is straightforward to check first that
\be
\bea
A_\b(\pm P_s,P_0)e^{ \pm\frac{ \b}{2}\partial_{P_s}} \left[\mathcal{P}^{(\text{m})}(P_0,P_s,P_t)\right]=\pm\frac{\cos{\left(2\pi\b\left(\pm P_s+\frac{P_0}{2}+\frac{\b}{4}\right)\right)}}{\sin{\left(2\pi \b P_s\right)}}\mathcal{P}^{(\text{m})}(P_0,P_s,P_t).
\eea
\ee
So overall we need to show
\be\label{eq:crazyidentity2}
\bea
&i\text{ $\cos$}{\left(2\pi\b\left(P_s+\frac{P_0}{2}+\frac{\b}{4}\right)\right)}e^{\frac{ \b}{2}\partial_{P_s}}\left[\mathcal{S}_1^{(\text{m})}(P_0,P_s,P_t)\right]\\
&\quad \quad  -i \text{ $\cos$}{\left(2\pi\b\left(-P_s+\frac{P_0}{2}+\frac{\b}{4}\right)\right)}e^{-\frac{ \b}{2}\partial_{P_s}}\left[\mathcal{S}_1^{(\text{m})}(P_0,P_s,P_t)\right]\\
& \ \quad \quad \quad \quad \quad \quad \quad \quad = 2 i \text{ $\sin{\left(2\pi \b P_s\right)}\cos{(2\pi \b P_t)}$} \ \mathcal{S}_1^{(\text{m})}(P_0,P_s,P_t).
\eea
\ee
Denoting $u^*_{(k)}\equiv \frac{\log{z^{(\text{m})}_1}}{2\pi i \s^2}+\frac{k}{\s^2}$, we notice that
\be
\bea
e^{\pm \frac{ \b}{2}\partial_{P_s}}\left[\mathcal{S}_1^{(\text{m})}(P_0,P_s,P_t)\right]=\sum_{k=0}^{\s^2-1}e^{\mp \pi i \b \times (2 \s u^{*}_{(k)})}\mathcal{M}_{\b}\left(u^*_{(k)}\right).
\eea
\ee
This yields
\be\label{eq:LHSgoal}
\bea
\text{LHS of (\ref{eq:crazyidentity2})}&=e^{2\pi i \b P_s}\sum_{k=0}^{\s^2-1}\mathcal{M}_{\b}\left(u^*_{(k)}\right)\times \sin{\left(\pi \b \left(2 \s u^{*}_{(k)}-P_0-\frac{\b}{2}\right)\right)}\\
& \ \ \ +e^{-2\pi i \b P_s}\sum_{k=0}^{\s^2-1}\mathcal{M}_{\b}\left(u^*_{(k)}\right)\times \sin{\left(\pi \b \left(2 \s u^{*}_{(k)}+P_0+\frac{\b}{2}\right)\right)}.
\eea
\ee
Using (\ref{eq:speccasestGshift}) we can derive the following identity for general $u\in\mathbb{C}$:
\be
\bea
e^{2\pi i \b P_s}&\mathcal{M}_{\b}\left(u+\frac{1}{n}\right) \left[\cos{\left(2\pi \b P_t\right)}+\sin{\left(\pi \b\left(-2\s (u+1/n)+P_0+\frac{\b}{2}\right)\right)}\right]\\
&=e^{-2\pi i \b P_s}\mathcal{M}_{\b}\left(u\right)\left[\cos{\left(2\pi \b P_t\right)}+\sin{\left(\pi \b\left(2\s u+P_0+\frac{\b}{2}\right)\right)}\right].
\eea
\ee
Now let us evaluate the last relation at $$u=u^*_{(k)}\equiv \frac{\log{z^{(\text{m})}_1}}{2\pi i \s^2}+\frac{k}{\s^2}$$
 and sum over $k$ as prescribed in $\mathbf{M}^{(+)}$. We get
 \be\label{eq:summstepproof}
 \bea
{}&e^{2\pi i \b P_s}\cos{\left(2\pi \b P_t\right)}\sum_{k=0}^{\s^2-1}\mathcal{M}_{\b}\left(u^*_{(k)}+\frac{1}{n}\right)\\
&\quad \quad \quad \ \ +e^{2\pi i \b P_s}\sum_{k=0}^{\s^2-1}\mathcal{M}_{\b}\left(u^*_{(k)}+\frac{1}{n}\right)\sin{\left[\pi \b\left(-2\s \left(u^*_{(k)}+\frac{1}{n}\right)+P_0+\frac{\b}{2}\right)\right]}\\
&=e^{-2\pi i \b P_s}\cos{\left(2\pi \b P_t\right)}\sum_{k=0}^{\s^2-1}\mathcal{M}_{\b}\left(u^*_{(k)}\right)\\
&\quad \quad \quad \quad \ \ \ +e^{-2\pi i \b P_s}\sum_{k=0}^{\s^2-1}\mathcal{M}_{\b}\left(u^*_{(k)}\right)\sin{\left[\pi \b\left(2\s \left(u^*_{(k)}\right)+P_0+\frac{\b}{2}\right)\right]}.
 \eea
 \ee
 Rearranging and using the periodicity property (\ref{eq:periodsumMStrong}) with the choice of periodic functions $f(x)\equiv 1$ and $f(x)\equiv \sin{(x)}=\sin{(x+2\pi m)}$ we arrive at
 \be\label{eq:finalstepproof}
 \bea
2i&\sin{(2\pi \b P_s)}\cos{\left(2\pi \b P_t\right)}\sum_{k=0}^{\s^2-1}\mathcal{M}_{\b}\left(u^*_{(k)}\right)=\\
 &\quad e^{2\pi i \b P_s}\sum_{k=0}^{\s^2-1}\mathcal{M}_{\b}\left(u^*_{(k)}\right)\sin{\left[\pi \b\left(2\s u^*_{(k)}-P_0-\frac{\b}{2}\right)\right]} \\
 &\quad +e^{-2\pi i \b P_s}\sum_{k=0}^{\s^2-1}\mathcal{M}_{\b}\left(u^*_{(k)}\right)\sin{\left[\pi \b\left(2\s u^*_{(k)}+P_0+\frac{\b}{2}\right)\right]}. \\
& \quad \quad \quad \quad \hfill  \quad \quad \quad \quad \hfill  \quad \quad \quad \quad \hfill  \quad \quad \quad \quad \hfill \quad \quad \quad \quad \hfill  \quad \quad \quad \quad \hfill  \quad \quad \quad \quad \hfill  \quad \hfill  \square
 \eea
 \ee
This is exactly (\ref{eq:crazyidentity2}) (c.f. (\ref{eq:LHSgoal})) and hence it concludes the derivation.
\par Two important remarks are in order. First, the summation step done in (\ref{eq:summstepproof}) is pivotal; without it  (and without the periodicities that the sum obeys, as described in (\ref{eq:periodsumMStrong})) we wouldn't be able to arrive at the final step (\ref{eq:finalstepproof}). Second, when we evaluated the expression at $u=u^*_{(k)}$ we didn't use anywhere the information that we chose $z_1$, instead of $z_2$. The only important point was that we had a solution of the quantum modular polynomial. Therefore, the proof carries through in an identical way for $\mathbf{M}^{(-)}$.
\par Exactly analogous steps and logic lead to the proof of (\ref{ammp}), (\ref{ebpz}). It is essential to mention that the shift relations determine the crossing kernels up to a momentum-independent constant. The fact that each of $\mathbf{M}^{(\pm)}$ as defined in (\ref{eq:splitmodmain2}) satisfies the crossing transformation of the blocks (\ref{eq:mpmmod}) with no additional overall coefficient is supported from the analytic derivation of the result as we describe it in Appendix \ref{appb1} as well as from the proof of the shift relations that we just presented.
\paragraph{Fusion kernel.} For general $b^2$ the basic shift relation that fixes the Teschner-Vartanov fusion kernel  \cite{Eberhardt:2023mrq} is
\begin{multline} 
\sum_{\eta=\pm} \frac{ \Gamma(-b^2-2b \eta P_{t})\Gamma(1-2b \eta P_{t})}{\Gamma(\frac{1}{2}\pm b P_2+b P_3-b \eta P_{t})\Gamma(\frac{1}{2} \pm b P_1-b P_4- b \eta P_{t})} \, \mathbf{F}^{(b)}_{P_s,P_t+\frac{\eta b}{2}}\left[\begin{smallmatrix} P_2 & P_3 \\ P_1 & P_4 \end{smallmatrix}\right] \\
=\frac{1}{\Gamma(\frac{1+b^2}{2} \pm b P_{s}+b P_3-b P_4)} \, \mathbf{F}^{(b)}_{P_s,P_t}\left[\begin{smallmatrix} P_2 & P_3+\frac{b}{2} \\ P_1 & P_4-\frac{b}{2} \end{smallmatrix}\right] \ ,\label{eq:shiftrelationTV}
\end{multline}
and similarly with $b\rightarrow b^{-1}$.
Denoting the integral in the Teschner-Vartanov expression (\ref{eq:deffusionkernel}) as
\be
\bea
\int_{i\mathbb{R}} \frac{du}{i} \ f_b(u)\equiv \sixjnorm{P_1}{P_2}{P_{s}}{P_3}{P_4}{P_{t}}_{(b)},
\eea
\ee
a direct consequence of (\ref{eq:shiftrelationTV}) is the following shift relation for the integral \cite{Eberhardt:2023mrq}
\begin{multline}
    \sum_{\eta=\pm} \eta \cos\big(\pi b(P_1+\eta P_4+P_t)\big)\cos\big(\pi b(P_2+\eta P_3-P_t)\big) \sixjnorm{P_1}{P_2}{P_s}{P_3}{P_4}{P_t + \frac{ \eta b}{2}}_{(b)} \\
    =\sin(2\pi b P_t) \cos\big(\tfrac{\pi b^2}{2}+\pi b(-P_s+P_3-P_4)\big) \sixjnorm{P_1}{P_2}{P_s}{P_3+\frac{b}{2}}{P_4-\frac{b}{2}}{P_t}_{(b)}\ . \label{eq:mainshiftfusiona}
\end{multline}
We want to study this particular shift relation in the case $b=\b$, since as we explained the kernels $\mathbf{F}^{(\pm)}$ only differ in the contributions of this particular integral. According to our notation from section \ref{sec:fusionkernelcgeq25} we have
\be
\bea
\sixjnorm{P_1}{P_2}{P_{s}}{P_3}{P_4}{P_{t}}_{(b=\b)}&\propto \int_{i\mathbb{R}} \frac{du}{i} \ f_\b(u)\\
&=\frac{1}{2\sqrt{\Delta^{\text{(f)}}}}\sum_{k=0}^{\s^2-1}\mathcal{F}_{\b}\left(\frac{\log{z_1^{(\text{f})}}}{2\pi i \s^2}+\frac{k}{\s^2}\right)-\frac{1}{2\sqrt{\Delta^{\text{(f)}}}}\sum_{k=0}^{\s^2-1}\mathcal{F}_{\b}\left(\frac{\log{z_2^{(\text{f})}}}{2\pi i \s^2}+\frac{k}{\s^2}\right)\\
&\equiv \frac{\mathcal{S}_1^{(\text{f})}(P_i;P_s,P_t)}{4is_{mn}(P_t)s_{mn}(P_s)\sqrt{\mathfrak{D}^{(\text{f})}}}-\frac{\mathcal{S}_2^{(\text{f})}(P_i;P_s,P_t)}{4is_{mn}(P_t)s_{mn}(P_s)\sqrt{\mathfrak{D}^{(\text{f})}}}\\
&\equiv \sixjnorm{P_1}{P_2}{P_{s}}{P_3}{P_4}{P_{t}}_{(\b)}^{(1)}+\sixjnorm{P_1}{P_2}{P_{s}}{P_3}{P_4}{P_{t}}_{(\b)}^{(2)}.
\eea
\ee
The proportionality constant in the first line only depends on the coprime pair $(m,n)$ and hence it is irrelevant at the level of the shift relations. \par We will show explicitly that (\ref{eq:mainshiftfusiona}) is satisfied individually for $\sixjnorm{P_1}{P_2}{P_{s}}{P_3}{P_4}{P_{t}}_{(\b)}^{(i)}$, $i=1,2$ when $b=\b$. The proof is intricate and beautiful, and we relegate it to Appendix \ref{app:shiftrelFpm}. As in the case of the modular kernel, the essential ingredient is the main periodicity property (\ref{eq:periodsumFStrong}). 
\par This establishes that the non-meromorphic kernels $\mathbf{F}^{(\pm)}$
satisfy the shift relations of the fusion kernel, and therefore implement the crossing transformation (\ref{eq:fpmfus}) for the sphere four-point blocks.
\subsection{Liouville theory at rational $c\geq25$}\label{sec:liouvillec25}
The Virasoro crossing kernels were originally constructed, among other reasons, to prove crossing symmetry and modular covariance of Liouville theory \cite{Teschner:2001rv}. 
For $c\in \mathbb{C}\backslash(-\infty,1]$ (or $b^2\in\mathbb{C}\backslash(-\infty,0)$), there is a field normalization such that the two- and three-point structure constants of Liouville theory read \cite{Ribault:2014hia}
\begin{align}\label{eq:liouvillesc}
 B^{(b)}_{P} := \prod_\pm \Gamma_b(\pm 2P)\Gamma_b(Q\pm 2P) , \ \ \ \ C^{(b)}_{P_1,P_2,P_3} := \prod_{\pm,\pm,\pm}\Gamma_b\left(\tfrac{Q}{2}\pm P_1\pm P_2\pm P_3\right)\ .
\end{align}
These expressions define meromorphic functions in the Liouville momenta and are manifestly reflection-symmetric in each variable and invariant under $b\leftrightarrow b^{-1}$. The structure constant is also manifestly permutation-symmetric in the three momenta.
\par Using the fact that the primary operator spectrum in Liouville theory is diagonal with $P=\bar{P}\in i \mathbb{R}$, the statement of crossing symmetry of the sphere four-point functions can be recast in terms of the Virasoro fusion kernel (\ref{eq:deffusionkernel}) as the following non-trivial identity (valid in principle for $b^2\in\mathbb{C}\backslash(-\infty,0)$)\cite{Teschner:2001rv}\footnote{In technical terms, one should properly view this identity as a \textit{distributional} equality valid when integrated against suitable test functions ($P_s$ being the integration variable on the LHS, and $P_t$ on the RHS). We will see the importance of this remark in section \ref{sec:liouvillecless1} when we discuss the case of $c\leq1$ Liouville theory.}
\be
\bea
\frac{C^{(b)}_{P_1,P_2,P_s}C^{(b)}_{P_s,P_3,P_4}}{B^{(b)}_{P_s}}\mathbf{F}^{(b)}_{P_s,P_t}\left[\begin{smallmatrix} P_2 & P_3 \\ P_1 & P_4 \end{smallmatrix}\right]=\frac{C^{(b)}_{P_2,P_3,P_t}C^{(b)}_{P_t,P_1,P_4}}{B^{(b)}_{P_t}}\mathbf{F}^{(b)}_{P_t,P_s}\left[\begin{smallmatrix} P_2 & P_1 \\ P_3 & P_4 \end{smallmatrix}\right].
\eea
\ee
When $b=\b$, we can rewrite this using our results 
\begin{align}\label{eq:crossLiouvbb}
\frac{C^{(\b)}_{P_1,P_2,P_s}C^{(\b)}_{P_s,P_3,P_4}}{B^{(\b)}_{P_s}}&\left(\mathbf{F}^{(+)}_{P_s,P_t}\left[\begin{smallmatrix} P_2 & P_3 \\ P_1 & P_4 \end{smallmatrix}\right]+\mathbf{F}^{(-)}_{P_s,P_t}\left[\begin{smallmatrix} P_2 & P_3 \\ P_1 & P_4 \end{smallmatrix}\right]\right)\\
&=\frac{C^{(\b)}_{P_2,P_3,P_t}C^{(\b)}_{P_t,P_1,P_4}}{B^{(\b)}_{P_t}}\left(\mathbf{F}^{(+)}_{P_t,P_s}\left[\begin{smallmatrix} P_2 & P_1 \\ P_3 & P_4 \end{smallmatrix}\right]+\mathbf{F}^{(-)}_{P_t,P_s}\left[\begin{smallmatrix} P_2 & P_1 \\ P_3 & P_4 \end{smallmatrix}\right]\right)\nonumber,
\end{align}
with (c.f. Appendix \ref{appendixA})
\be
\bea
 B^{(\b)}_{P} &= \gamma_{m,n}^4\times  \s^{-8P^2}\prod_{\pm}\left[G_{m,n}\left(\pm\frac{2P}{\s}\right)G_{m,n}\left(m^{-1}+n^{-1}\pm \frac{2P}{\s}\right)\right]^{-1},\\
 C^{(\b)}_{P_1,P_2,P_3} &=\gamma_{m,n}^8\times \s^{2+\b^2+\b^{-2}-4 \left(P_1^2+P_2^2+P_3^2\right)} \prod_{\pm,\pm,\pm}\left[G_{m,n}\left(\frac{m^{-1}+n^{-1}}{2}+\frac{\pm P_1\pm P_2\pm P_3}{\s}\right)\right]^{-1}.
\eea
\ee
\par We will now show that an identical relation is satisfied individually for either $\mathbf{F}^{(\pm)}$, and hence (\ref{eq:crossLiouvbb}) can be understood merely as a linear combination of the former. Indeed, it is straightforward to see that the prefactor $F_{\b}(P_i|P_s,P_t)$ in (\ref{eq:prefintegdFrational}) satisfies
\be
\bea
\frac{\left.F_{\b}(P_i|P_s,P_t)\right|_{P_1\leftrightarrow P_3,P_t\leftrightarrow P_s}}{F_{\b}(P_i|P_s,P_t)}=\frac{B^{(\b)}_{P_t}C^{(\b)}_{P_1,P_2,P_s}C^{(\b)}_{P_s,P_3,P_4}}{B^{(\b)}_{P_s}C^{(\b)}_{P_2,P_3,P_t}C^{(\b)}_{P_t,P_1,P_4}}.
\eea
\ee
On the other hand, from its definition the function $\mathcal{F}_\b$ is \textit{invariant} under such change for any $u\in\mathbb{C}$
\be
\bea
\left.\mathcal{F}_\b(u)\right|_{P_1\leftrightarrow P_3,P_t\leftrightarrow P_s}=\mathcal{F}_\b(u).
\eea
\ee
Due to the symmetries of the quantum fusion polynomial that we explained in section \ref{sec:fusionkernelcgeq25}, the fusion determinant $\mathfrak{D}^{(\text{f})}$ and the two sums (corresponding to the two roots $z^{(\text{f})}_{1,2}$ or, to $\mathbf{F}^{(\pm)}$) are also \textit{invariant} under exchanging $P_1\leftrightarrow P_3,P_t\leftrightarrow P_s$, and therefore we conclude
\be\label{eq:LiouvilleFpm}
\bea
\frac{C^{(\b)}_{P_1,P_2,P_s}C^{(\b)}_{P_s,P_3,P_4}}{B^{(\b)}_{P_s}}&\mathbf{F}^{(+)}_{P_s,P_t}\left[\begin{smallmatrix} P_2 & P_3 \\ P_1 & P_4 \end{smallmatrix}\right]=\frac{C^{(\b)}_{P_2,P_3,P_t}C^{(\b)}_{P_t,P_1,P_4}}{B^{(\b)}_{P_t}}\mathbf{F}^{(+)}_{P_t,P_s}\left[\begin{smallmatrix} P_2 & P_1 \\ P_3 & P_4 \end{smallmatrix}\right],\\
\frac{C^{(\b)}_{P_1,P_2,P_s}C^{(\b)}_{P_s,P_3,P_4}}{B^{(\b)}_{P_s}}&\mathbf{F}^{(-)}_{P_s,P_t}\left[\begin{smallmatrix} P_2 & P_3 \\ P_1 & P_4 \end{smallmatrix}\right]=\frac{C^{(\b)}_{P_2,P_3,P_t}C^{(\b)}_{P_t,P_1,P_4}}{B^{(\b)}_{P_t}}\mathbf{F}^{(-)}_{P_t,P_s}\left[\begin{smallmatrix} P_2 & P_1 \\ P_3 & P_4 \end{smallmatrix}\right].
\eea
\ee
\par Working similarly it is a small extension to show that the non-meromorphic modular kernels $\mathbf{M}^{(\pm)}$ also satisfy the torus one-point modular covariance statement in Liouville theory \cite{Hadasz:2009sw} when $b=\b$, namely
\be\label{eq:LiouvilleMpm}
\bea
\frac{C^{(\b)}_{P_0,P_s,P_s}}{B^{(\b)}_{P_s}}\mathbf{M}^{(+)}_{P_s,P_t}[P_0]=\frac{C^{(\b)}_{P_0,P_t,P_t}}{B^{(\b)}_{P_t}}\mathbf{M}^{(+)}_{P_t,P_s}[P_0], \quad 
\frac{C^{(\b)}_{P_0,P_s,P_s}}{B^{(\b)}_{P_s}}\mathbf{M}^{(-)}_{P_s,P_t}[P_0]=\frac{C^{(\b)}_{P_0,P_t,P_t}}{B^{(\b)}_{P_t}}\mathbf{M}^{(-)}_{P_t,P_s}[P_0].
\eea
\ee
\par The relations (\ref{eq:LiouvilleFpm}) and (\ref{eq:LiouvilleMpm}) reinforce the conclusion that the kernels $\mathbf{M}^{(\pm)}$, $\mathbf{F}^{(\pm)}$ individually satisfy the crossing transformation of the corresponding conformal blocks, where the support of the integral is given exactly by the Liouville spectrum $P_t\in i \mathbb{R}$. As we have explained, the kernels $\mathbf{M}^{(\pm)},\mathbf{F}^{(\pm)}$ possess square-root branch point singularities and therefore it is tempting to imagine deforming the contours in (\ref{eq:mpmmod}), (\ref{eq:fpmfus}) and picking up discontinuities across the various branch cuts as well as various poles. One could then ask whether the resulting support defines a crossing-symmetric (and possibly new?) theory. Such analytic continuations are common in Liouville theory and have been explored in the past, though the novelty here is exactly the presence of square-root branch cuts which is worth exploring in this context. We defer these investigations to future work.

\section{Virasoro Kernels at rational $c\in(-\infty,1]$}\label{sec:cleq1kernels}
In the present section we will discuss the Virasoro crossing kernels at $c\leq1$ and rational. In particular, we will use the symmetry of the shift relations under Virasoro-Wick Rotation (c.f. (\ref{eq:VWRdefn})) to derive the (physical) modular and fusion kernels valid at $b^2\in\mathbb{Q}_{<0}$ (or $c\in\mathbb{Q}_{(-\infty,1]}$) and for generic values of the Liouville momenta $P_i$. As described in the introduction here we will use the parameter $\hat{b}$ instead, which for the rational case takes values $\hat{b}=\b$. This means that whenever we use $(m,n)$ in the present section, these should be thought of as defining a central charge
\be
c=13-6\left(\frac{m}{n}+\frac{n}{m}\right)\leq1.
\ee
We will see that the modular and fusion kernels can again be expressed as linear combinations of two functions, each of which is an admissible crossing kernel for $c\leq1$. The main novelty compared to the Teschner and Teschner-Vartanov solutions for $c\geq 25$ is that the full kernels possess square-root branch point singularities, but are otherwise even functions in the involved momenta. Finally, we use these results to demonstrate that timelike Liouville theory \cite{Zamolodchikov:2005fy,Schomerus:2003vv,Kostov:2005kk} at $c\in\mathbb{Q}_{(-\infty,1]}$ is crossing symmetric and modular covariant.
\subsection{Modular Kernel}\label{sec:modkerncleq1}
In section \ref{sec:shiftrel} we showed that the non-meromorphic kernels $\mathbf{M}^{(\pm)}$ are both solutions to the modular kernel shift relations for $c\in\mathbb{Q}_{[25,\infty)}$. Using the Virasoro-Wick Rotation symmetry of the shift relations \cite{Ribault:2023vqs} we can therefore derive two other solutions $\hat{\mathbf{M}}^{(\pm)}$ valid at $c\in\mathbb{Q}_{(-\infty,1]}$. These take the form\footnote{The choice of the factors of $i$ can be understood from the Jacobian of the change of variables  $P\rightarrow iP$ as dictated from the VWR, see \cite{Ribault:2023vqs}.}
\be\label{eq:defnmhpm}
\bea
\hat{\mathbf{M}}^{(\pm)}_{P_s,P_t}[P_0]&:=\mp i \mathfrak{R}\mathbf{M}^{(\pm)}_{P_s,P_t}[P_0].
\eea
\ee
More explicitly,
\be\label{eq:splitmodleq1main2}
\boxed{
\bea
\hat{\mathbf{M}}^{(+)}_{P_s,P_t}[P_0]&=\left(\frac{P_t}{P_s}\right)\frac{M_\b(iP_0|iP_t,iP_s)}{2\sinh{\left(2\pi \s P_t\right)}\sinh{\left(2\pi \s P_s\right)}}\times \frac{1}{\sqrt{\hat{\mathfrak{D}}^{(m)}}}\sum_{k=0}^{\s^2-1}\hat{\mathcal{M}}_{\b}\left(\frac{\log{\hat{z}^{(\text{m})}_1}}{2\pi i \s^2}+\frac{k}{\s^2}\right),\\
\hat{\mathbf{M}}^{(-)}_{P_s,P_t}[P_0]&=\left(\frac{P_t}{P_s}\right)\frac{M_\b(iP_0|iP_t,iP_s)}{2\sinh{\left(2\pi \s P_t\right)}\sinh{\left(2\pi \s P_s\right)}}\times \frac{1}{\sqrt{\hat{\mathfrak{D}}^{(m)}}}\sum_{k=0}^{\s^2-1}\hat{\mathcal{M}}_{\b}\left(\frac{\log{\hat{z}^{(\text{m})}_2}}{2\pi i \s^2}+\frac{k}{\s^2}\right)
\eea
}.
\ee
$M_{\b}$ is defined in (\ref{eq:prefintegdMrational}), and
\be
\bea
\hat{\mathcal{M}}_{\b}(u)\equiv \frac{e^{4\pi  \s P_t u}\prod_{\pm}\tilde{G}_{m,n}\left(u-1+\frac{m^{-1}+n^{-1}}{4}-\frac{i}{2}\left(\pm\frac{2P_s}{\s}+\frac{P_0}{\s}\right)\right)}{\prod_{\pm}\tilde{G}_{m,n}\left(u-\frac{m^{-1}+n^{-1}}{4}-\frac{i}{2}\left(\pm\frac{2P_s}{\s}-\frac{P_0}{\s}\right)\right)}.
\eea
\ee
The data $\{\hat{z}_{1}^{(\text{m})},\hat{z}_{2}^{(\text{m})},\hat{\mathfrak{D}}^{(\text{m})}\}$ originate from a degree-two polynomial which is simply the VWR-ed quantum modular polynomial (\ref{eq:modpoly}). Denoting $\text{ch}_{mn}(x)\equiv \cosh{(2\pi \s x)}, \text{sh}_{mn}(x)\equiv \sinh{(2\pi \s x)}$, we have
\begin{align}\label{eq:VWRmoddet}
\hat{\mathfrak{D}}^{(m)}&\equiv 1+\left(\frac{\text{sh}_{mn}{\left(\frac{P_0}{2}+ \frac{i(m+n)}{4\s}\right)}}{\text{sh}_{mn}{\left(P_t\right)} \ \text{sh}_{mn}{\left(P_s\right)}}\right)^2,\\
\hat{\Delta}^{(m)}&\equiv -4 \ \text{sh}^2_{mn}(P_t)\text{sh}^2_{mn}(P_s) \ \hat{\mathfrak{D}}^{(m)},\label{eq:VWRactualDet}\\
\hat{z}_{1,2}^{(\text{m})}&\equiv \frac{\text{sh}_{mn}{\left(P_t\right)}}{\text{sh}_{mn}{\left(P_0/2-P_t+\frac{i(m+n)}{4\s}\right)}}\left[(-1)^{mn}\text{ch}_{mn}{\left(P_s\right)}\pm  \text{sh}_{mn}{\left(P_s\right)}\left(\hat{\mathfrak{D}}^{(\text{m})}\right)^{1/2}\right]\label{eq:VWRrootsmod}.
\end{align}
In parallel to the previous section, we can again define $4\times 4$ matrices whose determinants capture the VWR-ed modular determinant (\ref{eq:VWRmoddet}) (or the actual discriminant (\ref{eq:VWRactualDet})). Indeed, consider
\be\label{eq:modgram1hat}
\bea
\hat{\mathcal{O}}^{\text{(m)}}_{m,n}:=\begin{pmatrix}
1 &  -\text{ch}_{mn}\left(P_s\right) & 0 &  0 \\
 -\text{ch}_{mn}\left(P_s\right) & 1 & i\text{sh}_{mn}\left(\frac{P_0}{2}+\frac{i(m+n)}{4\s}\right) & 0 \\
0 & i\text{sh}_{mn}\left(\frac{P_0}{2}+\frac{i(m+n)}{4\s}\right) & 1 & -\text{ch}_{mn}\left(P_t\right) \\
0  & 0 & -\text{ch}_{mn}\left(P_t\right) & 1
\end{pmatrix},
\eea
\ee
and
\be\label{eq:modgram2hat}
\bea
\hat{\mathcal{G}}^{\text{(m)}}_{m,n}:=\begin{pmatrix}
1 &  i\text{sh}_{mn}\left(\frac{P_0}{2}+\frac{i(m+n)}{4\s}\right) & i\text{sh}_{mn}\left(\frac{P_0}{2}+\frac{i(m+n)}{4\s}\right) &  -\text{ch}_{mn}\left(2P_s\right) \\
 i\text{sh}_{mn}\left(\frac{P_0}{2}+\frac{i(m+n)}{4\s}\right) & 1 & -\text{ch}_{mn}\left(2P_t\right) & i\text{sh}_{mn}\left(\frac{P_0}{2}+\frac{i(m+n)}{4\s}\right) \\
i\text{sh}_{mn}\left(\frac{P_0}{2}+\frac{i(m+n)}{4\s}\right) & -\text{ch}_{mn}\left(2P_t\right) & 1 & i\text{sh}_{mn}\left(\frac{P_0}{2}+\frac{i(m+n)}{4\s}\right) \\
-\text{ch}_{mn}\left(2P_s\right)  & i\text{sh}_{mn}\left(\frac{P_0}{2}+\frac{i(m+n)}{4\s}\right) & i\text{sh}_{mn}\left(\frac{P_0}{2}+\frac{i(m+n)}{4\s}\right) & 1
\end{pmatrix}.
\eea
\ee
Then,
\be
\bea
\text{det}\left[\hat{\mathcal{O}}^{\text{(m)}}_{m,n}\right]=-\frac{\hat{\Delta}^{(m)}}{4}, \quad \quad  \text{det}\left[\hat{\mathcal{G}}^{\text{(m)}}_{m,n}\right]=-\left(1+\text{ch}_{mn}\left(2P_s\right)\right)\left(1+\text{ch}_{mn}\left(2P_t\right)\right)\hat{\Delta}^{(m)}.
\eea
\ee
\par Similar to what we discussed in section \ref{sec:modkernelcgeq25}, we recognize the matrix $\hat{\mathcal{O}}^{\text{(m)}}_{m,n}$ as the standard \textit{Gram matrix} that encodes the six dihedral angles $\hat{\psi}_i$ of a \textit{3-orthoscheme}, denoted as $OT(\hat{\psi}_1,\hat{\psi}_2,\hat{\psi}_3)$. The three (non-right) angles in our case read (c.f. (\ref{eq:genangleGram}))
\be
\bea
\hat{\psi}_1\equiv2\pi i \s P_s , \qquad \hat{\psi}_2\equiv\pi i \s  P_0+\frac{\pi}{2}(1-m-n) , \qquad \hat{\psi}_3\equiv2\pi i \s P_t.
\eea
\ee 
These are simply the ``$i$ rotated'' angles that we had in (\ref{eq:anglesOT}) for the $c\in\mathbb{Q}_{[25,\infty)}$ case, as expected from the prescription of the VWR. However if one insists on the angle-interpretation, in the present case we need to take $P_s,P_t,P_0\in i \mathbb{R}$, such that $\hat{\psi}_i\in(0,\pi)$. Analogously, the matrix $\hat{\mathcal{G}}^{\text{(m)}}_{m,n}$ can be interpreted as the \textit{vertex Gram matrix} that encodes the six edge lengths of a 3-tetrahedron that we denote as $T(\hat{\ell}^{(\text{m})}_i)$ with (c.f. (\ref{eq:genlengthGram}))
\be
\bea
\hat{\ell}^{(\text{m})}_1&=\hat{\ell}^{(\text{m})}_2=\hat{\ell}^{(\text{m})}_4=\hat{\ell}^{(\text{m})}_5\equiv \hat{\ell}^{(\text{m})}_0 \equiv\pi  \s P_0+\frac{i\pi}{2}(m+n-1),  \\
\hat{\ell}^{(\text{m})}_3&\equiv \hat{\ell}^{(\text{m})}_s\equiv 4\pi \s P_s, \qquad \hat{\ell}^{(\text{m})}_6\equiv \hat{\ell}^{(\text{m})}_t\equiv  4\pi \s P_t.
\eea
\ee
The interpretation in terms of lengths here imposes $P_s,P_t,P_0\in \mathbb{C}$ with $\Im{P_s},\Im{P_t}=\frac{\mathbb{Z}}{2\s}$ and $\Im{P_0}=\frac{1-m-n}{2\s}+\frac{2\mathbb{Z}}{\s}$. There is no indication however from our general expressions (\ref{eq:splitmodleq1main2}) that one should restrict to these particular values of the momenta. We will comment more on these interesting connections of our kernels with the three-polyhedra in Section \ref{sec:discussion}.
\par Since the shift relations of the modular kernel are linear, we can take linear combinations of the solutions $\hat{\mathbf{M}}^{(\pm)}$ and still form another solution valid at $c\in\mathbb{Q}_{(-\infty,1]}$. The \textit{positive} linear combination defined as
\be\label{eq:fullmodkerncleq1}
\boxed{
\hat{\mathbf{M}}^{(\mathsf{b})}_{P_s,P_t}[P_0]:=\frac{1}{2}\left(\hat{\mathbf{M}}^{(+)}_{P_s,P_t}[P_0]+\hat{\mathbf{M}}^{(-)}_{P_s,P_t}[P_0]\right)
}
\ee
is distinguished for two reasons:
\begin{itemize}
\item It satisfies the crossing transformations of the $c\leq1$ torus 1-point blocks:
\be\label{eq:integrmodkerncleq1}
(-i\tau)^{h_0}\mathcal{F}^{(i\mathsf{b}), \tau}_{P_s} = \int_{i\mathbb{R}+\Lambda}\frac{dP_t}{i}\ \hat{\mathbf{M}}^{(\mathsf{b})}_{P_s,P_t}[P_0] \ \mathcal{F}^{(i\mathsf{b}), -1/\tau}_{P_t}
\ee
since both $\hat{\mathbf{M}}^{(\pm)}$ satisfy the same relation (due to the fact that they are solutions to the respective shift relations).
\item It is \textit{reflection-symmetric} under $P_s\rightarrow -P_s$ or $P_t\rightarrow -P_t$. Indeed, from the definition of $\hat{\mathbf{M}}^{(\pm)}$ in (\ref{eq:defnmhpm}) and the reflections properties of $\mathbf{M}^{(\pm)}$ that we discussed in section \ref{sec:properties}, we see right away
\be
\hat{\mathbf{M}}^{(\pm)}_{-P_s,P_t}[P_0]=\hat{\mathbf{M}}^{(\mp)}_{P_s,P_t}[P_0] \quad \text{and} \quad \hat{\mathbf{M}}^{(\pm)}_{P_s,-P_t}[P_0]=\hat{\mathbf{M}}^{(\mp)}_{P_s,P_t}[P_0].
\ee
\end{itemize}
These properties signify that we should view (\ref{eq:fullmodkerncleq1}) as defining the \textit{physical modular kernel} in the regime $c\in\mathbb{Q}_{(-\infty,1]}$. This is the first main result of the present section. 
\par Note that, unlike the Teschner modular kernel (\ref{eq:defmodularkernel}) valid at $c\geq25$, the kernel (\ref{eq:fullmodkerncleq1}) possesses square-root branch point singularities (and \textit{no} logarithmic singularities, for the reasons explained in section \ref{sec:properties}) as dictated by the zeroes of the modular determinant $\hat{\mathfrak{D}}^{(m)}$ in (\ref{eq:splitmodleq1main2})\footnote{Because of the fact that $\hat{\mathbf{M}}^{(\pm)}$ encode the two branches of the square-root $\sqrt{\hat{\mathfrak{D}}^{(m)}}$ with the \textit{same} sign in the overall factor (contrary to what we had in the $c\in\mathbb{Q}_{[25,\infty)}$ case), we can imagine rewriting compactly the physical modular kernel (\ref{eq:fullmodkerncleq1}) as
\be
\bea
\hat{\mathbf{M}}^{(\mathsf{b})}_{P_s,P_t}[P_0]=\frac{1}{2}\text{disc}_{P_t^*}\left[\hat{\mathbf{M}}^{(+)}_{P_s,P_t}[P_0]\right].
\eea
\ee
Here by $\text{disc}_{P_t^*}$ we collectively mean the discontinuities across all branch points specified by the solutions (in $P_t$) to the equation $\hat{\mathfrak{D}}^{(m)}=0$. It would be extremely interesting to understand what this means concretely in terms of the integration contour in the crossing transformation of conformal blocks or whether such form could generalize for the modular kernel at irrational $c\leq1$. Similar remarks hold for the $c\leq1$ fusion kernel that we discuss in the next section.
}. Uncovering the structure of these branch points is a very important task, though we will not say more about this here. Also in that regard, the prescription for choosing the real constant $\Lambda$ in the integration contour of (\ref{eq:integrmodkerncleq1}) should be that it is sufficiently large in mangitude such that all the branch cuts or poles are evaded by the vertical line.
\par It is worth mentioning that we could also consider the opposite linear combination
\be
\bea
\frac{1}{2}\left(\hat{\mathbf{M}}^{(+)}_{P_s,P_t}[P_0]-\hat{\mathbf{M}}^{(-)}_{P_s,P_t}[P_0]\right).
\eea
\ee
This expression defines a non-trivial function that solves the shift relations for $c\in\mathbb{Q}_{(-\infty,1]}$, however it has \textit{neither} of the above two properties: its integral against the torus 1-point blocks yields \textit{zero}, and it is an \textit{odd} function in the Liouville momenta $P_s,P_t$. Nevertheless, it overall defines a \textit{meromorphic} function in the momenta and this is exactly the unphysical solution that arises from Virasoro-Wick-Rotating the Teschner modular kernel from the regime $c\geq 25$, as discussed originally in \cite{Ribault:2023vqs}. 
\par Before proceeding to the case of the fusion kernel, as a concrete example we exhibit in its full form the modular kernel at $c=1$ that is reflection-symmetric in all the involved momenta\footnote{Recently the modular kernel at $c=1$ was also investigated from a different perspective in \cite{DelMonte:2025vqv}. It is not clear to us whether their kernel is reflection symmetric in the various momenta, or whether it is just proportional to $\hat{\mathbf{M}}^{(\pm)}$ (i.e. to either of the non-reflection symmetric kernels). It will be extremely interesting to unify the set up that \cite{DelMonte:2025vqv} used for the derivation of the $c=1$ modular kernel with our present construction.}:
\be\label{eq:modkernc1}
\bea
\hat{\mathbf{M}}^{(c=1)}_{P_s,P_t}[P_0]&=\frac{-\pi^2(2\pi)^{iP_0}P_t \ e^{-2\pi P_t}\tilde{G}\left(iP_0\right)G\left(\pm2iP_s\right)G\left(1-iP_0\pm2iP_t\right)}{\sqrt{2}P_s\sinh{(2\pi P_s)}\sinh{(2\pi P_t)}G\left(2\pm2iP_t\right)G\left(1-iP_0\pm2iP_s\right)}\\
&\quad \quad \times\left(\hat{\mathfrak{D}}^{(\text{m})}|_{\b=1}\right)^{-1/2}\left[\hat{\mathcal{M}}_{1}\left(\frac{\log{\left(\hat{z}_1^{(\text{m})}|_{\b=1}\right)}}{2\pi i}\right)+\hat{\mathcal{M}}_{1}\left(\frac{\log{\left(\hat{z}_2^{(\text{m})}|_{\b=1}\right)}}{2\pi i}\right)\right], 
\eea
\ee
where
\be
\bea
\hat{\mathcal{M}}_{1}\left(u\right)&=\frac{e^{4\pi  P_t u}\prod_{\pm}\tilde{G}\left(u-\frac{1}{2}-\frac{i}{2}\left(\pm2P_s+P_0\right)\right)}{\prod_{\pm}\tilde{G}\left(u-\frac{1}{2}-\frac{i}{2}\left(\pm2P_s-P_0\right)\right)},\\
\hat{\mathfrak{D}}^{(\text{m})}|_{\b=1}&=1+\left(\frac{\sinh{(\pi P_0)}}{\sinh{(2\pi P_s)}\sinh{(2\pi P_t)}}\right)^2,\\
\hat{z}_{1,2}^{(\text{m})}|_{\b=1}&=\frac{\sinh{(2\pi P_t)}}{\sinh{(\pi (P_0-2P_t))}}\left[\cosh{(2\pi P_s)}\mp\sinh{(2\pi P_s)}\left(\hat{\mathfrak{D}}^{(\text{m})}|_{\b=1}\right)^{1/2}\right].
\eea
\ee
\subsection{Fusion Kernel}\label{sec:fuskerncleq1}
We work similarly for the case of the fusion kernel. In section \ref{sec:shiftrel} and Appendix \ref{app:shiftrelFpm} we showed that the non-meromorphic kernels $\mathbf{F}^{(\pm)}$ are both solutions to the fusion kernel shift relations for $c\in\mathbb{Q}_{[25,\infty)}$. Using the Virasoro-Wick Rotation symmetry we can therefore derive two other solutions $\hat{\mathbf{F}}^{(\pm)}$ valid at $c\in\mathbb{Q}_{(-\infty,1]}$ that take the form
\be\label{eq:defnfhpm}
\bea
\hat{\mathbf{F}}^{(\pm)}_{P_s,P_t}\left[\begin{smallmatrix} P_2 & P_3 \\ P_1 & P_4 \end{smallmatrix}\right]&:=\mp i \mathfrak{R}\mathbf{F}^{(\pm)}_{P_s,P_t}\left[\begin{smallmatrix} P_2 & P_3 \\ P_1 & P_4 \end{smallmatrix}\right].
\eea
\ee
More explicitly,
\be\label{eq:splitfusleq1main2}
\boxed{
\bea
\hat{\mathbf{F}}^{(+)}_{P_s,P_t}\left[\begin{smallmatrix} P_2 & P_3 \\ P_1 & P_4 \end{smallmatrix}\right]&=\left(\frac{P_t}{P_s}\right)\frac{F_\b(iP_3,iP_2,iP_1,iP_4|iP_t,iP_s)}{2\sinh{\left(2\pi \s P_t\right)}\sinh{\left(2\pi \s P_s\right)}}\times \frac{1}{\sqrt{\hat{\mathfrak{D}}^{(f)}}}\sum_{k=0}^{\s^2-1}\hat{\mathcal{F}}_{\b}\left(\frac{\log{\hat{z}^{(\text{f})}_1}}{2\pi i \s^2}+\frac{k}{\s^2}\right),\\
\hat{\mathbf{F}}^{(-)}_{P_s,P_t}\left[\begin{smallmatrix} P_2 & P_3 \\ P_1 & P_4 \end{smallmatrix}\right]&=\left(\frac{P_t}{P_s}\right)\frac{F_\b(iP_3,iP_2,iP_1,iP_4|iP_t,iP_s)}{2\sinh{\left(2\pi \s P_t\right)}\sinh{\left(2\pi \s P_s\right)}}\times \frac{1}{\sqrt{\hat{\mathfrak{D}}^{(f)}}}\sum_{k=0}^{\s^2-1}\hat{\mathcal{F}}_{\b}\left(\frac{\log{\hat{z}^{(\text{f})}_2}}{2\pi i \s^2}+\frac{k}{\s^2}\right)
\eea
}.
\ee
$F_{\b}$ is defined in (\ref{eq:prefintegdFrational}), and
\be
\bea
\hat{\mathcal{F}}_{\b}(u)\equiv \frac{\prod_{J\in\{\sigma_E=-1\}}\tilde{G}_{m,n}\left(u-1+\frac{m^{-1}+n^{-1}}{4}-\frac{i}{2}\sum_{i\in E} \sigma^{(J)}_i \frac{P_i}{\s}\right)}{\prod_{I\in\{\sigma_E=+1\}}\tilde{G}_{m,n}\left(u-\frac{m^{-1}+n^{-1}}{4}-\frac{i}{2}\sum_{i\in E} \sigma^{(I)}_i \frac{P_i}{\s}\right)}. 
\eea
\ee
The data $\{\hat{z}_{1}^{(\text{f})},\hat{z}_{2}^{(\text{f})},\hat{\mathfrak{D}}^{(\text{f})}\}$ originate from a degree-two polynomial which is simply the VWR-ed quantum fusion polynomial (\ref{eq:fuspoly}). Its coefficients read
\be
\bea
\hat{\alpha}^{(\text{f})}_{mn}&=\sum_{\substack{\sigma\in \mathbb{Z}_2^E|\\ \sigma_V=1}}\sigma_E \ e^{2\pi i \s^2\left(\sigma_E\frac{m^{-1}+n^{-1}}{4}+\frac{i}{2}\sum_{i\in E}\sigma_i\frac{P_i}{\s}\right)},\\
\hat{\beta}^{(\text{f})}_{mn}&=-4\sum_{ij\in P}\text{sh}_{mn}{(P_i)}\text{sh}_{mn}{(P_j)},\\
\hat{\gamma}^{(\text{f})}_{mn}&=\sum_{\substack{\sigma\in \mathbb{Z}_2^E|\\ \sigma_V=1}}\sigma_E \ e^{-2\pi i \s^2\left(\sigma_E\frac{m^{-1}+n^{-1}}{4}+\frac{i}{2}\sum_{i\in E}\sigma_i\frac{P_i}{\s}\right)},
\eea
\ee
and therefore

\be
\bea
\hat{z}_{1,2}^{(\text{f})}\equiv \frac{-\hat{\beta}^{(\text{f})}_{mn}\pm2i\text{sh}_{mn}{\left(P_t\right)}\text{sh}_{mn}{\left( P_s\right)}\sqrt{\hat{\mathfrak{D}}^{(f)}}}{2\hat{\alpha}^{(\text{f})}_{mn}}, \qquad \hat{\mathfrak{D}}^{(\text{f})}\equiv \frac{-4 \ \text{det} \left[\hat{\mathcal{G}}^{\text{(f)}}_{m,n}\right]}{\text{sh}_{mn}{\left( P_s\right)}^2\text{sh}_{mn}{\left( P_t\right)}^2},
\eea
\ee
with
\be\label{eq:gramfusionahat}
\bea
\hat{\mathcal{G}}^{\text{(f)}}_{m,n}:=\begin{pmatrix}
1 & -\text{ch}_{mn}\left(\hat{a}_2\right) & -\text{ch}_{mn}(\hat{a}_3) & \text{ch}_{mn}(\hat{a}_s) \\
 -\text{ch}_{mn}(\hat{a}_2) & 1 & \text{ch}_{mn}(\hat{a}_t) & -\text{ch}_{mn}(\hat{a}_1) \\
-\text{ch}_{mn}(\hat{a}_3) & \text{ch}_{mn}(\hat{a}_t) & 1 & -\text{ch}_{mn}(\hat{a}_4) \\
 \text{ch}_{mn}(\hat{a}_s) & -\text{ch}_{mn}(\hat{a}_1) & -\text{ch}_{mn}(\hat{a}_4) & 1
\end{pmatrix}, \quad \hat{a}_i\equiv P_i+\frac{i(m+n)}{2\s}.
\eea
\ee
The matrix $\hat{\mathcal{G}}^{\text{(f)}}_{m,n}$ can again be interpreted geometrically as the standard \textit{vertex Gram matrix} that encodes the six edge lengths of a \textit{3-tetrahedron} that we denote $T(\hat{\ell}^{(\text{f})}_i)$, $i=1,\cdots,6$. From (\ref{eq:gramfusionahat}), (\ref{eq:genlengthGram}), we read the following lengths:
\be
\bea
\hat{\ell}^{(\text{f})}_1&\equiv 2\pi  \s P_4+i\pi(m+n), \ \ \hat{\ell}^{(\text{f})}_2\equiv2\pi  \s P_1+i\pi(m+n), \ \ \hat{\ell}^{(\text{f})}_3\equiv2\pi  \s P_s+i\pi(1+m+n), \\
\hat{\ell}^{(\text{f})}_4&\equiv2\pi  \s P_2+i\pi(m+n) , \ \ \hat{\ell}^{(\text{f})}_5\equiv 2\pi  \s P_3+i\pi(m+n), \ \ \hat{\ell}^{(\text{f})}_6\equiv 2\pi  \s P_t+i\pi(1+m+n).
\eea
\ee
 The length interpretation requires $P_s, P_t, P_1,P_2,P_3,P_4\in\mathbb{C}$ with $\Im{P_i}=-\frac{m+n}{2\s}+\frac{\mathbb{Z}}{\s}$ for $i=1,\cdots,4$, and $\Im{P_s},\Im{P_t}=-\frac{1+m+n}{2\s}+\frac{\mathbb{Z}}{\s}$, though our expressions for the fusion kernel are in principle more general than these restrictions. We will return to comment on these interesting relations in Section \ref{sec:discussion}.
\par The shift relations of the fusion kernel are linear, and hence we can take linear combinations of the solutions $\hat{\mathbf{F}}^{(\pm)}$ and still form another solution valid at $c\in\mathbb{Q}_{(-\infty,1]}$. The \textit{positive} linear combination defined as
\be\label{eq:fullfuskerncleq1}
\boxed{
\hat{\mathbf{F}}^{(\mathsf{b})}_{P_s,P_t}\left[\begin{smallmatrix} P_2 & P_3 \\ P_1 & P_4 \end{smallmatrix}\right]:=\frac{1}{2}\left(\hat{\mathbf{F}}^{(+)}_{P_s,P_t}\left[\begin{smallmatrix} P_2 & P_3 \\ P_1 & P_4 \end{smallmatrix}\right]+\hat{\mathbf{F}}^{(-)}_{P_s,P_t}\left[\begin{smallmatrix} P_2 & P_3 \\ P_1 & P_4 \end{smallmatrix}\right]\right)
}
\ee
is again distinguished for two reasons:
\begin{itemize}
\item It satisfies the crossing transformations of the $c\leq1$ sphere four-point blocks:
\be
\mathcal{F}^{(i\b), s-\text{channel}}_{P_s} = \int_{i\mathbb{R}+\Lambda} \frac{dP_t}{i}\ \hat{\mathbf{F}}^{(\b)}_{P_s,P_t}\left[\begin{smallmatrix} P_2 & P_3 \\ P_1 & P_4 \end{smallmatrix}\right]  \mathcal{F}^{(i\b), t-\text{channel}}_{P_t}
\ee
since both $\hat{\mathbf{F}}^{(\pm)}$ satisfy the same relation (due to the fact that they are solutions to the respective shift relations).
\item It is \textit{reflection-symmetric} under $P_s\rightarrow -P_s$ or $P_t\rightarrow -P_t$. This comes from the definition of $\hat{\mathbf{F}}^{(\pm)}$ in (\ref{eq:defnfhpm}) and the reflections properties of $\mathbf{F}^{(\pm)}$ that we discussed in section \ref{sec:properties}, which lead to the relations
\be
\hat{\mathbf{F}}^{(\pm)}_{-P_s,P_t}\left[\begin{smallmatrix} P_2 & P_3 \\ P_1 & P_4 \end{smallmatrix}\right]=\hat{\mathbf{F}}^{(\mp)}_{P_s,P_t}\left[\begin{smallmatrix} P_2 & P_3 \\ P_1 & P_4 \end{smallmatrix}\right] \quad \text{and} \quad \hat{\mathbf{F}}^{(\pm)}_{P_s,-P_t}\left[\begin{smallmatrix} P_2 & P_3 \\ P_1 & P_4 \end{smallmatrix}\right]=\hat{\mathbf{F}}^{(\mp)}_{P_s,P_t}\left[\begin{smallmatrix} P_2 & P_3 \\ P_1 & P_4 \end{smallmatrix}\right].
\ee
\end{itemize}
Just as in the case of the modular kernel, these properties make it natural to interpret (\ref{eq:fullfuskerncleq1}) as defining the \textit{physical fusion kernel} in the regime $c\in\mathbb{Q}_{(-\infty,1]}$. This is the second main result of this section.
\par The fusion kernel (\ref{eq:fullfuskerncleq1}), unlike the Teschner-Vartanov solution, possesses square-root branch point singularities determined by the zeroes of the fusion determinant $\hat{\mathfrak{D}}^{(\text{f})}$. Understanding in detail the structure of these singularities for any coprime pair $(m,n)$ is an important task that falls beyond the scope of this paper and will be addressed in future work.
\par Similar to the analysis of the modular kernel, the opposite linear combination
\be\label{vwrkernel}
\bea
\frac{1}{2}\left(\hat{\mathbf{F}}^{(+)}_{P_s,P_t}\left[\begin{smallmatrix} P_2 & P_3 \\ P_1 & P_4 \end{smallmatrix}\right]-\hat{\mathbf{F}}^{(-)}_{P_s,P_t}\left[\begin{smallmatrix} P_2 & P_3 \\ P_1 & P_4 \end{smallmatrix}\right]\right)
\eea
\ee
is also a solution to the fusion kernel shift relations at $c\in\mathbb{Q}_{(-\infty,1]}$, but it neither implements the crossing transformation of $c\leq1$ conformal blocks (the net result is zero), nor is it an even function of the internal momenta $P_s,P_t$ (instead, it is odd). It is therefore simply a by-product solution of the shift relations with no immediate physical meaning. Nevertheless, equation (\ref{vwrkernel}) defines a non-trivial \textit{meromorphic} function in the Liouville momenta that it can be shown to be equal to the VWR of the Teschner-Vartanov solution from $c\in\mathbb{Q}_{[25,\infty)}$, as it was originally observed in \cite{Ribault:2023vqs}.
\par A final point worth noting is that the case $m=n=1$ of the $c=1$ fusion kernel has been studied previously in \cite{Iorgov:2013uoa}, where the authors associated it to the connection coefficient of the Painlev\'e VI tau function.  One can check that the fusion kernel of \cite{Iorgov:2013uoa} coincides exactly with our $\hat{\mathbf{F}}^{(+)}|_{\b=1}$, but \textit{not} with the physical fusion kernel (\ref{eq:fullfuskerncleq1}) which is even under reflections of the internal momenta. Given the interesting connection of \cite{Iorgov:2013uoa} with the Painlev\'e VI (see also \cite{Gamayun:2012ma,Iorgov:2014vla}), it is interesting to ask whether the fusion kernels (\ref{eq:splitfusleq1main2}) for general $m,n$ have an interesting role to play within the framework of Painlevé VI and its generalizations. 
\subsection{Application: crossing symmetry of $c\leq 1$ Liouville theory}\label{sec:liouvillecless1}
Liouville theory at $c \leq1$ (sometimes known as ``timelike'' Liouville theory) is a non-unitary 2d CFT with many interesting applications (see e.g. \cite{Polyakov:1981rd,Harlow:2011ny,Bautista:2019jau,Anninos:2021ene,Chatterjee:2025yzo,Usciati:2025cdn}) that is \textit{not} simply the analytic continuation of usual Liouville theory valid at $c\geq25$ (or more generally at $c\in\mathbb{C}\backslash(-\infty,1]$). More specifically, the structure constants of the theory have been derived from the bootstrap in \cite{Zamolodchikov:2005fy,Schomerus:2003vv,Kostov:2005kk} and, in a particular normalization, can be chosen as (c.f. $\hat{b}\in \mathbb{R}_{(0,1]}$)
\be\label{eq:liouvillesccleq1}
\bea
\hat{B}^{(\hat{b})}_{P}:=\frac{1}{4P^2B^{(\hat{b})}_{iP}}, \ \ \ \ \hat{C}^{(\hat{b})}_{P_1,P_2,P_3} :=\frac{1}{C^{(\hat{b})}_{iP_1,iP_2,iP_3} },
\eea
\ee
where $B^{(b)}_P, C^{(b)}_{P_1,P_2,P_3}$ are defined in (\ref{eq:liouvillesc}).
\par In \cite{Ribault:2015sxa}, rather compelling numerical evidence was presented indicating that the theory with structure constants given by (\ref{eq:liouvillesccleq1}) is crossing symmetric and, similarly to Liouville theory at $c\geq25$, possesses a diagonal and continuous spectrum of primary operators with $P\in i\mathbb{R}+\varepsilon$. However no analytic proof of the sort that we described in section \ref{sec:liouvillec25} was available to date due to the lack of knowledge of the corresponding fusion kernel. 
\par In this section we will fill this gap for the case $\hat{b}=\b$ using the knowledge of the fusion kernel (\ref{eq:fullfuskerncleq1}) and the Virasoro-Wick Rotation. In particular we will prove the identity
\be\label{eq:crossingsymcleq1Liouv}
\bea
\frac{\hat{C}^{(\b)}_{P_1,P_2,P_s}\hat{C}^{(\b)}_{P_s,P_3,P_4}}{\hat{B}^{(\b)}_{P_s}}\hat{\mathbf{F}}^{(\b)}_{P_s,P_t}\left[\begin{smallmatrix} P_2 & P_3 \\ P_1 & P_4 \end{smallmatrix}\right]=\frac{\hat{C}^{(\b)}_{P_2,P_3,P_t}\hat{C}^{(\b)}_{P_t,P_1,P_4}}{\hat{B}^{(\b)}_{P_t}}\hat{\mathbf{F}}^{(\b)}_{P_t,P_s}\left[\begin{smallmatrix} P_2 & P_1 \\ P_3 & P_4 \end{smallmatrix}\right].
\eea
\ee
\par Indeed, starting from the statements of crossing symmetry for $\mathbf{F}^{(\pm)}$ at $c\in\mathbb{Q}_{[25,\infty)}$ (\ref{eq:LiouvilleFpm}), we can freely relabell the momenta as follows 
\be
P_1\rightarrow iP_3, \ P_2\rightarrow iP_2, \ P_3\rightarrow iP_1, \ P_4\rightarrow iP_4, \ P_s\rightarrow iP_t,  \ P_t\rightarrow iP_s.
\ee
This yields for $\mathbf{F}^{(+)}$
\be
\bea
\frac{P_s^2B^{(\b)}_{iP_s}}{C^{(\b)}_{iP_2,iP_1,iP_s}C^{(\b)}_{iP_s,iP_3,iP_4}}\left(\frac{P_t}{iP_s}\mathbf{F}^{(+)}_{iP_t,iP_s}\left[\begin{smallmatrix} iP_2 & iP_1 \\ iP_3 & iP_4 \end{smallmatrix}\right]\right)=\frac{P_t^2B^{(\b)}_{iP_t}}{C^{(\b)}_{iP_3,iP_2,iP_t}C^{(\b)}_{iP_t,iP_1,iP_4}}\left(\frac{P_s}{iP_t}\mathbf{F}^{(+)}_{iP_s,iP_t}\left[\begin{smallmatrix} iP_2 & iP_3 \\ iP_1 & iP_4 \end{smallmatrix}\right]\right)
\eea
\ee
or, equivalently
\be\label{eq:crossingtimelikep}
\bea
\frac{\hat{C}^{(\b)}_{P_1,P_2,P_s}\hat{C}^{(\b)}_{P_s,P_3,P_4}}{\hat{B}^{(\b)}_{P_s}}\hat{\mathbf{F}}^{(+)}_{P_s,P_t}\left[\begin{smallmatrix} P_2 & P_3 \\ P_1 & P_4 \end{smallmatrix}\right]=\frac{\hat{C}^{(\b)}_{P_2,P_3,P_t}\hat{C}^{(\b)}_{P_t,P_1,P_4}}{\hat{B}^{(\b)}_{P_t}}\hat{\mathbf{F}}^{(+)}_{P_t,P_s}\left[\begin{smallmatrix} P_2 & P_1 \\ P_3 & P_4 \end{smallmatrix}\right].
\eea
\ee
Working similarly for $\mathbf{F}^{(-)}$, we find
\be\label{eq:crossingtimelikem}
\bea
\frac{\hat{C}^{(\b)}_{P_1,P_2,P_s}\hat{C}^{(\b)}_{P_s,P_3,P_4}}{\hat{B}^{(\b)}_{P_s}}\hat{\mathbf{F}}^{(-)}_{P_s,P_t}\left[\begin{smallmatrix} P_2 & P_3 \\ P_1 & P_4 \end{smallmatrix}\right]=\frac{\hat{C}^{(\b)}_{P_2,P_3,P_t}\hat{C}^{(\b)}_{P_t,P_1,P_4}}{\hat{B}^{(\b)}_{P_t}}\hat{\mathbf{F}}^{(-)}_{P_t,P_s}\left[\begin{smallmatrix} P_2 & P_1 \\ P_3 & P_4 \end{smallmatrix}\right].
\eea
\ee
Therefore taking the \textit{positive} linear combination of the two equations we arrive at the desired equation (\ref{eq:crossingsymcleq1Liouv}). This establishes the long-anticipated crossing symmetry of $c\leq1$ Liouville theory, at least for all the rational values in this range.
\par Similar manipulations hold for the modular kernel where, starting from the modular covariance statements for $\mathbf{M}^{(\pm)}$ at $c\in\mathbb{Q}_{[25,\infty)}$ (\ref{eq:LiouvilleMpm}) and performing the relabel
\be
P_0\rightarrow iP_0, \ P_s\rightarrow i P_t, \ P_t\rightarrow i P_s,
\ee
one can show the following relations
\be\label{eq:modcovcless1}
\bea
\frac{\hat{C}^{(\b)}_{P_0,P_s,P_s}}{\hat{B}^{(\b)}_{P_s}}\hat{\mathbf{M}}^{(+)}_{P_s,P_t}[P_0]=\frac{\hat{C}^{(\b)}_{P_0,P_t,P_t}}{\hat{B}^{(\b)}_{P_t}}\hat{\mathbf{M}}^{(+)}_{P_t,P_s}[P_0], \quad 
\frac{\hat{C}^{(\b)}_{P_0,P_s,P_s}}{\hat{B}^{(\b)}_{P_s}}\hat{\mathbf{M}}^{(-)}_{P_s,P_t}[P_0]=\frac{\hat{C}^{(\b)}_{P_0,P_t,P_t}}{\hat{B}^{(\b)}_{P_t}}\hat{\mathbf{M}}^{(-)}_{P_t,P_s}[P_0].
\eea
\ee
Taking again the positive linear combination leads to the modular covariance identity involving the physical kernel $\hat{\mathbf{M}}^{(\b)}$, which establishes modular covariance of $c\leq1$ Liouville theory for all rational values in this range.
\par Let us make one final remark. Note that we could have equally well considered the \textit{opposite} linear combination of (\ref{eq:crossingtimelikep}), (\ref{eq:crossingtimelikem}) which leads to\footnote{This equation was also observed in \cite{Ribault:2023vqs} for \textit{general} $\hat{b}\in\mathbb{R}_{[0,1)}$, where the corresponding (fiducial) kernel is the VWR-ed Teschner-Vartanov kernel. As we have explained, when $\hat{b}=\b$ the kernel $\hat{\mathbf{F}}^{(+)}-\hat{\mathbf{F}}^{(-)}$ is exactly that. The general $\hat{b}\in\mathbb{R}_{[0,1)}$ analog of the \textit{physical} kernel $\hat{\mathbf{F}}^{(+)}+\hat{\mathbf{F}}^{(-)}$ is still not properly understood to date (see however \cite{Roussillon:2024wmr} for some attempts in that direction).} 

\begin{align}
&\frac{\hat{C}^{(\b)}_{P_1,P_2,P_s}\hat{C}^{(\b)}_{P_s,P_3,P_4}}{\hat{B}^{(\b)}_{P_s}}\left(\hat{\mathbf{F}}^{(+)}_{P_s,P_t}\left[\begin{smallmatrix} P_2 & P_3 \\ P_1 & P_4 \end{smallmatrix}\right]-\hat{\mathbf{F}}^{(-)}_{P_s,P_t}\left[\begin{smallmatrix} P_2 & P_3 \\ P_1 & P_4 \end{smallmatrix}\right]\right)=\nonumber\\
&\qquad \frac{\hat{C}^{(\b)}_{P_2,P_3,P_t}\hat{C}^{(\b)}_{P_t,P_1,P_4}}{\hat{B}^{(\b)}_{P_t}}\left(\hat{\mathbf{F}}^{(+)}_{P_t,P_s}\left[\begin{smallmatrix} P_2 & P_1 \\ P_3 & P_4 \end{smallmatrix}\right]-\hat{\mathbf{F}}^{(-)}_{P_t,P_s}\left[\begin{smallmatrix} P_2 & P_1 \\ P_3 & P_4 \end{smallmatrix}\right]\right)\label{eq:fiducialliouville}.
\end{align}
As it stands, this identity is a non-trivial relation between meromorphic functions and, from the point of view of crossing symmetry, it seems to suggest that the kernel $\hat{\mathbf{F}}^{(+)}-\hat{\mathbf{F}}^{(-)}$ also implements the crossing transformation of $c\leq1$ blocks. However as we have explained this is not correct, because  the integral of this (fiducial) kernel against a conformal block yields to zero. This is telling us that as far as crossing symmetry is concerned, equations such as (\ref{eq:crossingsymcleq1Liouv}), (\ref{eq:crossingtimelikep}), (\ref{eq:modcovcless1}), (\ref{eq:fiducialliouville}) should really be thought of in the distributional sense, that are valid when integrated against a suitable space of test functions (of which conformal blocks are part). Only when these equations include the \textit{physical} kernels (such as (\ref{eq:crossingsymcleq1Liouv}), (\ref{eq:crossingtimelikep}), (\ref{eq:crossingtimelikem}), (\ref{eq:modcovcless1})) there is no harm in viewing the equalities as equalities between usual functions.

\section{Discussion}\label{sec:discussion}
A refined understanding of the Virasoro crossing kernels goes a long way toward providing us with concrete insights into versatile areas of mathematical physics, from the Virasoro analytic bootstrap, to low-dimensional quantum topology, and the representation theory of quantum groups. 
\par In the present work we have unravelled a novel structure of these kernels when the parameter $b^2$ (that controls the central charge) takes on rational values. We have explained how the original forms of the modular and fusion kernels can be seen as particular instances of a \textit{state integral}, of the sort that appear in quantum topology \cite{hikami,Dimofte:2009yn,EllegaardAndersen:2011vps,EllegaardAndersen:2013pze,Kashaev:2012cz,Dimofte:2014zga,Dimofte:2015kkp}. An important aspect of the crossing kernels --contrary to e.g. the Virasoro conformal blocks on which they act -- is the fact that they only depend on the Liouville momenta and \textit{not} on the moduli of the corresponding Riemann surface associated to conformal blocks. This makes them relatively easier to understand, especially via the shift relations that they satisfy. A proper understanding of the shift relations  and the space of their solutions (or more generally, the solutions to the Moore-Seiberg consistency conditions)  is pivotal, and still not a completely solved problem. Here we have explained how, if one looses the meromorphicity assumption, we can generate new solutions to those equations. What was even more striking was the fact that these new solutions (at $b^2\in\mathbb{Q}$) are associated with an algebraic variety -- the quantum modular and fusion polynomials -- and possess, via the discriminant of these polynomials, geometric-like features similar to those of three-dimensional tetrahedra. We will comment more on this particular feature next.
\par It is known since the work of Teschner and Vartanov \cite{Teschner:2012em} that the modular and fusion kernels have an interesting ``semiclassical limit'', which is defined as $b\rightarrow0$ with the product $bP_i\equiv \frac{\ell_i}{4\pi i}$ being held fixed\footnote{Recently, these semiclassical limits were revisited in \cite{Hartman:2025ula,Liu:2025tzv}. Here we will basically follow \cite[Appendix A]{Hartman:2025ula}.}. We will be especially interested in the semiclassical limit of the integrals
\be
\mathcal{I}^{(\text{m})}_b\equiv \int_{i\mathbb{R}} \frac{du}{i} \ m_b(u), \qquad \mathcal{I}^{(\text{f})}_b\equiv \int_{i\mathbb{R}} \frac{du}{i} \ f_b(u)
\ee
that define the modular and fusion kernel respectively via (\ref{eq:defmodularkernel}), (\ref{eq:deffusionkernel}). It can be shown from a saddle point analysis (including the one-loop determinant term) that \cite{Hartman:2025ula}\footnote{The limit for $\mathcal{I}^{(\text{f})}_b$ includes also an additional overall phase that gets cancelled when combined with the prefactor of the fusion kernel. Here we omit it because it is irrelevant for our discussion. See \cite{Hartman:2025ula} for the precise behaviour.}
\begin{align}\label{eq:saddlepointmod}
\mathcal{I}^{(\text{m})}_b&\sim \frac{1}{\sqrt{-D}} \ \text{exp}\left[-\frac{1}{\pi b^2}\left(\frac{1}{2}V_L\begin{psmallmatrix}
\frac{1}{4}\ell_0 &\frac{1}{4}\ell_0 & \ell_s \\ \frac{1}{4}\ell_0 & \frac{1}{4}\ell_0 &\ell_t 
\end{psmallmatrix}+\frac{1}{4}\sum_{i=\{s,t,0\}}\ell_i\psi_i\right)\right],\\
\mathcal{I}^{(\text{f})}_b&\sim\frac{1}{\sqrt{-\text{det}\tilde{\mathcal{G}}}} \ \text{exp}\left[-\frac{1}{\pi b^2}\left(V_L\begin{psmallmatrix}
\frac{1}{2}\ell_2 & \frac{1}{2}\ell_3& \frac{1}{2}\ell_t \\
\frac{1}{2}\ell_4 & \frac{1}{2}\ell_1 & \frac{1}{2}\ell_s
\end{psmallmatrix}+\frac{1}{4}\sum_{i=\{s,t,1,2,3,4\}}\ell_i\psi_i\right)\right].\label{eq:saddlepointfus}
\end{align}
We will explain to some extent the various terms; we refer to \cite{Hartman:2025ula} for the more detailed expressions and discussion. First, it is important to emphasize that the saddle point equations that lead to both of these behaviours are \textit{quadratic} in the relevant variable, and hence one needs to find two roots of the associated polynomial. In both (\ref{eq:saddlepointmod}), (\ref{eq:saddlepointfus}) we have the contribution only from \textit{one} of these two roots, since it can be seen that the other one is subdominant in the limit. The one-loop determinant contributions are therefore exactly the discriminants of these second order polynomials, which we called $-D$ and $-\text{det}\mathcal{G}$ respectively. Moreover, these discriminants are proportional to the determinants of the $4\times4$ vertex Gram matrices associated to the edge lengths of particular hyperbolic tetrahedra, one associated to the modular kernel and one to the fusion kernel \cite{Hartman:2025ula}. $V_L$ is the volume of the corresponding (generalized) hyperbolic tetrahedron \cite{Ushijima} with the notation being indicative of the corresponding edge lengths in each case. Finally, $\psi_i$ are the associated dihedral angles on the edges of length $\ell_i$ (and should be viewed as functions of the lengths $\ell_i$).
\par It is noteworthy how this computation resembles our expressions for the modular and fusion kernels at any $b^2\in\mathbb{Q}$. Indeed, as we described in sections \ref{sec:cgeq25rationalkernels} and \ref{sec:cleq1kernels}, there is a hidden geometric structure in the discriminants of the quantum modular and fusion polynomials which can be realized as determinants of Gram matrices associated to particular hyperbolic tetrahedra. To appreciate even more how closely related the cases $b\rightarrow0$ and $b^2\in\mathbb{Q}$ are, one can manipulate the functions $\mathcal{M}_\b$ and $\mathcal{F}_b$ which enter in the summand of our expressions for the kernels (defined in (\ref{eq:modularaction}), (\ref{eq:fusionaction})) to show that
\begin{align}
\mathcal{M}_{\b}\left(\frac{\log{z^{(\text{m})}_{1}}}{2\pi i \s^2}\right)&\approx \text{exp}\left[-\frac{1}{2\pi \s^2}V_L\begin{psmallmatrix}
\ell^{(\text{m})}_0 &\ell^{(\text{m})}_0 & \ell^{(\text{m})}_s \\ \ell^{(\text{m})}_0 & \ell^{(\text{m})}_0 &\ell^{(\text{m})}_t 
\end{psmallmatrix}\right], \\ \mathcal{M}_{\b}\left(\frac{\log{z^{(\text{m})}_{2}}}{2\pi i \s^2}\right)&\approx \text{exp}\left[+\frac{1}{2\pi \s^2}V_L\begin{psmallmatrix}
\ell^{(\text{m})}_0 &\ell^{(\text{m})}_0 & \ell^{(\text{m})}_s \\ \ell^{(\text{m})}_0 & \ell^{(\text{m})}_0 &\ell^{(\text{m})}_t 
\end{psmallmatrix}\right],
\end{align}
and
\begin{align}
\mathcal{F}_{\b}\left(\frac{\log{z^{(\text{f})}_{1}}}{2\pi i \s^2}\right)&\approx  \text{exp}\left[-\frac{1}{\pi \s^2}V_L\begin{psmallmatrix}
\ell^{(\text{f})}_1 & \ell^{(\text{f})}_2& \ell^{(\text{f})}_3 \\
\ell^{(\text{f})}_4 & \ell^{(\text{f})}_5 & \ell^{(\text{f})}_6
\end{psmallmatrix}\right] ,\\
 \mathcal{F}_{\b}\left(\frac{\log{z^{(\text{f})}_{2}}}{2\pi i \s^2}\right)&\approx \text{exp}\left[+\frac{1}{\pi\s^2}V_L\begin{psmallmatrix}
\ell^{(\text{f})}_1 & \ell^{(\text{f})}_2& \ell^{(\text{f})}_3 \\
\ell^{(\text{f})}_4 & \ell^{(\text{f})}_5 & \ell^{(\text{f})}_6
\end{psmallmatrix}\right].
\end{align}
The symbol $\approx$ here means that we have an exact equality of RHS and LHS (i.e. there is no limit) up to terms that do not depend on $V_L$. Quite remarkably, the volumes $V_L$ are exactly the volumes of the (generalized) hyperbolic tetrahedra associated to the \textit{vertex} Gram matrices that we discussed in sections \ref{sec:modkernelcgeq25} and \ref{sec:fusionkernelcgeq25}, namely (\ref{eq:grammodular2}), (\ref{eq:lengthsmod}) for the modular kernel, and (\ref{eq:gramf}), (\ref{eq:lengthsfusion}) for the fusion kernel.  To arrive at these relations we used an identity that we discuss in Appendix \ref{appendixA} (c.f. (\ref{alternativetGmn})) which relates the function $\tilde{G}_{m,n}$ with the Lobachevsky function $\Lambda(x)$ (c.f. (\ref{defLobachevsky})), the latter being a building block for the volume of the hyperbolic tetrahedra \cite{chokim,murakamiyano,Ushijima}. Similar results hold for the functions $\hat{\mathcal{M}_\b}$ and $\hat{\mathcal{F}}_\b$ for the $c\leq1$ crossing kernels.
\par It is therefore tempting to interpret our formulae for the full modular and fusion kernels at $b^2\in\mathbb{Q}$ as a sum of $m\times n$ number of ``instanton'' contributions -- specified respectively by either $z_1^{(\text{m})},z_1^{(\text{f})}, \hat{z}_1^{(\text{m})},\hat{z}_1^{(\text{f})}$-- plus another $m\times n$ number of ``anti-instanton'' contributions -- specified respectively by either $z_2^{(\text{m})},z_2^{(\text{f})}, \hat{z}_2^{(\text{m})},\hat{z}_2^{(\text{f})}$, with an overall \textit{one-loop exact} factor captured by the corresponding quantum modular and fusion determinants. 
\par It is certainly worth putting such an interpretation on firmer grounds, since this could have far reaching consequences. For example, one could naturally ask whether the corresponding Virasoro conformal blocks could take a similar ``instanton-anti-instanton'' form when $b^2\in\mathbb{Q}$, or what this means exactly for the space of solutions to the Virasoro analytic bootstrap (namely, spectra of primary operators and associated OPE coefficients) at $b^2\in\mathbb{Q}$. Alternatively, in the spirit of AdS/CFT it has been shown recently that the Virasoro crossing kernels at $b\in\mathbb{R}_{(0,1]}$ constitute an essential ingredient of the Hilbert space of pure 3d gravity with negative cosmological constant (at fixed topology) via a particular 3d TQFT called Virasoro TQFT \cite{Collier:2023fwi} or its dual formulation called Conformal Turaev-Viro theory \cite{Hartman:2025cyj}. In this setup it has been shown \cite{Hartman:2025ula} that one can build the 3d gravity path integral from triangulations via (generalized) hyperbolic tetrahedra. Given our results, it is tempting to consider the possibility that the relation between the Virasoro crossing kernels at $b^2\in\mathbb{Q}$ and the associated \textit{quantum} hyperbolic tetrahedra that we described in sections \ref{sec:cgeq25rationalkernels} and \ref{sec:cleq1kernels} may sharpen our understanding of what it truly means to construct a holographic 2d CFT, that is, a 2d CFT secretly encoding purely gravitational degrees of freedom in AdS$_3$. We leave these interesting questions to future work.
\par We conclude this discussion with a sharp research goal for the future that we believe will concretely address the aforementioned questions. The fascinating resemblance between the $b\rightarrow0$ and $b^2\to\mathbb{Q}$ limits of the modular and fusion kernels, as well as intuition from the state integrals that have appeared repeatedly in quantum topology, naturally leads us to advocate the following:
\begin{quote}
\begin{em}
	Conjecture:\\The Virasoro modular and fusion kernels $\mathbf{M}^{(b)},\mathbf{F}^{(b)}$ are associated to (quantum) modular forms with respect to the parameter $\tau\equiv b^2$.
\end{em}
\end{quote}
The notion of a quantum modular form is a generalization of the usual modular (and mock modular) forms that was put forward originally in the seminal paper by D.Zagier \cite{zagiermod}. In that paper, several examples of functions evaluated at the roots of unity were given that have the following property: their failure of modularity has improved analytic properties compared to the original function (which, for example, can have absolutely no continuity). After the paper by Zagier, the statement of quantum modularity has been sharpened a lot with many applications (see e.g. \cite{Garoufalidis:2021lcp,Garoufalidis:2023wez} and references therein) and roughly speaking quantum modular forms can be thought of as functions from $\mathbb{H}_-\cup \mathbb{Q}\cup\mathbb{H}_+$ ($\mathbb{H}_-,\mathbb{H}_+$ being the lower and upper half-planes) to some matrix group such that some multiplicative failure of modularity defines a function that has an analytic extension to some simply connected cut plane in $\mathbb{C}$ \cite{Wheeler:2023cye}. A prominent example of such function is the (log of the) Kashaev invariant of the figure eight knot $4_1$ (which was discussed in the original paper \cite{zagiermod}) and later was shown that many other state-integrals have similar properties\cite{Garoufalidis:2021lcp,Garoufalidis:2023wez,Garoufalidis:2022wij,Wheeler:2023cht}.
\par Given our present results for the Virasoro crossing kernels at $b^2\in\mathbb{Q}^{\times}$, their structural similarity to state integrals, and the expectation that Virasoro TQFT \cite{Collier:2023fwi}—within whose Hilbert space the said kernels act unitarily—is equivalent to the Andersen–Kashaev theory \cite{EllegaardAndersen:2011vps}, it seems worthwhile for our conjecture to be at least checked experimentally. We should mention though that a concrete step towards approaching the proposed problem is to first understand the behaviour of the crossing kernels at irrational and negative $b^2$, which is not fully understood to date. In any case, should this be true, its implications would be far-reaching for both pure 3D gravity with a negative cosmological constant and the Virasoro analytic bootstrap as we currently understand it. We hope to report on these important aspects in future publications.

\section*{Acknowledgements}

We would like to thank Matijn François, Stavros Garoufalidis, Alba Grassi,  Cristoforo Iossa, Oleg Lisovyy, Sylvain Ribault, Davide Saccardo, Jörg Teschner, and Campbell Wheeler for useful conversations. IT is particularly thankful to Matijn François, Stavros Garoufalidis, Alba Grassi, Cristoforo Iossa, Davide Saccardo, and Campbell Wheeler for illuminating discussions on state integrals, and to the University of Geneva for its hospitality, during which the idea for this project emerged. While at IPhT Saclay, IT was supported by the ERC Starting Grant 853507. The current work of IT is co-funded by the
European Union (ERC, FUNBOOTS, project number 101043588). IT is also forever indebted to Marianna K.S. for the recent birth of our daughter, and to Isadora T.K. for inspiration ever since. Finally, while at Aalto University, JR was supported by the Academy of Finland Centre of Excellence Programme grant number 346315 “Finnish centre of excellence
in Randomness and STructures” (FiRST). Views and opinions
expressed are however those of the author(s) only and do not necessarily reflect those of the
European Union or the European Research Council. Neither the European Union nor the
granting authority can be held responsible for them.

\begin{appendix}
\section{Special functions at $b^2\in\mathbb{Q}$} \label{appendixA}
In this appendix we record the special functions that enter the Virasoro crossing kernels when $b^{2}\in\mathbb{Q}$. The basic point is that the usual Barnes double gamma function $\Gamma_b$ reduces to a (finite) product of Barnes $G$ functions as we review below.
\paragraph{Double gamma function $\Gamma_b(x)$.} 
The Barnes double gamma function is a meromorphic  function with no zeroes and simple poles at $x=-pb-qb^{-1}$, where $p,q$ are non-negative integers. As a function of $b$, it is analytic in the whole $b^2-$complex
plane 
except for the negative part of the real axis, where it meets with a natural boundary of
analyticity. It also has the property $\Gamma_b=\Gamma_{b^{-1}}$,
 and satisfies the following shift relations in $b,b^{-1}$ which are incommensurable for $b^2\notin \mathbb{Q}$:
\begin{equation}
\begin{aligned}\label{eq:Gammabasicshift}
\frac{\Gamma_b(x+b^{\pm1})}{\Gamma_b(x)}&=\sqrt{2\pi}\frac{b^{\pm b^{\pm1}x\mp \frac{1}{2}}}{\Gamma(b^{\pm 1}x)}.
\end{aligned}
\end{equation}
When $b=\sqrt{\frac{m}{n}}$, for $m,n$ coprime integers, the two equations contain the same information and read
\be\label{eq:shiftrgammaational}
\bea
\left.\frac{\Gamma_b(x+b)}{\Gamma_b(x)}\right|_{b=\sqrt{\frac{m}{n}}}&=\sqrt{2\pi}\frac{\left(\frac{m}{n}\right)^{\frac{1}{2}\sqrt{\frac{m}{n}}x-\frac{1}{4}}}{\Gamma\left(\sqrt{\frac{m}{n}}x\right)} \ .
\eea
\ee
\par It is straightforward to show that the following ansatz satisfies the shift relation (\ref{eq:shiftrgammaational})
\be\label{eq:gbmn}
\boxed{
\bea
\Gamma_{b}(x)=\gamma_{m,n}\frac{(mn)^{\frac{x}4(Q-x)} (2\pi)^{\frac{x \sqrt{mn}}2 - \frac{m+n}4}}{G_{m,n}\left(\frac{x}{\sqrt{mn}}\right)} \ , \ \ \ \ \ \ \  b=\sqrt{\frac{m}{n}} 
\eea
} \ ,
\ee
where
\be \label{defGmn}
\bea
G_{m,n}(x):=\prod_{r=0}^{m-1}\prod_{s=0}^{n-1}G\left(x+\frac{r}{m}+\frac{s}{n}\right).
\eea
\ee
Here $G$ is the Barnes $G$ function, which is an entire function that captures the analytic continuation of the superfactorial and satisfies the shift relation $G(x+1)=\Gamma(x)G(x)$. In particular, it has $(\mathsf{p}+1)-$order zeroes at all non-positive integers $x_{\text{zeroes}}=-\mathsf{p}$, $\mathsf{p}\in\mathbb{Z}_{\geq0}$. The normalization constant $\gamma_{m,n}$ is chosen in such a way that it ensures the condition $\Gamma_b\left(\frac{Q}{2}\right)=1$\footnote{This constant will not be important for us (since the fusion kernels contain ratios of equal number of $\Gamma_b$'s in the numerator and denominator), but just for the record it reads
\be
\bea
\gamma_{m,n}:= (mn)^{-\frac{\mathsf{Q}^2}{16}} \; G_{m,n} \left(\frac{1}{2m} + \frac{1}{2n}\right) .
\eea
\ee
}. One way to understand (\ref{eq:gbmn}) is simply from the \textit{multiplication formula} of the $\Gamma_b$ for generic $b$, which reads (see e.g. \cite{Eberhardt:2023mrq})
\be\label{eq:mfgb}
\bea
\Gamma_b(z)=\lambda_{m,n,b} \, (mn)^{\frac{z}{4}(Q-z)} \prod_{r=0}^{n-1} \prod_{s=0}^{m-1} \Gamma_{\frac{b \sqrt{n}}{\sqrt{m}}}\left(\frac{z+s b+r b^{-1}}{\sqrt{mn}} \right) \ , \ \ \ \ \ \ m,n\in\mathbb{Z}_{>0} \ ,
\eea
\ee
for some appropriate constant $\lambda$ (independent of $z$). Evaluating this relation for $b=\sqrt{\frac{m}{n}}$ and using $\Gamma_{b\rightarrow1}(z)= \frac{(2\pi)^{\frac{z-1}{2}}}{G(z)}$, gives us back (\ref{eq:gbmn}).
\par Finally we note that (\ref{eq:shiftrgammaational}) implies the following shift relation
\be
\bea
\frac{G_{m,n}(x+1/n)}{G_{m,n}(x)}=m^{\frac{1}{2}-mx}(2\pi)^{\frac{m-1}{2}}\Gamma(mx),
\eea
\ee
and the exactly same equation with $n\leftrightarrow m$ (c.f. $G_{m,n}(x)=G_{n,m}(x)$ from its definition).
\paragraph{Double sine function $S_b(x)$.}The function $S_b(x):=\frac{\Gamma_b(x)}{\Gamma_b(Q-x)}$ is a meromorphic function with simple poles at $x=-pb-qb^{-1}$ and (simple) zeroes at $x=Q+pb+qb^{-1}$. It is also invariant under $b\rightarrow b^{-1}$ and satisfies the following basic shift relations:
\begin{equation}
\begin{aligned}\label{eq:Sbasicshift}
\frac{S_b(x+b^{\pm1})}{S_b(x)}&=2\sin{(\pi b^{\pm1} x)} .
\end{aligned}
\end{equation}
Again, when $b=\sqrt{\frac{m}{n}}$ for $m,n$ co-prime integers, the two equations contain the same information, and one can verify that the following ansatz satisfies the shift relations\footnote{This formula was also written down in \cite{Collier:2024mgv}, cf. eqn (4.12).}:
\be
\bea
S_b(z)=\left.\frac{\Gamma_b(z)}{\Gamma_b(Q-z)}\right|_{b=\sqrt{\frac{m}{n}}}=(2 \pi )^{z \sqrt{m n}-\frac{m+n}{2}} \prod _{k=0}^{m-1} \prod _{l=0}^{n-1} \frac{G\left(\frac{k+1}{m}+\frac{l+1}{n}-\frac{z}{\sqrt{m n}}\right)}{G\left(\frac{k}{m}+\frac{l}{n}+\frac{z}{\sqrt{m n}}\right)}  .
\eea
\ee
We can render this formula a bit simpler after using the meromorphic function $$\tilde{G}(x):=\frac{G(1+x)}{G(1-x)}$$
which has poles of order $\mathsf{p}$ at all positive integers $x_{\text{poles}}=\mathsf{p}\in\mathbb{Z}_{>0}$ and zeroes of the same order for all negative integers $x_{\text{zeroes}}=-\mathsf{p}$. We then rewrite the above compactly as
\be\label{eq:sbmn}
\boxed{
\bea
S_b(z)=(2 \pi )^{\sqrt{m n}\left(z-\frac{Q}{2}\right)} \ \tilde{G}_{m,n}\left(\frac{1}{\sqrt{mn}}\left(\frac{Q}{2}-z\right)\right)\ \ , \ \ \ \ \   b=\sqrt{\frac{m}{n}} 
\eea
}  \ ,
\ee
where 
\be\label{eq:tg}
\tilde{G}_{m,n}(x):=\prod _{k=0}^{m-1} \prod _{l=0}^{n-1} \tilde{G}\left(x-x_{(k,l)}^{m,n}\right), \quad \quad x_{(k,l)}^{m,n}\equiv 1-\left(\frac{k+1/2}{m}+\frac{l+1/2}{n}\right).
\ee
For $b=1$ we recover the known expression: $S_{b\rightarrow1}(z)=(2\pi)^{z-1}\tilde{G}(1-z)$, since $\tilde{G}_{1,1}(x)=\tilde{G}(x)$ (and $x^{1,1}_{(0,0)}=0$). In general, for a given coprime pair $(m,n)$ we have $m\times n$ distinct `roots' $x_{(k,l)}^{m,n}$ that satisfy 
\be\label{eq:ineqroots}
\bea
\left|x_{(k,l)}^{m,n}\right|\leq x_{(0,0)}^{m,n}<1, \quad \quad \forall k\in[0,m-1],\forall l\in[0,n-1].
\eea
\ee
Also, for any $m,n$, we can write the compact expression
\be\bea
\tilde{G}_{m,n}(x)=\frac{G_{m,n}\left(\frac{Q}{2\sqrt{mn}}+x\right)}{G_{m,n}\left(\frac{Q}{2\sqrt{mn}}-x\right)}.
\eea
\ee
We find it instructive to mention another equivalent formula to \eqref{eq:sbmn}. From \cite[Theorem 1.9]{Garoufalidis:2014ifa} we can infer\footnote{We use the known relation between the quantum dilogarithm $\Phi_b$ and the function $S_b$:
$$
\Phi_b(z) = e^{\frac{\pi i z^2}{2}} e^{\frac{\pi i}{24}(b^2+b^{-2})} S_b(\tfrac{Q}2+iz)~.
$$.}
\begin{equation}\label{eq:Sbalt0}
    S_b(z)=e^{\frac{\pi i\left(z-Q/2\right)^2}{2}-\frac{\pi i(m^2+n^2)}{24mn}} \times \frac{e^{\frac{i}{2\pi mn}\text{Li}_2\left(e^{-2\pi i \sqrt{mn}z}\right)}\left(1-e^{-2\pi i \sqrt{mn}z}\right)^{1+\frac{z}{\sqrt{mn}}}}{D_n\left(e^{-2\pi i \sqrt{\frac{m}n}z}; e^{2\pi i \frac{m}n}\right)D_m\left(e^{-2\pi i \sqrt{\frac{n}m}z}; e^{2\pi i \frac{n}m}\right)}~,
\end{equation}
whenever $b^2=\frac{m}{n}$ with $m,n$ coprime integers. Here
\begin{equation} \label{defDn}
    D_n(x;q) := \prod_{k=1}^{n} (1-xq^k )^{\frac{k}n}
    \end{equation}
and $\text{Li}_2(z)$ is the usual dilogarithm with the property
\begin{equation}\label{eq:dil} 
    \text{Li}_2(z) + \text{Li}_2(z^{-1}) = -\frac{\pi^2}6 - \frac{\operatorname{log}(-z)^2}{2}.
\end{equation}
It is quite remarkable that the ratio of multivalued functions appearing on the RHS of (\ref{eq:Sbalt0}) ends up giving an overall meromorphic function. 
\par Starting from (\ref{eq:Sbalt0}) we can further rewrite $S_b$ in terms of the Lobachevsky function that appears in the volume of the hyperbolic tetrahedron (note that our definition differs from the standard one by a factor of $\pi$ in the argument of $\Lambda$):
\begin{equation} \label{defLobachevsky}
    \Lambda(z):=\frac{1}{4i}(\text{Li}_2(e^{2\pi i z}) - \text{Li}_2(e^{-2\pi i  z})) = \frac{1}{2} \Im\left(\text{Li}_2(e^{2\pi i  z})\right)~, ~~~ z\in\mathbb{R}.
\end{equation}
Using (\ref{eq:dil}), for $z \in \mathbb R$ we get,
\begin{align} \label{SbintermsofLi2}
  S_b(z) = \; & e^{\frac{\pi i\left(z-Q/2\right)^2}{2}-\frac{\pi i(m^2+n^2+1)}{24mn}} \times \frac{e^{\frac{\Lambda(\sqrt{mn}z)}{\pi mn}} e^{\frac{1}{8\pi i mn}\log^2{\left(-e^{2\pi i \sqrt{mn}z}\right)}}\left(1-e^{-2\pi i \sqrt{mn}z}\right)^{1+\frac{z}{\sqrt{mn}}}}{D_n\left(e^{-2\pi i \sqrt{\frac{m}n}z}; e^{2\pi i \frac{m}n}\right)D_m\left(e^{-2\pi i \sqrt{\frac{n}m}z}; e^{2\pi i \frac{n}m}\right)}~.
\end{align}
Finally, the function $\tilde G_{m,n}(z) = \left.S_b(-z\sqrt{mn} + \tfrac{Q}2)(2\pi)^{mnz}\right|_{b=\sqrt{m/n}}$ can be brought to the following form
\begin{equation} \label{alternativetGmn}
     \tilde G_{m,n}(z) = (2\pi)^{mnz} e^{\frac{\pi i m n z^2}{2} - \frac{\pi i(m^2+n^2+1)}{24mn}}  \frac{e^{\frac{\Lambda(-mnz + \frac{m+n}2)}{\pi m n}} L(z)}{D_n\left(-e^{2\pi i m(z-\frac{1}{2n})}; e^{2\pi i \frac{m}n}\right)D_m\left(-e^{2\pi i n(z-\frac{1}{2m})}; e^{2\pi i \frac{n}m}\right)},
\end{equation}
where
\begin{equation} \label{defL}
    L(z) := e^{\frac{1}{8\pi i mn}\log^2{\left((-1)^{mn}e^{-2\pi i mnz}\right)}}\left(1+(-1)^{mn}e^{2\pi imnz}\right)^{1+\frac{1}{2m}+\frac{1}{2n}-z}.
\end{equation}

\paragraph{Properties of $\tilde{G}_{m,n}$.} Since $b^2\in\mathbb{Q}$ throughout this work, all the expressions for the crossing kernels are written via (\ref{eq:sbmn}) in terms of the meromorphic function $\tilde{G}_{m,n}(z)=\left.S_b(-z\sqrt{mn} + \tfrac{Q}2) (2\pi)^{mnz}\right|_{b=\sqrt{m/n}}$. We will now list some of its useful properties (for any co-prime pair $(m,n)$) that we repeatedly use in the main text.
\begin{itemize}
\item \textit{Complex conjugation:}
\begin{equation}
    \overline{\tilde{G}_{m,n}(x)}=\tilde{G}_{m,n}(\overline{x})~
\end{equation}
which follows from the same property of the Branes' $G$ function. 
\item \textit{Self-duality:} 
\begin{equation}
    \tilde{G}_{m,n}=\tilde{G}_{n,m}~.
\end{equation}
\item \textit{Inverse relation:}
\begin{equation}
   \tilde{G}^{-1}_{m,n}\left(x\right)=\tilde{G}_{m,n}(-x)~.
\end{equation}
\item \textit{Analytic structure:}
\\
The poles and zeroes are dictated from the definition (\ref{eq:tg}), and the fact that $\tilde{G}(x)=\frac{G(1+x)}{G(1-x)}$. Therefore, for a given $(m,n)$ we have
\begin{center}

\begin{tabular}{|c||c|c|}
  \hline
  $\tilde{G}_{m,n}(x)$ & location & order \\
  \hline\hline
  $x_{\text{poles}}$ & $\mathsf{p}+x_{(k,l)}^{m,n}$ & $\mathsf{p}$ \\
  \hline
  $x_{\text{zeroes}}$ & $-\mathsf{p}+x_{(k,l)}^{m,n}$ & $\mathsf{p}$ \\
  \hline
\end{tabular}~
\end{center}
for any $\mathsf{p}\in\mathbb{Z}_{> 0}$ and $\forall k\in[0,m-1],\forall l\in[0,n-1]$.
\item \textit{Asymptotics at large argument:}\\
The asymptotics of $\tilde{G}_{m,n}$ are determined by the ones of $\tilde{G}(x)=\frac{G(1+x)}{G(1-x)}$. For the Barnes' $G$ function we have the following asymptotics as $|z|\rightarrow \infty$ with $|\text{Arg}(z)|<\pi$ \cite{LopezFerreiraDoubleGamma}
\be
\bea
\log{G(1+z)}\sim z\log{\Gamma(z+1)}+\frac{z^2}{4}-\left(\frac{1}{2}z(z+1)+\frac{1}{12}\right)\log{z}-\log{A}+O(1/z^2).
\eea
\ee
$A$ is the so-called Glaisher–Kinkelin constant. Using that we can deduce 
\be
\bea
\log{\tilde{G}(z)}\sim \frac{i\pi}{2}z^2+z\log{(2\pi)}-\frac{i\pi}{12}+O(z^{-1}), \quad \quad \text{as} \ \text{Im}z\rightarrow +\infty.
\eea
\ee
The asymptotics for $\text{Im}z\rightarrow-\infty$ follow easily from $\tilde{G}(-x)=\tilde{G}(x)^{-1}$. Therefore, using the definition (\ref{eq:tg}) and for fixed $m,n$, we find
\begin{align}\label{eq:asymptTGmn}
&\log{\tilde{G}_{m,n}(z)}=\sum_{k=0}^{m-1}\sum_{l=0}^{n-1}\log{\tilde{G}\left(z-x^{m,n}_{(k,l)}\right)}\nonumber\\
& \sim\frac{i\pi mn}{2}z^2+z\left(mn\log{(2\pi)}-i\pi H_{m,n}\right)-\frac{i\pi}{24} \left(\frac{m}{n}+\frac{n}{m}\right)-H_{m,n}\log{(2\pi)}+O(z^{-1}), \nonumber\\
& \quad \quad \quad \quad \quad \quad \quad \quad \quad \quad \quad \quad \quad \quad \quad \quad \quad \quad \quad \quad \quad \quad \quad \quad \ \ \text{as} \ \text{Im}z\rightarrow +\infty.
\end{align}
The constant is defined as
\be
H_{m,n}:=\sum_{k=0}^{m-1}\sum_{l=0}^{n-1}x_{(k,l)}^{m,n}=\frac{mn}{2}+n(1-2^{-m^{-1}})\zeta(-m^{-1})+n\zeta_{H}(-m^{-1};m+1/2)
\ee
where $\zeta(s)$ is the usual Riemann zeta function, and $\zeta_{H}(s;a)=\sum_{n=0}^{\infty}(n+a)^{-s}$ is the Hurwitz zeta function. Again, the asymptotics for $ \text{Im}z\rightarrow -\infty$ are deduced easily from the fact that $\tilde{G}_{m,n}(-x)=\tilde{G}^{-1}_{m,n}\left(x\right)$.
\item \textit{Basic shift relations:}
\begin{align}\label{eq:shifttot}
   & \frac{\tilde{G}_{m,n}(x + \frac{k}{n})}{\tilde{G}_{m,n}(x)} = s_{m,n}(x;k), \qquad \qquad \text{for any} ~ k \in \mathbb Z,
\end{align}
where
\begin{equation}
   s_{m,n}(x;k) =  \frac{(2\pi)^{mk }e^{\pi i m k\left( x+\frac{ k}{2n}\right)}}{\left(-e^{2\pi i m(x+\frac{1}{2n})} ; e^{2\pi i \frac{m}{n}} \right)_k},
\end{equation}
and $(.;.)_k$ is the $q$-Pochhammer symbol $ (x;q)_k \equiv \prod_{l=0}^{k-1} \left(1-x q^l\right)$. We can obtain a similar relation for shifts by $l/m$ (for any $l\in\mathbb{Z}$) simply by exchanging $m\leftrightarrow n$ in (\ref{eq:shifttot}) (and using the self-duality $\tilde{G}_{n,m}=\tilde{G}_{m,n}$). 
\par The function $s_{m,n}(x;k)$ obeys the periodicity $s_{m,n}(x\pm\frac{1}{m};k)=(-1)^ks_{m,n}(x;k)$. Note also that $s_{m,n}\neq s_{n,m}$. Its form is in general a ratio of products of trigonometric functions, e.g. for $k=\pm1$ and $k=\pm n$ we get
\begin{align}\label{eq:speccasestGshift}
    s_{m,n}(x;\pm1)&=\left[\frac{2\cos{\left(\pi m \left(x\pm\frac{1}{2n}\right)\right)}}{(2\pi)^m}\right]^{\mp1},  \\
    s_{m,n}(x;\pm n)&=(-1)^{mn}(2\pi)^{\pm mn}\left[2\cos{mn\pi(x\mp1/2)}\right]^{\mp1}\label{eq:splusn}.
\end{align}
In particular, the latter relation implies the following indentity for the $q$-Pochhammer symbol when $(m,n)$ are co-prime positive integers
\be\label{eq:redidentity}
\bea
1+(-1)^{mn}e^{2\pi i mnx}=\left[\left(-e^{2\pi i m(x+\frac{1}{2n})} ; e^{2\pi i \frac{m}{n}} \right)_{\varepsilon \times n}\right]^{\varepsilon}, \quad \text{for both $\varepsilon=\pm1$}.
\eea
\ee

\item \textit{Almost quasi-periodicity:}\\
From (\ref{eq:shifttot}) we deduce 
\begin{equation}
    \tilde{G}_{m,n}\left(x+\frac{k}{n}\right)\tilde{G}_{m,n}\left(x-\frac{k}{n}\right)=\tilde{G}_{m,n}(x)^2 \times \theta_{m,n}(x;k), \qquad \text{for any} ~ k \in \mathbb Z,
\end{equation} 
where
\begin{equation}
    \theta_{m,n}(x;k):= s_{m,n}(x;k)s_{m,n}(x;-k)=\frac{e^{\frac{\pi i k^2m}{n}}}{\left(-e^{2\pi i m(x+\frac{1}{2n})} ; e^{2\pi i \frac{m}{n}} \right)_k\left(-e^{2\pi i m(x+\frac{1}{2n})} ; e^{2\pi i \frac{m}{n}} \right)_{-k}}~.
\end{equation}
We call this ``almost quasi-periodicity'' as opposed to the actual ``quasi-periodicity'' which holds when $\theta_{m,n}=1$. We record two important examples for $k=1$ and $k=n$:
\begin{equation}
    \theta_{m,n}(x;1)=\frac{\cos{\left(\pi m \left(x-\frac{1}{2n}\right)\right)}}{\cos{\left(\pi m \left(x+\frac{1}{2n}\right)\right)}} \ , \ \  \ \ \ \theta_{m,n}(x;n)=(-1)^{mn}.
\end{equation}

\item \textit{Shift relations (v2):}\\
From (\ref{eq:shifttot}) it is straightforward to deduce
\begin{equation} \label{shifteqkl}
    \frac{\tilde{G}_{m,n}(x + \frac{k}{n} + \frac{l}{m})}{\tilde{G}_{m,n}(x)} = (-1)^{kl} s_{m,n}(x;k) s_{n,m}(x;l), \qquad (k,l) \in \mathbb Z^2.
\end{equation}
An interesting special case of the above is (recall $\mathsf{Q}\equiv \left.Q\right|_{b=\sqrt{\frac{m}{n}}}=\frac{m+n}{\sqrt{mn}}$):
\be\label{eq:qmnshiftgtild}
\bea
\frac{\tilde{G}_{m,n}(x\pm\frac{\mathsf{Q}}{\sqrt{mn}})}{\tilde{G}_{m,n}(x)}=-\left[\frac{4\cos{\pi m (x\pm\frac{1}{2n})}\cos{\pi n (x\pm\frac{1}{2m})}}{(2\pi)^{m+n}}\right]^{\mp1}. 
\eea
\ee
\item \textit{Shift relations (v3):}\\
Finally, notice that for $m$ and $n$ coprime integers, we can always find a pair of integers $(p,q)$ such that $mp + nq = 1$. Hence we can write
\begin{equation}
    \frac{\tilde{G}_{m,n}(x + \frac{k}{mn})}{\tilde{G}_{m,n}(x)} = \frac{\tilde{G}_{m,n}(x + \frac{k(mp + nq)}{mn})}{\tilde{G}_{m,n}(x)} = \frac{\tilde{G}_{m,n}(x + \frac{kp}{n} + \frac{kq}{m})}{\tilde{G}_{m,n}(x)}.
\end{equation}
Applying \eqref{shifteqkl}, we then obtain
\begin{equation} \label{shifteqbykovermn}
    \frac{\tilde{G}_{m,n}(x + \frac{k}{mn})}{\tilde{G}_{m,n}(x)} = (-1)^{k^2 p q} s_{m,n}(x;kp) s_{n,m}(x;kq).
\end{equation}
Remarkably, although there exists an infinite number of pairs $(p,q)$ satisfying $mp+nq=1$, it can be proved that the RHS of \eqref{shifteqbykovermn} does not depend on the choice of pair $(p,q)$ \cite[Lemma 2.2]{Garoufalidis:2014ifa}.
\end{itemize}

\section{Analytic derivations at rational $c\in[25,\infty)$}\label{AppendixB}
In this appendix we will discuss in detail the analytic derivation of the expressions (\ref{eq:splitModmain}), (\ref{eq:splitmodmain2}) and (\ref{eq:splitfusmain}), (\ref{eq:splitfusmain2}) for the modular and fusion kernels when $b=\b$. The main observation is to view the corresponding integrals
\be
\bea
\int_{i\mathbb{R}} \frac{du}{i} \ m_\b(u), \qquad \int_{i\mathbb{R}} \frac{du}{i} \ f_\b(u)
\eea
\ee
as ``state integrals'', and apply the Garoufalidis-Kashaev (GK) lemma described in the introduction.
\subsection{Modular kernel}\label{appb1}
\par
\begin{figure}[h]
\centering
\begin{minipage}{0.42\textwidth}
    \centering
    \input{fig1Contours.tex}
\end{minipage}
\hspace{0.2cm} $\Rightarrow$
\begin{minipage}{0.45\textwidth}
    \centering
    \input{fig2Contours.tex}
\end{minipage}

\vspace{0.5cm} 
$\Rightarrow$\begin{minipage}{0.6\textwidth}
    \centering
    \input{fig3Contours.tex}
\end{minipage}

\caption{Breaking of the integration contour for the integrals (\ref{eq:modkenrbrat}), (\ref{eq:fuskenrbrat}) defining the modular and fusion kernels at $b^2\in\mathbb{Q}$ using the Garoufalidis-Kashaev lemma. The red and green dots represent the two roots of the quantum modular and fusion polynomials defined in (\ref{eq:modpoly}), (\ref{eq:fuspoly}).}\label{fig1}
\end{figure}
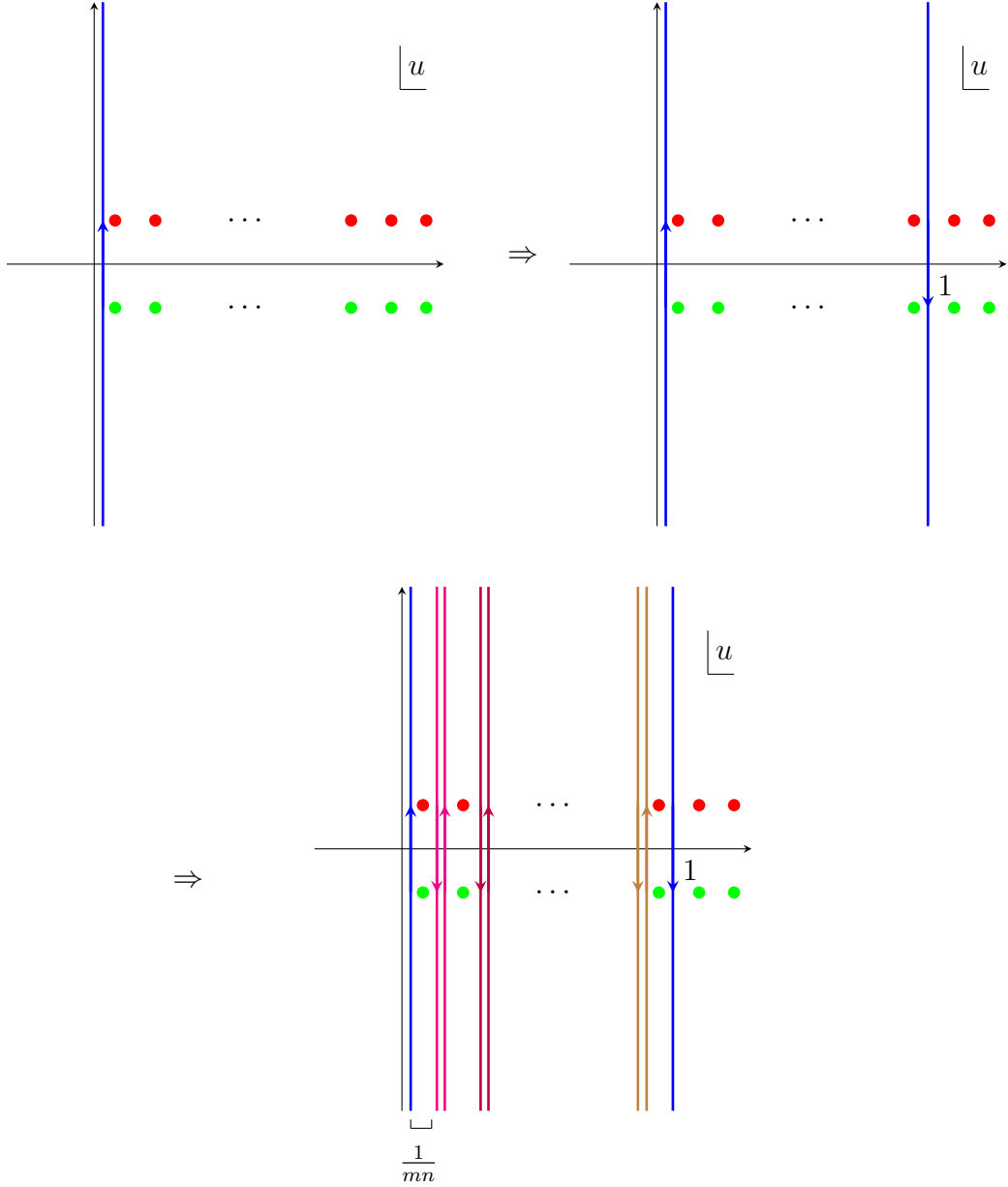

 We start by showing that $m_\b(u)$ defined in (\ref{eq:prefintegdMrational}) is a quasi-periodic function with quasi-period 1, and hence it satisfies
\be\label{eq:qpmod}
\bea
m_\b(u+1)m_\b(u-1)=m_\b(u)^2.
\eea
\ee
This relies on the following shift relation obeyed by $\tilde{G}_{m,n}$ which is discussed in Appendix \ref{appendixA} (see (\ref{eq:shifttot}), (\ref{eq:splusn})),
\be
\bea
\frac{\tilde{G}_{m,n}(x\pm 1)}{\tilde{G}_{m,n}(x)}=(-1)^{mn}(2\pi)^{\pm mn}\left[2\cos{\left(mn\pi(x\mp1/2)\right)}\right]^{\mp1}.
\eea
\ee
Using that one finds
\be
\bea
\frac{m_\b(u\pm1 )}{m_\b(u)}=\left[\frac{m_1(z(u))}{m_2(z(u))}\right]^{\pm1}, \quad z(u)\equiv e^{2\pi i \s^2 u}
\eea
\ee
where
\be\label{eq:m1m2}
\bea
m_1(z):=\frac{e^{\pi i \s \left(a_0-2P_s\right)}}{2i}\prod_{\pm}\left(1+z\times (-1)^{\s^2}e^{2\pi i \s^2 \left(-\frac{m^{-1}+n^{-1}}{4}-\frac{1}{2}\left(\pm\frac{2P_t}{\s}-\frac{P_0}{\s}\right)\right)}\right),\\
m_2(z):=\frac{e^{-\pi i \s \left(a_0-2P_s\right)}}{2i}\prod_{\pm}\left(1+z\times  (-1)^{\s^2}e^{2\pi i \s^2 \left(\frac{m^{-1}+n^{-1}}{4}-\frac{1}{2}\left(\pm\frac{2P_t}{\s}+\frac{P_0}{\s}\right)\right)}\right)
\eea
\ee
and hence (\ref{eq:qpmod}) follows. Crucially, the difference of $m_1,m_2$ is exactly equal to the quantum modular polynomial (\ref{eq:modpoly}),
\be\label{eq:diffm1m2}
\bea
m_2(z)-m_1(z)=P^{(\text{m})}_{mn}(z;\vec{P}).
\eea
\ee
Since (\ref{eq:qpmod}) is satisfied, we can apply the GK lemma to the integral in (\ref{eq:modkenrbrat}), which yields
\be\bea
\int_{i\mathbb{R}} \frac{du}{i} \ m_\b(u)&=\left(\int_{i\mathbb{R}}-\int_{1+i\mathbb{R}}\right)\frac{du}{i} \ \frac{m_\b(u)}{1-\frac{m_1(z(u))}{m_2(z(u))}}\\
&=\left(\int_{i\mathbb{R}}-\int_{1+i\mathbb{R}}\right)\frac{du}{i} \ \frac{\frac{m_1(z(u))}{m_2(z(u))}\times m_{\b}(u-1)}{1-\frac{m_1(z(u))}{m_2(z(u))}}\\
&=\left(\int_{i\mathbb{R}}-\int_{1+i\mathbb{R}}\right)\frac{du}{i} \ \frac{m_1(z(u))\times m_{\b}(u-1)}{P^{(\text{m})}_{mn}(z(u);\vec{P})}.
\eea
\ee
The numerator in the last expression can be simpified to
\be\label{eq:relmM}
\bea
m_1(z(u))\times m_{\b}(u-1)&= \frac{\s^2}{2i}\ z(u) \mathcal{M}_{\b}(u)
\eea
\ee
where $\mathcal{M}_{\b}(u)$ was given in (\ref{eq:modularaction}). From this last relation it is immediate to derive the following useful identity
\be\label{eq:relmM2}
\bea
\frac{\mathcal{M}_{\b}(u\pm 1)}{\mathcal{M}_{\b}(u)}=\left[\frac{m_1(z(u))}{m_2(z(u))}\right]^{\pm1}=\frac{m_\b(u\pm1 )}{m_\b(u)}.
\eea
\ee
We then have
\be\label{eq:semifinalmodint}
\bea
\int_{i\mathbb{R}} \frac{du}{i} \ m_\b(u)&=\frac{\s^2}{2}\left(\int_{1+i\mathbb{R}}-\int_{i\mathbb{R}}\right)du \ \frac{z(u)\mathcal{M}_{\b}(u)}{P^{(\text{m})}_{mn}(z(u);\vec{P})}.
\eea
\ee
A couple of remarks are now in order.
\begin{itemize}
\item \textit{Good behaviour at large imaginary $u$}: We will show that we can safely close the contour in (\ref{eq:semifinalmodint}) in the counterclockwise (i.e. positive convention) orientation due to the fact that the integrand is well-behaved at large positive/negative imaginary values of $u$. Indeed, the factor $\frac{z(u)}{P^{(\text{m})}_{mn}(z(u);\vec{P})}$ is obviously well-behaved. Together with the function $\mathcal{M}_{\b}(u)$, and after using the asymptotics (\ref{eq:asymptTGmn}), we deduce
\be
\bea
\log\frac{z(u)\mathcal{M}_{\b}(u)}{P^{(\text{m})}_{mn}(z(u);\vec{P})}\sim \begin{cases}
-2\pi i \s  \left(2P_s-\frac{m+n}{2\sqrt{mn}}+P_0\right)u + \text{subleading}, \quad \Im(u)\rightarrow +\infty,\\
2\pi i \s \left(-2P_s-\frac{m+n}{2\sqrt{mn}}+P_0\right)u + \text{subleading}, \quad   \Im(u)\rightarrow -\infty.
\end{cases}
\eea
\ee
We see that the integrand decays exponentially at large imaginary $u$ if
\be\label{eq:condMb}
\bea
|\Re P_s|<\frac{1}{2}\left(\frac{m+n}{2\sqrt{mn}}-\Re{P_0}\right).
\eea
\ee
Unsurprisingly, this is precisely the condition ensuring convergence of the original integral representation of the modular kernel (c.f. (\ref{eq:condmodkernelintegral})), except evaluated at $b=\b$. The point then is that we can always arrange an initial choice of the parameters $P_s,P_0$ such that (\ref{eq:condMb}) is satisfied, proceed with the contour manipulations, and at the end of the calculation we will examine the analytic continuation to other values.
Therefore under (\ref{eq:condMb}) we can safely write
\be
\int_{i\mathbb{R}} \frac{du}{i} \ m_\b(u)=\frac{\s^2}{2}\ointctrclockwise_{\mathcal{C}_{[0,1]}} du \ \frac{z(u)\mathcal{M}_{\b}(u)}{P^{(\text{m})}_{mn}(z(u);\vec{P})}.
\ee
\item \textit{Contributing poles only from $P^{(\text{m})}_{mn}$}: We will next show that $\mathcal{M}_{\b}(u)$ does \textit{not} possess any poles in the strip $0\leq\text{Re}(u)\leq1$, and hence the only relevant singularities of the integrand come from the polynomial $P^{(\text{m})}_{mn}$. Indeed using the second expression in (\ref{eq:modularaction}) and the pole structure of $\tilde{G}_{m,n}$ discussed in Appendix \ref{appendixA} we see right away that there are four series\footnote{This is because of the two additional contributions $\pm\frac{2P_t}{\s}$.} of poles for $\mathcal{M}_\b(u)$:
\be\label{eq:zeroesmodkern}
\bea
u^{\mathsf{q};(k,l)}_{\text{poles}}&=\mathsf{q}+x_{(k,l)}^{m,n}-\frac{m^{-1}+n^{-1}}{4}+\frac{1}{2}\left(\pm\frac{2P_t}{\s}+\frac{P_0}{\s}\right),\\
u^{\mathsf{p};(k,l)}_{\text{poles}}&=-\mathsf{p}+x_{(k,l)}^{m,n}+\frac{m^{-1}+n^{-1}}{4}+\frac{1}{2}\left(\pm\frac{2P_t}{\s}-\frac{P_0}{\s}\right),
\eea
\ee
for any pair of positive integers $\mathsf{q}\in\mathbb{Z}_{\geq2}$, $\mathsf{p}\in\mathbb{Z}_{\geq1}$ and $\forall k\in[0,m-1],\forall l\in[0,n-1]$. 
\par We are interested in $\Re (u_{\text{poles}})$. Taking $P_t,P_0\in i \mathbb{R}$, the last two momentum-dependent contributions in (\ref{eq:zeroesmodkern}) are irrelevant in that regard\footnote{Note that this statement has some extended regime of validity for $P_t,P_s$ since we can also have non-zero $\Re P_t, \Re P_0$ such that the location of the poles are always outside the desired strip (without any conflict, also, with the convergence condition (\ref{eq:condMb})).}. It is now easy to see that the low-lying poles -- namely for $\mathsf{q}=2$ and $\mathsf{p}=1$ -- are already \textit{outside} the strip $0\leq \Re (u)\leq 1$, $\forall k\in[0,m-1],\forall l\in[0,n-1]$. This follows from the inequality (\ref{eq:ineqroots}). Using that we find
\be\label{eq:lowlyingmodkern}
\bea
1<1+\frac{m^{-1}+n^{-1}}{4}&\leq \Re\left(u^{\mathsf{q}=2;(k,l)}_{\text{poles}}\right) \leq 3-\frac{3(m^{-1}+n^{-1})}{4},\\
-2+\frac{3(m^{-1}+n^{-1})}{4}&\leq \Re\left(u^{\mathsf{p}=1;(k,l)}_{\text{poles}}\right) \leq -\frac{m^{-1}+n^{-1}}{4}<0
\eea
\ee
which concludes the argument.
\item \textit{Subdivision of the closed contour}: There is one last step to bring the derivation home. As it is clear by now the integral is determined essentially by the polynomial $P^{(\text{m})}_{mn}$. The relevant variable is
\be
\bea
z(u)=e^{2\pi i \s^2 u},
\eea
\ee
which is $\mathbb{Z}/{\s^2}$-periodic, i.e. $z(u+\frac{k}{mn})=z(u)$, for any $k\in\mathbb{Z}$. It therefore makes sense to subdivide further the closed contour into (see Figure \ref{fig1})
\be
\bea
\int_{i\mathbb{R}} \frac{du}{i} \ m_\b(u)&=\frac{\s^2}{2}\sum_{k=0}^{\s^2-1}\ointctrclockwise_{\mathcal{C}_{\left[\frac{k}{\s^2},\frac{k+1}{\s^2}\right]}}du \ \frac{z(u)\mathcal{M}_{\b}(u)}{P^{(\text{m})}_{mn}(z(u);\vec{P})}\\
&=\frac{\s^2}{2}\sum_{k=0}^{\s^2-1}\ointctrclockwise_{\mathcal{C}_{\left[0,\frac{1}{\s^2}\right]}}du\ \frac{z(u)\mathcal{M}_{\b}(u+k/\s^2)}{P^{(\text{m})}_{mn}(z(u);\vec{P})}\\
&=\frac{\s^2}{2}\ointctrclockwise_{\mathcal{C}_{\left[0,\frac{1}{\s^2}\right]}}du\ \frac{z(u)}{P^{(\text{m})}_{mn}(z(u);\vec{P})}\left(\sum_{k=0}^{\s^2-1}\mathcal{M}_{\b}(u+k/\s^2)\right).
\eea
\ee
The last integral is simply picking up the residues at the two roots of the polynomial located in the strip $0\leq \text{Re} u\leq \frac{1}{mn}$, whereas the rest of the singularities in the unit interval simply go along for the ride due to the periodicity in the $z$ variable.   We therefore get
\be
\bea
\int_{i\mathbb{R}} &\frac{du}{i} \ m_\b(u)\\
& \ =\frac{1}{2\sqrt{\Delta^{(\text{m})}}}\sum_{k=0}^{\s^2-1}\mathcal{M}_{\b}\left(\frac{\log{z^{(\text{m})}_1}}{2\pi i \s^2}+\frac{k}{\s^2}\right)-\frac{1}{2\sqrt{\Delta^{(\text{m})}}}\sum_{k=0}^{\s^2-1}\mathcal{M}_{\b}\left(\frac{\log{z^{(\text{m})}_2}}{2\pi i \s^2}+\frac{k}{\s^2}\right)
\eea
\ee
which concludes our derivation.
\end{itemize}

\subsection{Fusion kernel}\label{appb2}
We will next perform the same analysis for the fusion kernel. We start by showing that $f_\b(u)$ defined in (\ref{eq:prefintegdFrational}) is a quasi-periodic function with quasi-period 1, and hence it satisfies
\be\label{eq:qpfus}
\bea
f_\b(u+1)f_\b(u-1)=f_\b(u)^2.
\eea
\ee
This relies again on the identities (\ref{eq:shifttot}), (\ref{eq:splusn}) of $\tilde{G}_{m,n}$. Using that one finds
\be
\bea
\frac{f_\b(u\pm1)}{f_\b(u)}=\left[\frac{f_1(z(u))}{f_2(z(u))}\right]^{\pm1}, \quad z(u)\equiv e^{2\pi i \s^2 u}
\eea
\ee
where
\be
\bea
f_1(z)&:=\frac{(-1)^{mn+1}}{z}\prod_{I\in\{\sigma_E=+1\}}\left(1+z\times (-1)^{\s^2}e^{2\pi i \s^2\left(-\frac{m^{-1}+n^{-1}}{4}-\frac{1}{2}\sum_{i\in E} \sigma^{(I)}_i \frac{P_i}{\s}\right)}\right),\\
f_2(z)&:=\frac{(-1)^{mn+1}}{z}\prod_{J\in\{\sigma_E=-1\}}\left(1+z\times (-1)^{\s^2}e^{2\pi i \s^2\left(\frac{m^{-1}+n^{-1}}{4}-\frac{1}{2}\sum_{i\in E} \sigma^{(J)}_i \frac{P_i}{\s}\right)}\right),\\
\eea
\ee
and hence (\ref{eq:qpfus}) follows. Crucially, the difference of $f_1,f_2$ is exactly equal to the quantum fusion polynomial (\ref{eq:fuspoly})
\be\label{eq:fuspolyapp}
f_2(z)-f_1(z)=P^{(\text{f})}_{mn}(z;\vec{P}).
\ee
Since (\ref{eq:qpfus}) is satisfied, we can apply the GK lemma to the integral in (\ref{eq:fuskenrbrat}) which yields
\be
\bea
 \int_{i\mathbb{R}} \frac{du}{i} \ f_\b(u)&=\left(\int_{i\mathbb{R}}-\int_{1+i\mathbb{R}}\right)\frac{du}{i} \ \frac{f_\b(u)}{1-\frac{f_1(z(u))}{f_2(z(u))}}\\
 &=\left(\int_{i\mathbb{R}}-\int_{1+i\mathbb{R}}\right)\frac{du}{i} \ \frac{\frac{f_1(z(u))}{f_2(z(u))}\times f_\b(u-1)}{1-\frac{f_1(z(u))}{f_2(z(u))}}\\
 &=\left(\int_{i\mathbb{R}}-\int_{1+i\mathbb{R}}\right)\frac{du}{i} \ \frac{f_1(z(u))\times f_\b(u-1)}{P^{(\text{f})}_{mn}(z(u);\vec{P})}.
\eea
\ee
The numerator in the last expression can be simplified to
\be
f_1(z(u))\times f_\b(u-1)=\frac{\s^2}{2i} \ z(u)\mathcal{F}_\b(u)
\ee
where $\mathcal{F}_\b$ was given in (\ref{eq:fusionaction}). From this last relation it is immediate to derive the following useful identity
\be
\bea
\frac{\mathcal{F}_{\b}(u\pm 1)}{\mathcal{F}_{\b}(u)}=\left[\frac{f_1(z(u))}{f_2(z(u))}\right]^{\pm1}=\frac{f_\b(u\pm1 )}{f_\b(u)}.
\eea
\ee
We then have
\be\label{eq:fusionalmfinal}
 \int_{i\mathbb{R}} \frac{du}{i} \ f_\b(u)=\frac{\s^2}{2}\left(\int_{1+i\mathbb{R}}-\int_{i\mathbb{R}}\right)du \ \frac{z(u)\mathcal{F}_\b(u)}{P^{(\text{f})}_{mn}(z(u);\vec{P})}.
\ee
Just like in the case of the modular kernel, we next make the following remarks.
\begin{itemize}
\item \textit{Good behaviour at large imaginary $u$}: It is straightforward to check that the integrand in (\ref{eq:fusionalmfinal}) behaves exactly as in (\ref{eq:condfuskernelintegral}) (with $Q\equiv \mathsf{Q}=\frac{m+n}{\sqrt{mn}}$) as $u$ attains large positive/negative imaginary part. Therefore, it is safe to close the contour in the counterclockwise fashion (positive convention) at infinity, and write
\be
\bea
 \int_{i\mathbb{R}} \frac{du}{i} \ f_\b(u)=\frac{\s^2}{2}\ointctrclockwise_{\mathcal{C}_{[0,1]}} du \ \frac{z(u)\mathcal{F}_\b(u)}{P^{(\text{f})}_{mn}(z(u);\vec{P})}.
\eea
\ee
\item \textit{Contributing poles only from $P^{(\text{f})}_{mn}$}: Just like in the case of the modular kernel, we will next show that $\mathcal{F}_{\b}(u)$ does \textit{not} possess any poles in the strip $0\leq\text{Re}(u)\leq1$, and hence the only relevant singularities of the integrand come from the polynomial $P^{(\text{f})}_{mn}$. Indeed using the second expression in (\ref{eq:fusionaction}) and the pole structure of $\tilde{G}_{m,n}$ discussed in Appendix \ref{appendixA} we see right away that there are eight series of poles for $\mathcal{F}_\b(u)$:
\be\label{eq:polesmathcalFb}
\bea
u^{\mathsf{q};(k,l)|(J)}_{\text{poles}}&=\mathsf{q}+x_{(k,l)}^{m,n}-\frac{m^{-1}+n^{-1}}{4}+\frac{1}{2}\sum_{J\in\{\sigma_E=-1\}}\sigma^{(J)}_i \frac{P_i}{\s},\\
u^{\mathsf{p};(k,l)|(I)}_{\text{poles}}&=-\mathsf{p}+x_{(k,l)}^{m,n}+\frac{m^{-1}+n^{-1}}{4}+\frac{1}{2}\sum_{I\in\{\sigma_E=+1\}}\sigma^{(I)}_i \frac{P_i}{\s},
\eea
\ee
for any pair of positive integers $\mathsf{q}\in\mathbb{Z}_{\geq2}$, $\mathsf{p}\in\mathbb{Z}_{\geq1}$ and $\forall k\in[0,m-1],\forall l\in[0,n-1]$. Note that the label $(J)$ includes four terms associated to the first four rows of table \ref{tbl:2kalos}, and similarly the label $(I)$ includes four terms associated to the signs in the last four rows of the same table.\\
Again, what matters is $\Re (u_{\text{poles}})$. Taking all $P_{s,t,1,\cdots, 4}\in i \mathbb{R}$, the contributions from the sums in (\ref{eq:polesmathcalFb}) are irrelevant. With that, the low-lying poles behave exactly as in (\ref{eq:lowlyingmodkern}) for the modular kernel, and hence we conclude that there are \textit{no} poles of $\mathcal{F}_\b$ inside the contour of integration.
\item \textit{Subdivision of the closed contour}: Finally, we subdivide the contour exactly as in the case of the modular kernel to obtain
\be
\bea
\int_{i\mathbb{R}} \frac{du}{i} \ f_\b(u)&=\frac{\s^2}{2}\ointctrclockwise_{\mathcal{C}_{\left[0,\frac{1}{\s^2}\right]}}du\ \frac{z(u)}{P^{(\text{f})}_{mn}(z(u);\vec{P})}\left(\sum_{k=0}^{\s^2-1}\mathcal{F}_{\b}(u+k/\s^2)\right).
\eea
\ee
The last integral is simply picking up the residues at the two roots of the polynomial located in the strip $0\leq \text{Re} u\leq \frac{1}{mn}$ and, just like in the case of the modular kernel, the rest of the singularities in the unit interval simply come due to the periodicity in the $z$ variable. We therefore get
\be
\bea
\int_{i\mathbb{R}} &\frac{du}{i} \ f_\b(u)\\
& \ =\frac{1}{2\sqrt{\Delta^{\text{(f)}}}}\sum_{k=0}^{\s^2-1}\mathcal{F}_{\b}\left(\frac{\log{z_1^{(\text{f})}}}{2\pi i \s^2}+\frac{k}{\s^2}\right)-\frac{1}{2\sqrt{\Delta^{\text{(f)}}}}\sum_{k=0}^{\s^2-1}\mathcal{F}_{\b}\left(\frac{\log{z_2^{(\text{f})}}}{2\pi i \s^2}+\frac{k}{\s^2}\right)
\eea
\ee
which concludes the derivation.
\end{itemize}

\section{Proof of the shift relation for $\mathbf{F}^{(\pm)}$}\label{app:shiftrelFpm}
In this appendix we prove that the shift relation (\ref{eq:mainshiftfusiona}) at $b=\b$, which we repeat here for convenience
\begin{multline}
    \sum_{\eta=\pm} \eta \cos\big(\pi \b(P_1+\eta P_4+P_t)\big)\cos\big(\pi \b(P_2+\eta P_3-P_t)\big) \sixjnorm{P_1}{P_2}{P_s}{P_3}{P_4}{P_t + \frac{ \eta \b}{2}}_{(\b)} \\
    =\sin(2\pi \b P_t) \cos\big(\tfrac{\pi \b^2}{2}+\pi \b(-P_s+P_3-P_4)\big) \sixjnorm{P_1}{P_2}{P_s}{P_3+\frac{\b}{2}}{P_4-\frac{\b}{2}}{P_t}_{(\b)}\  \label{eq:mainshiftfusionapp},
\end{multline}
is satisfied for either
\begin{align}\label{6j1}
\sixjnorm{P_1}{P_2}{P_{s}}{P_3}{P_4}{P_{t}}_{(\b)}^{(1)}&\equiv \frac{1}{4is_{mn}(P_t)s_{mn}(P_s)\sqrt{\mathfrak{D}^{(\text{f})}}}\sum_{k=0}^{\s^2-1}\mathcal{F}_{\b}\left(\frac{\log{z_1^{(\text{f})}}}{2\pi i \s^2}+\frac{k}{\s^2}\right),\\
\sixjnorm{P_1}{P_2}{P_{s}}{P_3}{P_4}{P_{t}}_{(\b)}^{(2)} &\equiv -\frac{1}{4is_{mn}(P_t)s_{mn}(P_s)\sqrt{\mathfrak{D}^{(\text{f})}}}\sum_{k=0}^{\s^2-1}\mathcal{F}_{\b}\left(\frac{\log{z_2^{(\text{f})}}}{2\pi i \s^2}+\frac{k}{\s^2}\right)\label{6j2}.
\end{align}
\textit{Proof}: Without loss of generality we consider $m\in\mathbb{Z}_{\text{odd}}\geq 1$. From the corresponding definitions it is straightforward to check the following relations:
\be
\bea
\left[\alpha^{(\text{f})}_{mn}\right]_{P_t\rightarrow P_t+\frac{\eta\b}{2}}&=e^{\frac{\eta\pi im}{2}}\left(\alpha^{(\text{f})}_{mn}\right)^{'},\\
\left[\beta^{(\text{f})}_{mn}\right]_{P_t\rightarrow P_t+\frac{\eta\b}{2}}&=-\left(\beta^{(\text{f})}_{mn}\right)^{'},\\
\left[\mathfrak{D}^{(\text{f})}\right]_{P_t\rightarrow P_t+\frac{\eta\b}{2}}&=\left(\mathfrak{D}^{(\text{f})}\right)^{'}, \qquad \qquad\ \eta=\pm,
\eea
\ee
where for brevity we called $\left(\cdot \right)^{'}\equiv \left[\ \cdot \ \right]_{P_3\rightarrow P_3+\frac{\b}{2},P_4\rightarrow P_4-\frac{\b}{2}}.$
Therefore the roots transform as
\be
\bea
\left[z_{1,2}^{(\text{f})}\right]_{P_t\rightarrow P_t+\frac{\eta\b}{2}}&=e^{\pi i \left(1-\frac{\eta m}{2}\right)}\left(z_{1,2}^{(\text{f})}\right)^{'}.
\eea
\ee
In addition, the function $\mathcal{F}_\b$ for $u\in\mathbb{C}$ behaves as
\be\label{eq:Fbtransf}
\bea
\left.\mathcal{F}_\b\left(u\right)\right|_{P_t\rightarrow P_t+\frac{\eta \b}{2}}=-\mathcal{F}^{'}_\b\left(u-\frac{\eta}{4n}\right)\times g_\eta \left(u-\frac{\eta}{4n}\right)
\eea
\ee
with
\be
\bea
g_\eta (u)=\begin{cases}
&\frac{\cos{\left(\pi \b (\s u-\mathsf{Q}/4-\frac{1}{2}P_{13|24st})\right)}}{\cos{\left(\pi \b (\s u +\mathsf{Q}/4-\frac{1}{2}P_{23s|14t})\right)}}, \qquad \ \  \eta=+1 \\
\\
&\frac{\cos{\left(\pi \b (\s u+\mathsf{Q}/4-\frac{1}{2}P_{14s|23t})\right)}}{\cos{\left(\pi \b (\s u-\mathsf{Q}/4-\frac{1}{2}P_{24|13st})\right)}}, \qquad \ \ \eta=-1 .
\end{cases}
\eea
\ee
Here we adopted the convenient notation $P_{I|J}\equiv \sum_{i\in I}P_i-\sum_{j\in J}P_j$ that we mentioned in the introduction, and to arrive at (\ref{eq:Fbtransf}) we used the identity (\ref{eq:shifttot}) with $k=\pm1$ .
\par Let us next consider the case (\ref{6j1}) associated to the root $z_1^{(\text{f})}$. We denote
\be
\bea
u^*_{(k)}\equiv \frac{\log{\left(z_1^{(\text{f})}\right)^{'}}}{2\pi i \s^2}+\frac{k}{\s^2}.
\eea
\ee
Putting everything together (we strip off the overall factor $4is_{mn}(P_t)s_{mn}(P_s)\sqrt{\left(\mathfrak{D}^{(\text{f})}\right)^{'}}$), the LHS of (\ref{eq:mainshiftfusionapp}) yields
\begin{multline}
  \text{LHS}=\bigg[ \cos\big(\pi \b P_{14t}\big)\cos\big(\pi \b P_{23|t}\big) \sum_{k=0}^{\s^2-1}g_{+}\left(u^*_{(k)}-\frac{\mathsf{m}}{\s^2}\right)\mathcal{F}^{'}_{\b}\left(u^*_{(k)}-\frac{\mathsf{m}}{\s^2}\right) \\
  - \cos\big(\pi \b P_{1t|4}\big)\cos\big(\pi \b P_{2|3t}\big)\sum_{k=0}^{\s^2-1}g_{-}\left(u^*_{(k)}+\frac{\mathsf{m}+1}{\s^2}\right)\mathcal{F}^{'}_{\b}\left(u^*_{(k)}+\frac{\mathsf{m}+1}{\s^2}\right) \bigg],\label{eq:provingFa}
\end{multline}
where $\mathsf{m}\equiv \frac{m-1}{2}\in\mathbb{Z}_{\geq0}$. 
\par We now observe that we can use the main periodicity property (\ref{eq:periodsumFStrong}) with the choices $f(x)\equiv g_{+}(x/\pi m)$ and $f(x)\equiv g_{-}(x/\pi m)$ for the corresponding sums\footnote{Notice that there is no worry in using (\ref{eq:periodsumFStrong}) with $\mathcal{F}'_\b$ instead of the usual $\mathcal{F}_\b$, since we have just relabelled the momenta consistently throughout.}. Note that both of these choices are $\mathbb{Z}\pi-$periodic, i.e. $f(x+\mathbb{Z}\pi)=f(x)$. Therefore (\ref{eq:provingFa}) becomes
\begin{multline}
\text{LHS}=\sin(2\pi \b P_t) \cos\big(\tfrac{\pi \b^2}{2}+\pi \b P_{3|4s}\big) \sum_{k=0}^{\s^2-1}\mathcal{F}^{'}_{\b}\left(u^*_{(k)}\right)\\
 +\sum_{k=0}^{\s^2-1}\mathcal{F}^{'}_{\b}\left(u^*_{(k)}\right)\bigg[-\sin(2\pi \b P_t) \cos\big(\tfrac{\pi \b^2}{2}+\pi \b P_{3|4s}\big)\\
 -\cos\big(\pi \b P_{1t|4}\big)\cos\big(\pi \b P_{2|3t}\big)g_{-}\left(u^*_{(k)}\right)+\cos\big(\pi \b P_{14t}\big)\cos\big(\pi b P_{23|t}\big)g_{+}\left(u^*_{(k)}\right)\bigg]\label{eq:provingFb}. 
\end{multline}
Here we have added and subtracted a factor of $\sin(2\pi \b P_t) \cos\big(\tfrac{\pi \b^2}{2}+\pi \b P_{3|4s}\big) $ inside the sum, and hence the first line is exactly where we want to arrive. We will next show that the rest of the terms evaluate to zero. For that, we first make use of the following trigonometric identity valid for arbitrary $u,P_1,P_2\in\mathbb{C}$:
\be
-\sin(2\pi \b P_t) \cos\big(\tfrac{\pi \b^2}{2}+\pi \b P_{3|4s}\big)=T^{(\b)}_{34ts}(u,P_1,P_2)
\ee
where\footnote{A similiar trigonometric identity was used in \cite{Eberhardt:2023mrq} to prove exactly the same shift relation for general $b^2\in\mathbb{C}\backslash(-\infty,0))$ (see section 3.6. of the paper). Our identity is analogous to that one, except adjusted to the setting where $b=\b$. Indeed, to translate between our expression and the one described in eqns (3.52)-(3.54) of \cite{Eberhardt:2023mrq} one simply uses the change of variables $p\equiv -\s u+\frac{\b+\b^{-1}}{4}+\frac{1}{2}P_{1234st}$.}
\begin{multline}
T^{(\b)}_{34ts}(u,P_1,P_2):=\cos (\pi  \b P_{1t|4}) \cos (\pi  \b P_{2|3t})g_{-}(u)-\cos (\pi  \b P_{14t}) \cos (\pi  \b P_{23|t})g_{+}(u)\\
-\sin \left(\pi  \b \left(\s u -\mathsf{Q}/4-P_{34t|12s}/2\right)\right) \sin \left(\pi  \b \left(\s u -\mathsf{Q}/4-P_{12t|34s}/2\right)\right)g_{-}(u)\\
+\sin \left(\pi  \b \left(\s u +\mathsf{Q}/4-P_{1234st}/2\right)\right) \sin \left(\pi  \b \left(\s u +\mathsf{Q}/4+P_{1234|st}/2\right)\right) g_{+}(u).
\end{multline}
Evaluating this at $u=u^*_{(k)}$ and plugging it back to (\ref{eq:provingFb}) we get
\begin{multline}
\text{LHS}=\sin(2\pi \b P_t) \cos\big(\tfrac{\pi \b^2}{2}+\pi \b P_{3|4s}\big) \sum_{k=0}^{\s^2-1}\mathcal{F}^{'}_{\b}\left(u^*_{(k)}\right)+\sum_{k=0}^{\s^2-1}\mathcal{F}^{'}_{\b}\left(u^*_{(k)}\right)\bigg[t_{+}(u^*_{(k)})- \ t_{-}(u^*_{(k)})\bigg],
\end{multline}
where we defined the trigonometric combinations
\be
\bea
t_{+}(u)&:=\sin \left(\pi  \b \left(\s u +\mathsf{Q}/4-P_{1234st}/2\right)\right) \sin \left(\pi  \b \left(\s u +\mathsf{Q}/4-P_{st|1234}/2\right)\right)g_{+}(u),\\
t_{-}(u)&:=\sin \left(\pi  \b \left(\s u -\mathsf{Q}/4-P_{34t|12s}/2\right)\right) \sin \left(\pi  \b \left(\s u -\mathsf{Q}/4-P_{12t|34s}/2\right)\right)g_{-}(u).
\eea
\ee
To see now that the second sum is zero we define
\be
\bea
&\mathcal{F}_\b^{''}(u):=\frac{\tilde{G}_{m,n}\left(u-1+\frac{m^{-1}+n^{-1}}{4}-\frac{P_{34t|12s}}{2\s}\right)\tilde{G}_{m,n}\left(u-1+\frac{m^{-1}+n^{-1}}{4}-\frac{P_{12t|34s}}{2\s}\right)}{\tilde{G}_{m,n}\left(u+\frac{1}{n}-\frac{m^{-1}+n^{-1}}{4}-\frac{P_{st|1234}}{2\s}\right)\tilde{G}_{m,n}\left(u+\frac{1}{n}-\frac{m^{-1}+n^{-1}}{4}-\frac{P_{1234st}}{2\s}\right)}\\
&\quad\times \frac{\tilde{G}_{m,n}\left(u+\frac{1}{2n}-1+\frac{m^{-1}+n^{-1}}{4}-\frac{P_{23s|14t}}{2\s}\right)\tilde{G}_{m,n}\left(u+\frac{1}{2n}-1+\frac{m^{-1}+n^{-1}}{4}-\frac{P_{14s|23t}}{2\s}\right)}{\tilde{G}_{m,n}\left(u+\frac{1}{2n}-\frac{m^{-1}+n^{-1}}{4}-\frac{P_{13|24st}}{2\s}\right)\tilde{G}_{m,n}\left(u+\frac{1}{2n}-\frac{m^{-1}+n^{-1}}{4}-\frac{P_{24|13st}}{2\s}\right)}.
\eea
\ee
Using the functional relation of $\tilde{G}_{m,n}(x)$ under shifts by $1/n$ (c.f. (\ref{eq:shifttot}), (\ref{eq:speccasestGshift})) we observe that 
\begin{align}\label{id1}
t_{+}(u^*_{(k)})\mathcal{F}^{'}_{\b}\left(u^*_{(k)}\right)&=\frac{(-1)^m(2\pi)^{2m}}{4}\mathcal{F}_\b^{''}(u^*_{(k)}),\\
t_{-}(u^*_{(k)})\mathcal{F}^{'}_{\b}\left(u^*_{(k)}\right)&=\frac{(-1)^m(2\pi)^{2m}}{4}\mathcal{F}_\b^{''}(u^*_{(k)}-1/n)\label{id2}.
\end{align}
and hence
\be
\bea
\text{LHS}&=\sin(2\pi \b P_t) \cos\big(\tfrac{\pi \b^2}{2}+\pi \b P_{3|4s}\big) \sum_{k=0}^{\s^2-1}\mathcal{F}^{'}_{\b}\left(u^*_{(k)}\right)\\
&\qquad +\frac{(-1)^m(2\pi)^{2m}}{4}\sum_{k=0}^{\s^2-1}\left[\mathcal{F}^{''}_{\b}\left(u^*_{(k)}\right)-\mathcal{F}^{''}_{\b}\left(u^*_{(k)}-1/n\right)\right]. 
\eea
\ee
We are almost done since here we \textit{cannot} simply use the periodicity property (\ref{eq:periodsumFStrong}) to argue that the difference in the second line is zero; indeed, the argument $u^*_{(k)}$ is evaluated at the configuration with $P_3\rightarrow P_3+\b/2,P_4\rightarrow P_4-\b/2$, whereas $\mathcal{F}^{''}_{\b}$ is not associated with that configuration. To proceed, from (\ref{id1}) and the fact that $\mathcal{F}^{'}_{\b}\left(u^*_{(k)}\right)$ satisfies the periodicity (\ref{eq:periodicityFb}), we notice that $\mathcal{F}_\b^{''}$ satisfies the following periodicity property:
\be\label{eq:crucialid}
\bea
\mathcal{F}_\b^{''}(u^*_{(k)}+l)=\mathcal{F}_\b^{''}(u^*_{(k)}) , \qquad \forall l\in\mathbb{Z}.
\eea
\ee
We emphasize that this is true only for $u=u^*_{(k)}$, and not for arbitrary $u$.
\par It is now evident that, after rearranging the sum and using (\ref{eq:crucialid}), the second term is identically zero and hence
\be
\bea
{}&\text{LHS}=\sin(2\pi \b P_t) \cos\big(\tfrac{\pi \b^2}{2}+\pi \b P_{3|4s}\big) \sum_{k=0}^{\s^2-1}\mathcal{F}^{'}_{\b}\left(u^*_{(k)}\right). \quad \quad \quad \square
\eea
\ee
\par As in the case of the modular kernel, the proof does not distinguish between the two roots $z^{(\text{f})}_{1},z^{(\text{f})}_2$. The only important information was that we had a solution to the quantum fusion polynomial, and hence one works identically to show that (\ref{6j2}) also solves the same shift relation. This concludes our proof. 
\end{appendix}

\bibliographystyle{JHEP}
\bibliography{rationalkernels}
\end{document}

%% file: fig1Contours.tex
\begin{tikzpicture}[>=stealth, scale=1.2]

  \draw[->] (-3,0) -- (2,0) node[right] {};
  \draw[->] (-2,-3) -- (-2,3) node[above] {};

  \draw[line width=1pt, blue] (-1.9,-3) -- (-1.9,3);

 \draw (1.5,2) -- (1.8,2); 
  \draw (1.5,2) -- (1.5,2.5); 
  \node[anchor=west] at (1.47,2.25) {$u$};
 
  \draw[->, blue, line width=1pt] (-1.9,-0.5) -- (-1.9,0.5);

  \node[blue, anchor=west] at (0.1,2.8) {};

 \fill[red] (-1.76,0.5) circle (2pt);
\fill[green] (-1.76,-0.5) circle (2pt);

\fill[red] (-1.3,0.5) circle (2pt);
\fill[green] (-1.3,-0.5) circle (2pt);

\node[below right] at (-0.6,-0.3) {$\cdots$};
\node[below right] at (-0.6,0.7) {$\cdots$};

\fill[red] (0.94,0.5) circle (2pt);
\fill[green] (0.94,-0.5) circle (2pt);

\fill[red] (1.4,0.5) circle (2pt);
\fill[green] (1.4,-0.5) circle (2pt);

\fill[red] (1.8,0.5) circle (2pt);
\fill[green] (1.8,-0.5) circle (2pt);

\end{tikzpicture}

%% file: fig2Contours.tex
\begin{tikzpicture}[>=stealth, scale=1.2]

   \draw[->] (-3,0) -- (2,0) node[right] {};
  \draw[->] (-2,-3) -- (-2,3) node[above] {};

  \draw (1.5,2) -- (1.8,2); 
  \draw (1.5,2) -- (1.5,2.5); 
  \node[anchor=west] at (1.47,2.25) {$u$};
 
   \draw[line width=1pt, blue] (-1.9,-3) -- (-1.9,3);
  \draw[->, blue, line width=1pt] (-1.9,-0.5) -- (-1.9,0.5);
    \node[blue, anchor=west] at (0.1,2.8) {};

\draw (1.1,0.1) -- (1.1,-0.1);
  \node[below] at (1.3,-0.01) {$1$};
 
  \draw[line width=1pt, blue] (1.1,-3) -- (1.1,3);
  \draw[->, blue, line width=1pt] (1.1,0.5) -- (1.1,-0.5); 
  \node[red, anchor=west] at (1,2.8) {};

 \fill[red] (-1.76,0.5) circle (2pt);
\fill[green] (-1.76,-0.5) circle (2pt);

\fill[red] (-1.3,0.5) circle (2pt);
\fill[green] (-1.3,-0.5) circle (2pt);

\node[below right] at (-0.6,-0.3) {$\cdots$};
\node[below right] at (-0.6,0.7) {$\cdots$};

\fill[red] (0.94,0.5) circle (2pt);
\fill[green] (0.94,-0.5) circle (2pt);

\fill[red] (1.4,0.5) circle (2pt);
\fill[green] (1.4,-0.5) circle (2pt);

\fill[red] (1.8,0.5) circle (2pt);
\fill[green] (1.8,-0.5) circle (2pt);

\end{tikzpicture}

%% file: fig3Contours.tex
\begin{tikzpicture}[>=stealth, scale=1.2]

 \draw[->] (-3,0) -- (2,0) node[right] {};
  \draw[->] (-2,-3) -- (-2,3) node[above] {};

  \draw (1.5,2) -- (1.8,2); 
  \draw (1.5,2) -- (1.5,2.5); 
  \node[anchor=west] at (1.47,2.25) {$u$};
 
   \draw[line width=1pt, blue] (-1.9,-3) -- (-1.9,3);
  \draw[->, blue, line width=1pt] (-1.9,-0.5) -- (-1.9,0.5);
    \node[blue, anchor=west] at (0.1,2.8) {};

\draw (1.1,0.1) -- (1.1,-0.1);
  \node[below] at (1.3,-0.01) {$1$};
 
  \draw[line width=1pt, blue] (1.1,-3) -- (1.1,3);
  \draw[->, blue, line width=1pt] (1.1,0.5) -- (1.1,-0.5); 
  \node[red, anchor=west] at (1,2.8) {};

   \draw (-1.9,-3.2) -- (-1.66,-3.2); 
   \draw (-1.9,-3.2) -- (-1.9,-3.1);
   \draw (-1.66,-3.2) -- (-1.66,-3.1);
  \node[anchor=west] at (-2.15,-3.6) {$\frac{1}{mn}$};

  \draw[line width=1pt, magenta] (-1.6,-3) -- (-1.6,3);
  \draw[->, magenta, line width=1pt] (-1.6,0.5) -- (-1.6,-0.5); 
  \node[red, anchor=west] at (1,2.8) {};

  \draw[line width=1pt, magenta] (0.49-1.2-1.2+0.4,-3) -- (0.49-1.2-1.2+0.4,3);
  \draw[<-, magenta, line width=1pt] (0.49-1.2-1.2+0.4,0.5) -- (0.49-1.2-1.2+0.4,-0.5); 
  \node[red, anchor=west] at (1,2.8) {};

  \draw[line width=1pt, purple] (0.9-2,-3) -- (0.9-2,3);
  \draw[->, purple, line width=1pt] (0.9-2,0.5) -- (0.9-2,-0.5); 
  \node[red, anchor=west] at (1,2.8) {};

  \draw[line width=1pt, purple] (0.99-2,-3) -- (0.99-2,3);
  \draw[<-, purple, line width=1pt] (0.99-2,0.5) -- (0.99-2,-0.5); 
  \node[red, anchor=west] at (1,2.8) {};



\draw[line width=1pt, brown] (0.7,-3) -- (0.7,3);
  \draw[->, brown, line width=1pt] (0.7,0.5) -- (0.7,-0.5); 
  \node[red, anchor=west] at (1,2.8) {};

  \draw[line width=1pt, brown] (0.8,-3) -- (0.8,3);
  \draw[<-, brown, line width=1pt] (0.8,0.5) -- (0.8,-0.5); 
  \node[red, anchor=west] at (1,2.8) {};

  \fill[red] (-1.76,0.5) circle (2pt);
\fill[green] (-1.76,-0.5) circle (2pt);

\fill[red] (-1.3,0.5) circle (2pt);
\fill[green] (-1.3,-0.5) circle (2pt);

\node[below right] at (-0.6,-0.3) {$\cdots$};
\node[below right] at (-0.6,0.7) {$\cdots$};

\fill[red] (0.94,0.5) circle (2pt);
\fill[green] (0.94,-0.5) circle (2pt);

\fill[red] (1.4,0.5) circle (2pt);
\fill[green] (1.4,-0.5) circle (2pt);

\fill[red] (1.8,0.5) circle (2pt);
\fill[green] (1.8,-0.5) circle (2pt);

\end{tikzpicture}